\documentclass[sigconf]{acmart}

\AtBeginDocument{%
  \providecommand\BibTeX{{%
    \normalfont B\kern-0.5em{\scshape i\kern-0.25em b}\kern-0.8em\TeX}}}

\copyrightyear{2023}
\acmYear{2023}
\setcopyright{acmlicensed}\acmConference[CHI '23]{Proceedings of the 2023 CHI Conference on Human Factors in Computing Systems}{April 23--28, 2023}{Hamburg, Germany}
\acmBooktitle{Proceedings of the 2023 CHI Conference on Human Factors in Computing Systems (CHI '23), April 23--28, 2023, Hamburg, Germany}
\acmPrice{15.00}
\acmDOI{10.1145/3544548.3581290}
\acmISBN{978-1-4503-9421-5/23/04}

\newcommand{\rev}[1]{{{#1}}} 
\newcommand{\cut}[1]{{}} 

\newcommand{\revFinal}[1]{{{#1}}} 

\usepackage{multirow}
\usepackage{enumitem}
\usepackage{listings}
\usepackage{xcolor}
\usepackage{acmart-taps}

\definecolor{maroon}{cmyk}{0, 0.87, 0.68, 0.32}
\definecolor{halfgray}{gray}{0.55}
\definecolor{ipython_frame}{RGB}{207, 207, 207}
\definecolor{ipython_bg}{RGB}{247, 247, 247}
\definecolor{ipython_red}{RGB}{186, 33, 33}
\definecolor{ipython_green}{RGB}{0, 128, 0}
\definecolor{ipython_cyan}{RGB}{64, 128, 128}
\definecolor{ipython_purple}{RGB}{170, 34, 255}
\lstdefinelanguage{Markdown}{
    basicstyle=\ttfamily\footnotesize,
}
\lstdefinelanguage{iPython}{
    morekeywords={access,and,break,class,continue,def,del,elif,else,except,exec,finally,for,from,global,if,import,in,is,lambda,not,or,pass,print,raise,return,try,while},%
    morekeywords=[2]{abs,all,any,basestring,bin,bool,bytearray,callable,chr,classmethod,cmp,compile,complex,delattr,dict,dir,divmod,enumerate,eval,execfile,file,filter,float,format,frozenset,getattr,globals,hasattr,hash,help,hex,id,input,int,isinstance,issubclass,iter,len,list,locals,long,map,max,memoryview,min,next,object,oct,open,ord,pow,property,range,raw_input,reduce,reload,repr,reversed,round,set,setattr,slice,sorted,staticmethod,str,sum,super,tuple,type,unichr,unicode,vars,xrange,zip,apply,buffer,coerce,intern},%
    sensitive=true,%
    morecomment=[l]\#,%
    morestring=[b]',%
    morestring=[b]",%
    morestring=[s]{'''}{'''},
    morestring=[s]{"""}{"""},
    morestring=[s]{r'}{'},
    morestring=[s]{r"}{"},%
    morestring=[s]{r'''}{'''},%
    morestring=[s]{r"""}{"""},%
    morestring=[s]{u'}{'},
    morestring=[s]{u"}{"},%
    morestring=[s]{u'''}{'''},%
    morestring=[s]{u"""}{"""},%
    literate=
    {á}{{\'a}}1 {é}{{\'e}}1 {í}{{\'i}}1 {ó}{{\'o}}1 {ú}{{\'u}}1
    {Á}{{\'A}}1 {É}{{\'E}}1 {Í}{{\'I}}1 {Ó}{{\'O}}1 {Ú}{{\'U}}1
    {à}{{\`a}}1 {è}{{\`e}}1 {ì}{{\`i}}1 {ò}{{\`o}}1 {ù}{{\`u}}1
    {À}{{\`A}}1 {È}{{\'E}}1 {Ì}{{\`I}}1 {Ò}{{\`O}}1 {Ù}{{\`U}}1
    {ä}{{\"a}}1 {ë}{{\"e}}1 {ï}{{\"i}}1 {ö}{{\"o}}1 {ü}{{\"u}}1
    {Ä}{{\"A}}1 {Ë}{{\"E}}1 {Ï}{{\"I}}1 {Ö}{{\"O}}1 {Ü}{{\"U}}1
    {â}{{\^a}}1 {ê}{{\^e}}1 {î}{{\^i}}1 {ô}{{\^o}}1 {û}{{\^u}}1
    {Â}{{\^A}}1 {Ê}{{\^E}}1 {Î}{{\^I}}1 {Ô}{{\^O}}1 {Û}{{\^U}}1
    {œ}{{\oe}}1 {Œ}{{\OE}}1 {æ}{{\ae}}1 {Æ}{{\AE}}1 {ß}{{\ss}}1
    {ç}{{\c c}}1 {Ç}{{\c C}}1 {ø}{{\o}}1 {å}{{\r a}}1 {Å}{{\r A}}1
    {€}{{\EUR}}1 {£}{{\pounds}}1
    {^}{{{\color{ipython_purple}\^{}}}}1
    {=}{{{\color{ipython_purple}=}}}1
    {+}{{{\color{ipython_purple}+}}}1
    {*}{{{\color{ipython_purple}$^\ast$}}}1
    {/}{{{\color{ipython_purple}/}}}1
    {+=}{{{+=}}}1
    {-=}{{{-=}}}1
    {*=}{{{$^\ast$=}}}1
    {/=}{{{/=}}}1,
    literate=
    *{-}{{{\color{ipython_purple}-}}}1
     {?}{{{\color{ipython_purple}?}}}1,
    identifierstyle=\color{black}\ttfamily,
    commentstyle=\color{ipython_cyan}\ttfamily,
    stringstyle=\color{ipython_red}\ttfamily,
    keepspaces=true,
    showspaces=false,
    showstringspaces=false,
    basicstyle=\ttfamily\footnotesize,
    keywordstyle=\color{ipython_green}\ttfamily,
}

\begin{document}

\title[Model Sketching]{Model Sketching: Centering Concepts in Early-Stage\\Machine Learning Model Design}

\author{Michelle S. Lam}
\affiliation{%
  \institution{Stanford University}
  \city{Stanford}
  \state{CA}
  \country{USA}
}
\email{mlam4@cs.stanford.edu}

\author{Zixian Ma}
\affiliation{%
  \institution{Stanford University}
  \city{Stanford}
  \state{CA}
  \country{USA}
}
\email{zixianma@cs.stanford.edu}

\author{Anne Li}
\affiliation{%
  \institution{Stanford University}
  \city{Stanford}
  \state{CA}
  \country{USA}
}
\email{anne24@stanford.edu}

\author{Izequiel Freitas}
\affiliation{%
  \institution{Stanford University}
  \city{Stanford}
  \state{CA}
  \country{USA}
}
\email{ifreitas@stanford.edu}

\author{Dakuo Wang}
\authornote{Work was performed while author was visiting Stanford HAI.}
\affiliation{%
  \institution{Northeastern University}
  \city{Boston}
  \state{MA}
  \country{USA}
}
\email{d.wang@northeastern.edu}

\author{James A. Landay}
\affiliation{%
  \institution{Stanford University}
  \city{Stanford}
  \state{CA}
  \country{USA}
}
\email{landay@stanford.edu}

\author{Michael S. Bernstein}
\affiliation{%
  \institution{Stanford University}
  \city{Stanford}
  \state{CA}
  \country{USA}
}
\email{msb@cs.stanford.edu}

\renewcommand{\shortauthors}{M.S. Lam, Z. Ma, A. Li, I. Freitas, D. Wang, J.A. Landay, M.S. Bernstein}

\begin{abstract}
Machine learning practitioners often end up tunneling on low-level technical details like model architectures and performance metrics. Could early model development instead focus on high-level questions of which factors a model ought to pay attention to? Inspired by the practice of sketching in design, which distills ideas to their minimal representation, we introduce \textit{model sketching}: a technical framework for iteratively and rapidly authoring functional approximations of \rev{a machine learning model's decision-making logic}.
Model sketching refocuses practitioner attention on \rev{composing high-level, human-understandable concepts that the model is expected to reason over (e.g., profanity, racism, or sarcasm in a content moderation task) using zero-shot concept instantiation}. In an evaluation with 17 ML practitioners, model sketching reframed thinking from implementation to higher-level exploration, prompted iteration on a broader range of model designs, and helped identify gaps in the problem formulation---all in a fraction of the time ordinarily required to build a model.
\end{abstract}

\ccsdesc[500]{Human-centered computing~Human computer interaction (HCI)}
\ccsdesc[300]{Human-centered computing~Interactive systems and tools}
\ccsdesc[300]{Computing methodologies~Artificial intelligence}
\ccsdesc[300]{Computing methodologies~Machine learning}


\begin{teaserfigure}
  \includegraphics[width=\textwidth]{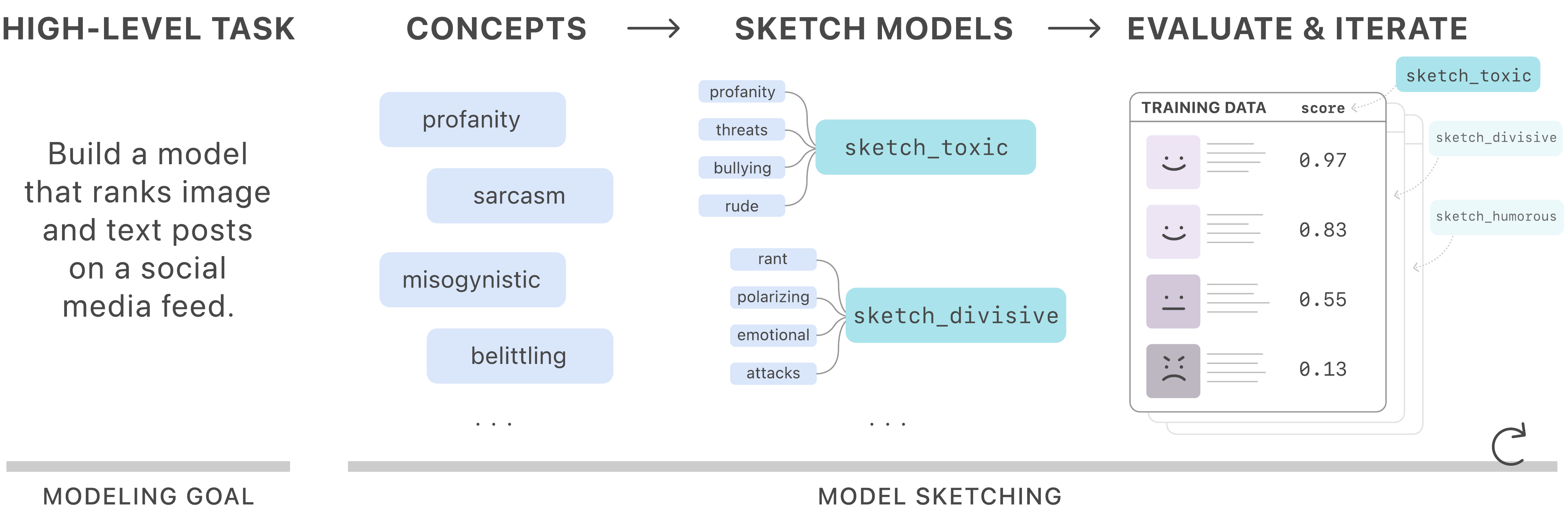}
  \caption{
      A conceptual diagram of the model sketching workflow.
      In just a few minutes, model sketching enables ML practitioners to author early-stage, functional models oriented around high-level concepts.
      We instantiate our approach in ModelSketchBook, a Python package for computational notebooks.
      \textit{Concepts}:~The user specifies {concepts} that are relevant to the decision-making task. Our system turns each concept into a functional model that can score unseen examples on that concept.
      \textit{Sketch Models}:~The user trains a sketch model, which aggregates concepts to predict ground-truth ratings for the overall modeling goal.
      \textit{Evaluate \& Iterate}:~The user evaluates the sketch model's predictions on training data to make sense of its behavior and gaps. They then iterate on concepts and sketch models to explore alternate approaches and better align the model with their goals.
  }
  \Description{Chart showing the progression to bring a modeling goal to life by creating concepts, which are then aggregated into various sketch models, which are evaluated and iterated on to create the best model that meets the modeling goal.}
  \label{fig:teaser}
\end{teaserfigure}

\maketitle

\section{Introduction}
Imagine that you are a machine learning practitioner tasked with creating a model to flag hateful image memes on a social media site. Once you dive into building and iterating on your model, the questions you are tackling quickly become ones like, ``How would a transformer model perform if we train on these examples and use these hyperparameters?'' Once you're in this mode, it is challenging to escape. How could we have helped your early model development work to instead stay focused on questions like, ``What factors---such as violence, racism, sexism, and profanity---should our model pay attention to when determining what content is hateful?'' Could machine learning tools help to center early efforts not on \textit{how to execute a solution}, but on \textit{what kinds of solutions to pursue} in the first place?

Current tools and approaches in machine learning (ML) promote a tunneling mindset that focuses attention on technical implementation~\cite{jun_hypothesisFormalization}. 
These tools pull us into a cascade of low-level tasks---gathering a dataset, extracting features, planning out data pipelines, considering tradeoffs in model architectures, deciding on a performance metric---all to arrive at a first version of a model. 
Such tunneling is dangerous because it locks us into a single problem formulation, which often persists even as we iterate~\cite{passi_barocas_problemFormulationFairness}. 
In contrast, it is the problem formulation itself that is the most important to iterate on early in the process, when high-level, underspecified decision-making goals such as ``detect hateful posts'' are translated into one of many concrete problem definitions~\cite{passi_barocas_problemFormulationFairness, jacobs_wallach_measurementAndFairness, passi_jackson_dataVision}. 
The more value-laden and ethically challenging a problem area, the more ML practitioners need to take care during the model design process to get this translation right. Unfortunately, this translation work tends to get buried and de-emphasized~\cite{zhang_howDoDataScienceWorkersCollaborate}.

To counter similar technical tunneling in their work, designers have long adopted a \textit{sketching practice} that purposefully foregrounds fluid exploration of high-level design possibilities.
Buxton defines sketching as any design process whose output is quick, plentiful, ambiguous, and minimal in detail~\cite{buxton2010sketching}.
Sketches distill an idea to its minimal representation, focusing attention on the most important aspects of a design direction. In doing so, sketches act as valuable cognitive tools for practitioners themselves to understand nascent designs~\cite{fallman_designOrientedHCI}. 
What is the right minimal representation for ML model development?
Interactive machine learning approaches (like Wekinator, Create ML, and Teachable Machine~\cite{wekinator, create_ml, teachable_machine}) and prompt-based approaches (using Large Language Models~\cite{GPT-3, wu_AI_chains} and Stable Diffusion~\cite{Rombach_2022_CVPR}) allow for rapid model development, but they are centered around inputs and outputs, largely abstracting away the decision-making logic in between.
For problem formulation tasks, it's precisely that \textit{decision-making logic} that we need to directly iterate on: what are the factors that our model should consider to make its decision?
Today's tools do not yet allow ML practitioners to sketch a model's decision-making logic to rapidly answer these kinds of critical model design questions.

In this paper, we introduce \textit{model sketching}, a technical framework that allows ML practitioners to create expressive, sketch-like versions of machine learning models.
To allow for direct iteration on a model's core logic, we use \textit{concepts}---the human-understandable ideas or factors that are relevant to a decision-making task---as the minimal representation for ML sketches.
For example, concepts in a hateful image meme detection task could range from concrete factors such as ``profanity,'' ``violence,'' and ``sexism'' to more abstract factors such as ``sarcasm'' and ``humor.''
While prior work has explored concept-oriented abstractions in ML~\cite{koh2020ConceptBottleneck, kim2018TCAV, ramos2020IMT, ng_IMTknowledgeDecomp}, a key difference in our approach is that we not only center concepts, but we also enable rapid instantiation and iteration of arbitrary concepts by leveraging zero-shot methods. This combination of concepts and flexible, rapid authoring unlocks new interactions where users can directly sketch model logic.
We implement our model sketching framework in a Python package called \textit{ModelSketchBook}, which leverages recent developments in zero-shot learning with pretrained models like GPT-3 and CLIP~\cite{GPT-3, CLIP} to allow users to turn open-ended concept descriptions into functioning building blocks.
With our tool loaded in a computational notebook environment, users can specify a concept like ``sexism'' or ``profanity'' and provide several image or text examples. 
ModelSketchBook draws on pretrained models to score each input on how sexist it might be or how likely it is to include profanity, and then the user combines sets of concepts together to train a holistic \textit{sketch model}. 
By interactively authoring concepts, incorporating them into sketch models, and inspecting resultant modeling outcomes, users can flexibly explore different modeling approaches and their ramifications.

\begin{figure*}[!tb]
  \includegraphics[width=0.95\textwidth]{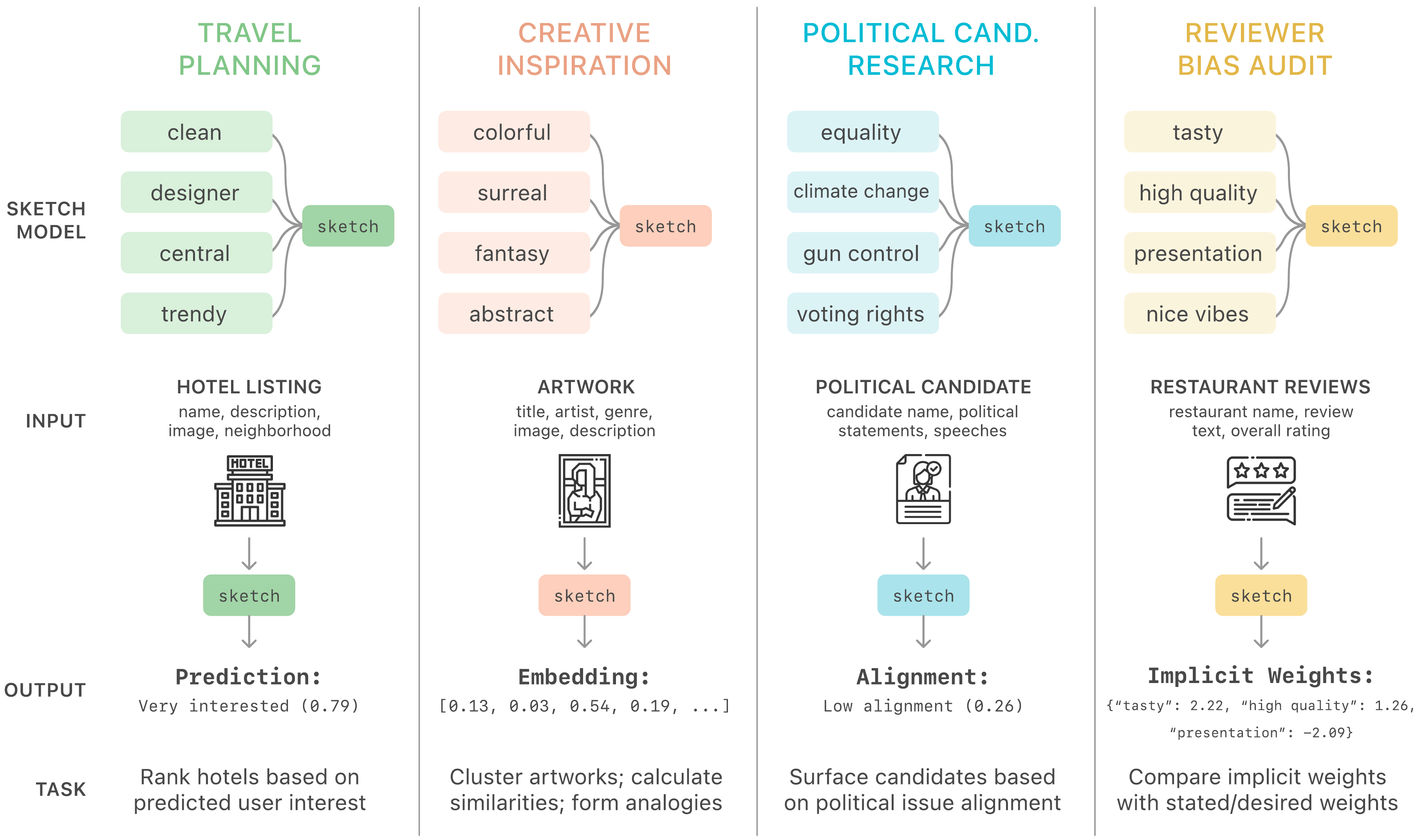}
  \caption{
    Model sketching case studies: (1)~Travel Planning, a personalized recommendation task, (2)~Creative Inspiration, an unsupervised clustering task, (3)~Political Candidate Research, a personalized search task, and (4)~Restaurant Reviews, a reviewer bias auditing task.
  }
  \label{fig:case_studies}
  \Description{Chart showing summary of model sketching case studies; chart contains four columns (Travel Planning, Creative Inspiration, Political Candidate Research, and Reviewer Bias Audit) and four rows (Sketch model, Input, Output, and Task)}
\end{figure*}

We illustrate the generalizability of our model sketching framework through demonstration case studies and a field evaluation. We provide case studies ranging from political candidate research to creative inspiration to reviewer bias auditing~(\autoref{fig:case_studies}).
Then, in a field evaluation, we asked 17 ML practitioners to author sketches that would determine whether internet memes sourced from the Hateful Memes Challenge~\cite{kiela_hatefulMemes} should be removed from a social media site. This task sets a high bar for technical difficulty (involving interacting text and image modalities), involves subjective judgement (providing sufficient interpretive flexibility to create a broad model design space), and involves ethical tradeoffs (requiring a consideration of multiple competing viewpoints).

Model sketching successfully enabled participants to shift from a technical mindset to a higher-level, design-oriented mindset that reasoned about the concepts that the model should and should not capture.
Before being introduced to model sketching, participants initially focused on technical details (e.g., model types like convolutional neural networks and transformers; performance metrics like F1 and AUC) when proposing a modeling plan for this task. After using ModelSketchBook, participants shifted to describe their approach in terms of higher-level design and problem formulation decisions.
Participants iteratively experimented with concepts to address gaps exposed by their early sketch models: all participants branched out beyond their initial set of concepts, eventually covering concepts ranging from ``racism'' and ``Islamophobia'' to ``sarcasm,'' ``irony,'' and ``controversial'', that would have not arisen if the model were developed with traditional methods.
With this expressive, lightweight interaction mode, participants turned their attention beyond modeling and identified gaps in the dataset, labeling method, and problem formulation; they noted factors such as inconsistent labels, class imbalance, lack of representation of certain types of harm, and subtleties of interaction between text and image semantics.
While participants initially estimated that it would take from days to months to produce a model for this task, after using our tool for 20-30 minutes, the vast majority of participants (14 of 17) were satisfied with the models they created. 

We view model sketching as an important step that moves beyond the mode of retroactively fixing broken models and towards one of helping ML practitioners to become better model designers from the start.
To summarize, this paper makes the following contributions:
\begin{itemize}
    \item \textbf{Model sketching}. 
    \rev{We introduce a framework for authoring zero-shot concept building blocks that can be flexibly aggregated into composite models, allowing machine learning practitioners to rapidly iterate over concepts to sketch a model's decision-making logic.}
    \item \textbf{The ModelSketchBook tool}. 
    \rev{We instantiate our model sketching framework by introducing an open-source Python package that allows users to interactively author concepts and sketch models in computational notebooks.}
    \item \textbf{An evaluation of model sketching with ML practitioners}. We perform a field evaluation with 17 machine learning practitioners who use our tool to author models for a challenging hateful meme detection task. We find that ML practitioners are successfully able to scope out from technical implementation details to higher-level decision-making logic. This cognitive shift helps ML practitioners to more broadly explore the model design space and to quickly discover gaps in the dataset, labeling method, and problem formulation.
\end{itemize}

\section{Related Work}

We survey research on sketching and prototyping in design and HCI, the work practices of machine learning practitioners, and current approaches for early-stage model design, all of which motivate our model sketching approach.

\subsection{Sketching and prototyping in design \& HCI}
The fields of design and HCI have long upheld the practice of sketching as a valuable tool for thought~\cite{fallman_designOrientedHCI,buxton2010sketching}. A key trait of a successful sketching or prototyping tool is that it refocuses cognitive processes on high-level iteration, empowering the practitioner to engage in ``design thinking rather than implementation tinkering''~\cite{hartmann_dTools}. A sketch is similar to a prototype in that both are instantiations of a design concept, but a sketch serves a different purpose: to explore and raise questions about design approaches rather than to propose answers or solutions~\cite{buxton2010sketching}. Thus, sketches are suitable for the early ideation stages of a design process while prototypes are intended for the later stages. 

\subsubsection*{What characterizes a sketch?}
\label{section:sketch_characteristics}
Buxton characterizes sketches as quick, plentiful, ambiguous, and minimal in detail~\cite{buxton2010sketching}.
The \textit{quick} and \textit{plentiful} attributes go hand in hand: methods with quick turnaround times enable practitioners to explore parallel design paths, an approach that supports more iterations, combats harmful design fixation, and leads to more diverse, higher-quality designs~\cite{dow_parallelPrototyping, tohidi_buxton_gettingTheRightDesign}. 
Sketches are powerful tools because they allow a practitioner to quickly iterate to test dramatically different high-level ideas.

The \textit{ambiguous} and \textit{minimal detail} attributes are also closely tied: much of the value of a sketch lies in both the cognitive task of distilling an idea into a stylized, minimal representation \cite{lim_anatomyOfPrototypes} and in leaving enough ambiguity to suggest---to oneself and others---multiple interpretations (and thus multiple future directions)~\cite{goel1995sketches}. Early work in user interface design demonstrated that overly-detailed prototyping approaches detract from this valuable cognitive work~\cite{landay_interactiveSketching} and do not substantially improve feedback quality~\cite{walker_hifiOrLofi}; in fact, simpler prototypes have been correlated with better design outcomes~\cite{yang_aStudyOfPrototypes}.

Finally, a sketch is \textit{expressive}: it must provide enough flexibility and control to allow a practitioner to communicate their idea and iterate upon it~\cite{buxton2010sketching,landay_interactiveSketching}. Traditional freehand sketches are highly expressive because designers can draw upon concrete or abstract visual and textual representations to hone their idea, allowing others to perceive both perceptual features and functional relations~\cite{ suwa_tversky_whatArchitectsSeeInTheirSketches}.

\revFinal{The machine learning domain presents distinct challenges for traditional prototyping methods due to AI's uncertain capabilities and complex outputs~\cite{yang_reexaminingWhetherWhy, yang_sketchingNLP}.}
However, we find that concept-based abstractions in particular hold promise for sketch-like approaches to ML model authoring.
Prior work on concept evolution in data labeling demonstrated that many decision-making goals are malleable and subjective, but these goals can be clarified through an iterative process of concept elicitation and refinement~\cite{kulesza_conceptEvolution, chang_revolt}. Researchers have even found that for real-life decision-makers such as judges, doctors, and managers, simple, statistically-derived rules based on a small number of human-interpretable features substantially outperform human experts~\cite{jung_goel_simpleRulesExpertClassification}. Concepts map well to the way that humans reason over decision-making tasks, and methods that statistically aggregate concepts work surprisingly well, even beyond prototyping contexts.

\subsection{ML practitioners: work practices and cognitive focus}
In our work, we use the terms ML practitioner and model developer interchangeably to refer to users who perform work that primarily deals with designing, developing, and iterating on machine learning models~\cite{muller_HowDataScienceWorkersWorkWithData,zhang_howDoDataScienceWorkersCollaborate}. 
Recent literature frames the data science and ML project workflow as a multi-stage collaborative lifecycle where domain experts focus on high-level problem definition and results interpretation, while ML practitioners place their cognitive focus on low-level implementation details~\cite{wang2019human, zhang_howDoDataScienceWorkersCollaborate,piorkowski_AIDevelopersCommChallenges}.
However, a major problem with the current workflow is that much of the translational work that ML practitioners perform to bridge between high-level goals and low-level implementation (i.e., via problem formulation, feature extraction, feature engineering) is rendered invisible~\cite{zhang_howDoDataScienceWorkersCollaborate, mao2019data,passi_trustInDataScience, muller_HowDataScienceWorkersWorkWithData}. 
As a result, ML teams lose important knowledge about the nature and meaning of features and the assumptions and reasoning that underlie modeling decisions~\cite{mao2019data,hou2017hacking}, which is especially problematic if ML practitioners need to revisit those decisions to counter harmful model behavior.
Similar patterns play out in statistical model authoring, which has prompted calls for high-level libraries to ``[express] conceptual hypotheses to bootstrap model implementations'' and bidirectional conceptual modeling to ``[co-author] conceptual models and model implementations''~\cite{jun_hypothesisFormalization}. 
Our work directly seeks to carry out this vision in the domain of ML model development by allowing ML practitioners to instantiate models using high-level concept-based abstractions that bridge to functional model implementations.

\subsection{Early-stage machine learning model design}
Work in interactive machine learning (IML) has long been interested in how we might engage humans in the model development process. In reaction to the slow and effortful process of building models, work in IML demonstrated that by enabling interactive model training and refinement, users can more rapidly develop models and keep them aligned with their goals~\cite{fails_olsen_IML,amershi_powerToThePeople, patel2010gestalt, francoise2021marcelle}.
In most IML systems, interactions take the form of labeled examples and demonstrations, such as positive and negative image examples in CueFlik~\cite{cueflik} and paired input gestures and output actions in Wekinator~\cite{wekinator}. 

Building on IML, work on interactive machine teaching (IMT) emphasizes the metaphor of \textit{teaching} to focus on the human's role as a teacher rather than just a label-provider~\cite{ramos2020IMT, simard2017IMT}. Through this metaphor, IMT highlights new challenges like the need for richer teaching strategies beyond data labeling.
A wizard-of-oz study along this vein explored how users might teach knowledge to a model, and this work found that users decomposed knowledge in terms of concepts, relationships between concepts, and rules that combine and apply these concepts~\cite{ng_IMTknowledgeDecomp}. However, without actually instantiating and evaluating their concepts, users struggled to form a mental model of how the ML model works, which made it especially difficult for them to know how to articulate subtle, abstract forms of knowledge. 

\begin{table*}[!tb]
  \centering
  \footnotesize
    \begin{tabular}{p{0.18\textwidth}p{0.07\textwidth} p{0.25\textwidth} p{0.2\textwidth} p{0.15\textwidth}}
    \toprule
    \textbf{Paradigm} & \textbf{Emphasis} & \textbf{Paradigm shift} & \textbf{Strategy to improve ML models} & \textbf{Design requirements}\\
    \midrule
    {Interactive Machine Learning} &
    {\textit{interactive}} &
    {ML models don't need to exist in a vacuum---through human \textit{interaction} and feedback}, we can achieve user alignment and rapid model authoring&
    {Better feedback loops}&
    {Speed}\\
    \midrule
    {Interactive Machine Teaching} &
    {\textit{teaching}} &
    {ML isn't just about learning---we need to focus on \textit{teaching}: \textit{who} is developing ML models and \textit{how} they're sharing their knowledge} &
    {Better teaching}&
    {Expressivity}\\
    \midrule
    {Model Sketching} &
    {\textit{sketching}} &
    {ML models don't need to be high-fidelity---in fact, \textit{low-fidelity}, \textit{transient} models are better at helping us to explore the multiverse of model designs} &
    {Better design space exploration}&
    {Minimal detail, Speed, Expressivity}\\
    \bottomrule
    \end{tabular}
    \caption{
    \rev{A summary of key ideas put forward by Interactive Machine Learning and Interactive Machine Teaching, alongside those of Model Sketching. The three paradigms are complementary, but emphasize different perspectives and strategies.}
    }
    \label{table:related_work_comparison}
\end{table*}

While model sketching shares many of the same motivations that underlie IML and IMT, it has slightly different goals---to explore a vague, unformulated model design space (problem formulation) rather than execute on a crisp modeling vision (model implementation). 
This means that in addition to the speed and expressivity required by IML and IMT, model sketching requires a \textit{minimal representation of a model's decision-making logic} to facilitate effective design space exploration~(Table~\ref{table:related_work_comparison}). 
To support model sketching, we need to provide a mechanism for the user to directly iterate on the decision-making primitives that control a model's actions. Thus, we cannot use black-box methods that opaquely map from inputs to outputs, such as example-based IML approaches and prompt-based model authoring. 
Concepts are an ideal representation that allows for iteration on decision-making strategy and alignment with the ways that humans prefer to decompose and express their knowledge~\cite{ng_IMTknowledgeDecomp}. 

The metaphor of sketching also informs how we must instantiate these concept-based representations.
Past work in IML and IMT has explored concept instantiation with examples, demonstrations, predefined features, and formal knowledge bases~\cite{cueflik, wekinator, ramos2020IMT, brooks2015featureInsight}. However, given that sketches are meant to explore a broad model design space that is not known ahead of time, users need to have the flexibility to author arbitrary concepts, so we cannot rely on static features or knowledge bases. 
An added challenge of sketching is that there is a steep cost associated with diverting the user's attention to different tasks (e.g., switching to label examples, develop a feature pipeline, or train a new helper model), which risk pulling the user out of the sketching mindset and into technical tunneling. 
\revFinal{Thus, we build off of zero-shot prompt-based modeling approaches~\cite{wu_AI_chains, jiang_PromptMaker} to instantiate concepts flexibly and rapidly without diverting attention.}


\section{Model Sketching}

In this section, we describe our model sketching framework, which aids ML practitioners in moving beyond technical implementation to explore the conceptual model design space during early phases of ML model development.

\subsection{The Model Sketching Framework}
Model sketching aims to facilitate the translational, interpretational work that goes in between a high-level description and any particular low-level implementation~\cite{passi_barocas_problemFormulationFairness,jacobs_wallach_measurementAndFairness}.
This work charts a path through the multitude of plausible strategies for any one high-level modeling goal, each of which ossifies a set of human values and ethical implications that are encoded into the model.
We address the gap between high-level modeling goals and low-level implementation with the intermediary representation of \textit{concepts}---human-understandable ideas or factors that are relevant to a decision-making task. By foregrounding this translational step between goals and their implementation, we aim to scaffold ML practitioners' reasoning and help them to more effectively sketch out and explore parallel solution paths. 

\subsubsection{Data prerequisites}
We require only a small amount of data in table form (e.g., spreadsheet, CSV, Pandas Dataframe) where each row represents a single example and each column represents a data field. These data fields can contain text, image, or numeric data. 
We recommend that the dataset has at least 40-50 rows to aid sensemaking during model development, while also allowing for a split between training and test data.
If the user provides ground-truth labels, our system can help with sensemaking around model accuracy; however, such labels are not required.

\subsubsection{Concepts: Crafting functional building blocks of decision-making}
\begin{figure*}[!tb]
  \includegraphics[width=\textwidth]{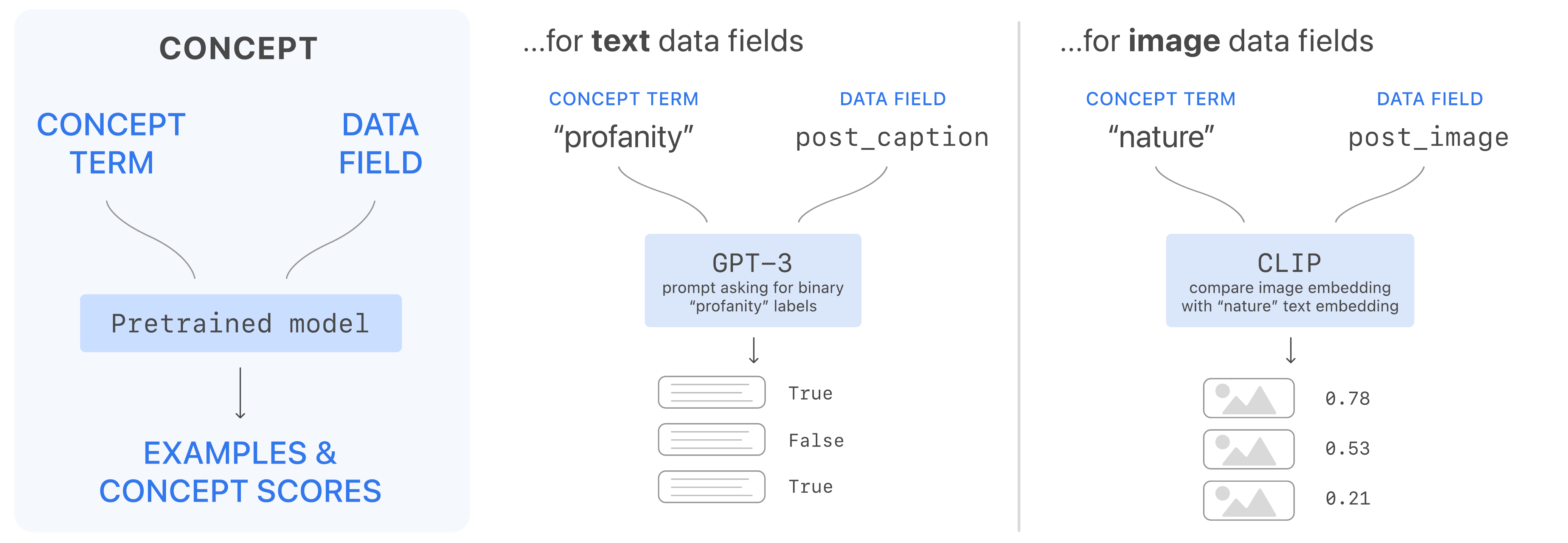}
  \caption{
      Users can turn human-understandable concepts into functional building blocks. A concept takes in a concept term and a data field as input. Then, a pretrained zero-shot model produces a score for each example indicating the extent to which the example relates to the specified concept. We adapt our underlying approach depending on the type of the data field. For text, we use GPT-3 prompts for binary label prediction. For images, we use CLIP embeddings for continuous score prediction.
  }
  \label{fig:sys_concepts}
  \Description{Figure demonstrating how a concept is created by providing a concept term (e.g., ``profanity'' or ``nature'') and the data field of interest (e.g., post_caption or post_image)into a pretrained model (GPT-3 for text data fields or CLIP for image data fields), which then outputs examples and their corresponding concept scores.}
\end{figure*}

Concepts are how we decompose the larger and more ambitious task of a sketch model into more manageable, human-understandable components.
We turn concepts into functional components by mapping from a data example to a score that indicates its relevance to the specified concept, or how well it exemplifies that concept~(\autoref{fig:sys_concepts}). For example, a concept for ``colorful'' might produce scores for a set of images indicating roughly how colorful they are; a concept for ``profanity'' might produce scores for a set of text comments that estimate how profane they are. Thus, concepts bear similarities to feature extractors, but they are not limited to lower-level, objective attributes, and they require much less time and effort to produce, as a new concept can be created in several seconds. Once created, a concept can be applied to any arbitrary example, seen or unseen.

We build upon advances in zero-shot modeling to instantiate concepts. Zero-shot modeling describes any machine learning modeling approach that does not require any training examples to perform a task. To create a concept, the user provides an input concept term (a brief word or phrase) and a data field (column name) from their dataset. Our framework then authors a zero-shot model that outputs a numeric score for that concept for every item in the dataset.
For our concepts, we implement two methods: 1)~zero-shot text classification via large language model (LLM) prompting using GPT-3 and 2)~zero-shot image classification via embedding similarity with a pre-trained image-to-text model using CLIP. We describe our technical approach in greater depth in Section~\ref{section:concept_technical}. While we use these two zero-shot methods in our work, our model sketching framework is compatible with any current or future zero-shot modeling method. 

\subsubsection{Sketch models: Putting concepts together to envision decision-making outcomes}
\begin{figure*}[!tb]
  \includegraphics[width=0.85\textwidth]{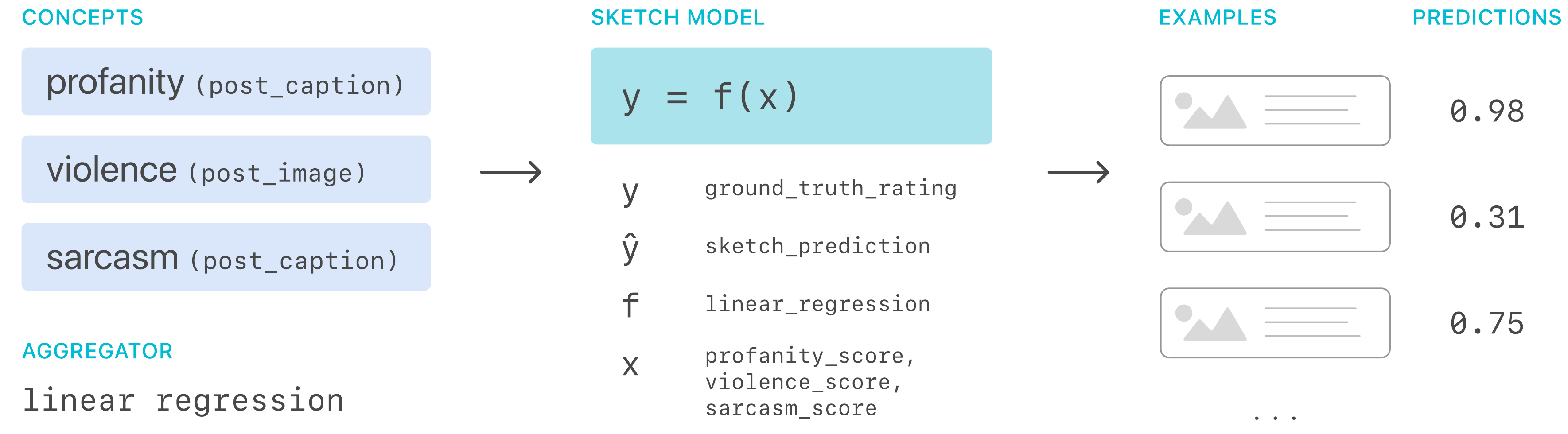}
  \caption{
      Sketch models combine concepts together to create an overall model for the decision-making task. First, the user chooses from among their authored concepts and decides on an aggregation method. We train a sketch model using the concept scores, the specified aggregator, and the user's ground-truth ratings (e.g., ratings of the hatefulness of posts). Then, the sketch model can produce predictions (e.g., scores estimating the hatefulness of posts) for any arbitrary example.
  }
  \label{fig:sys_sketches}
  \Description{Figure demonstrating the creation of a sketch model, by aggregating a set of concepts (e.g., profanity, violence, and sarcasm), which will produce a visual output of the model’s predictions on a set of arbitrary examples.}
\end{figure*}

The second component of our framework involves authoring sketch models, lightweight ML models that aggregate the signals produced by concepts~(\autoref{fig:sys_sketches}). A sketch model is designed to accomplish the same task as the full model that it is meant to emulate, so it expects the same inputs and produces the same class of outputs.

Once the user has authored a set of constituent concepts that are relevant to the decision-making task, they select which concepts they would like to combine into a sketch model. We provide an initial set of aggregators that can combine concept scores to produce a single sketch model score: linear regression, logistic regression, decision tree classification, random forest classification, and multi-layer perceptron (MLP) classification. 
Rather than require a complex deep learning model to integrate concepts, we build on prior research that demonstrated how simple linear models with small sets of human-interpretable features can achieve performance exceeding that of human decision-makers and rivaling that of complex prediction models~\cite{jung_goel_simpleRulesExpertClassification}.
However, we are not bound to this initial set of aggregators.
We use the ground-truth labels provided in the training dataset to train these aggregators when a sketch model is created; all subsequent runs of the sketch model will evaluate the trained model on new input data. If the dataset does not have labels, then the user can opt to provide manual weights for the concepts using the linear regression aggregator.

With the sketch-model layer of abstraction, ML practitioners can focus their attention on experimenting with different combinations of concepts to create sketches that might capture different sets of values in decision-making. For example, when authoring a hate speech detection model, an ML practitioner may experiment with sketches that focus on race-related or gender-related hate speech, or approaches that explicitly account for intersectionality. They may try out sketches that take a relatively lenient stance by only adding a concept related to speech that incites violence, or they may experiment with sketches that take a more strict stance on profanity by adding concepts for swearing and slurs.
While sketch models in their most stylized form are comprised entirely of concepts, sketch models are also compatible with more traditional features produced through custom feature extractors or existing ML models. 

\subsubsection{From sketch models to production models}
\label{section:sketch_to_production}
While sketch models initially serve as tools for thinking and iterating during early model development phases, we envision that they could inform the design of full production models. 

First, a key part of full-scale model design involves deciding how to \textit{decompose the model} into smaller parts. A notable characteristic of successful production ML systems---consistent across use cases like feed ranking and content moderation and across platforms like Facebook, LinkedIn, and Reddit---is that they are composed of smaller, more targeted ML models whose outputs are synthesized into a single score, often via weighted linear combination~\cite{eckles_2022, reddit_moderator_tools, linkedin_dwell_time}. 
In other words, the sub-models that ML practitioners produce in industry are more complex versions of our concepts, but they typically still boil down to human-understandable decision-making factors that will be synthesized into a single prediction via an aggregation process. 

Second, a substantial part of model development (up to 80\% by some estimates) is actually consumed by data cleaning and wrangling, and ML practitioners play an active role in shaping, curating, and designing data~\cite{muller_HowDataScienceWorkersWorkWithData}. Model sketching could assist in refining \textit{data requirements} by drawing out what kind of data is necessary for the task based on the concepts that the model must reason over. By testing out concepts on different data fields (e.g., different text or image fields), ML practitioners might gain a quick understanding of what forms of data are most informative to serve different concepts and, ultimately, what data fields and what kinds of data examples should be required for the modeling task.

\subsection{The ModelSketchBook tool}
To instantiate our model sketching framework, we introduce the {ModelSketchBook} API,\footnote{\url{https://github.com/StanfordHCI/ModelSketchBook}} a Python package that allows ML practitioners to create concepts and sketches of their own. 

\subsubsection{Environment and setup}
\label{section:sys-sketchbook-setup}
Our API can be loaded in any Python environment, but it is optimized for use in a computational notebook such as Jupyter or Colab. 
We chose to tailor our API design to the notebook environment because this is where many ML practitioners already carry out much of their exploratory modeling work~\cite{rule_explorationExplanationCompNotebooks,zhang_howDoDataScienceWorkersCollaborate}. 

\begin{lstlisting}[language=iPython]
    import model_sketch_book as msb
    
    # Set up the sketchbook
    sb = msb.create_model_sketchbook(
        goal="Detect hateful memes on social media.",
        datasets={  # Pass in datasets to use
            "train": df_train,
            "test": df_test,
        },
        schema = {  # Specify the types of data fields
            "text": InputType.Text,
            "img_url": InputType.Image,
            "overall_rating": InputType.GroundTruth,
        },
    )
\end{lstlisting}

After loading their dataset as a Pandas dataframe and importing the {ModelSketchBook} package in their notebook, the user sets up a sketchbook object. The sketchbook contains all of the datasets, concepts, and sketches that the user will author during their session, and it saves other metadata such as model predictions, performance metrics, and a history of concepts and sketches.

\subsubsection{Creating concepts: under the hood}
\label{section:concept_technical}
As mentioned previously, we specifically support text concepts powered by GPT-3 and image concepts powered by CLIP.
To achieve zero-shot text classification, we author prompts that list a batch of examples for the model to classify and ask the model to decide on a binary label (``<concept term>'' or ``not <concept term>'') for each provided example. Large language models like GPT-3 are designed to perform text completion on provided input prompts, so plausible completions provided by the model capture a rough sense of binary labels that might be reasonable based on the knowledge available in the model. We include an example prompt below.

\begin{lstlisting}[language=Markdown]
    Decide whether these comments are "motivational" 
    or "not motivational".
    1. You can do it!!
    2. Ugh, you're the worst.
    3. I'm so tired...

    Comment results:
\end{lstlisting}

The GPT-3 model will then return a text completion response similar to this that provides a label for the examples. We parse and post-process this GPT-3 response to retrieve binary labels.
\begin{lstlisting}[language=Markdown]
    1. motivational
    2. not motivational
    3. not motivational
\end{lstlisting}

To carry out zero-shot image classification, we use CLIP (Contrastive Language-Image Pre-Training)~\cite{CLIP}, a neural network trained on (image, text) pairs that allows images and text to be represented in a shared embedding space. Given a data field containing images and a concept term, we use CLIP to encode both the image and the concept term in that shared image-text embedding space. Then, we calculate and return the cosine similarity between the embedded image and concept term. Thus, the output score approximately captures the extent to which the given image relates to the specified concept term. We also allow users to normalize and re-calibrate these similarity scores to match their internal understanding of the concept, and we also allow users to apply custom thresholds to binarize continuous concept scores.

\begin{figure*}[!tb]
  \includegraphics[width=\textwidth]{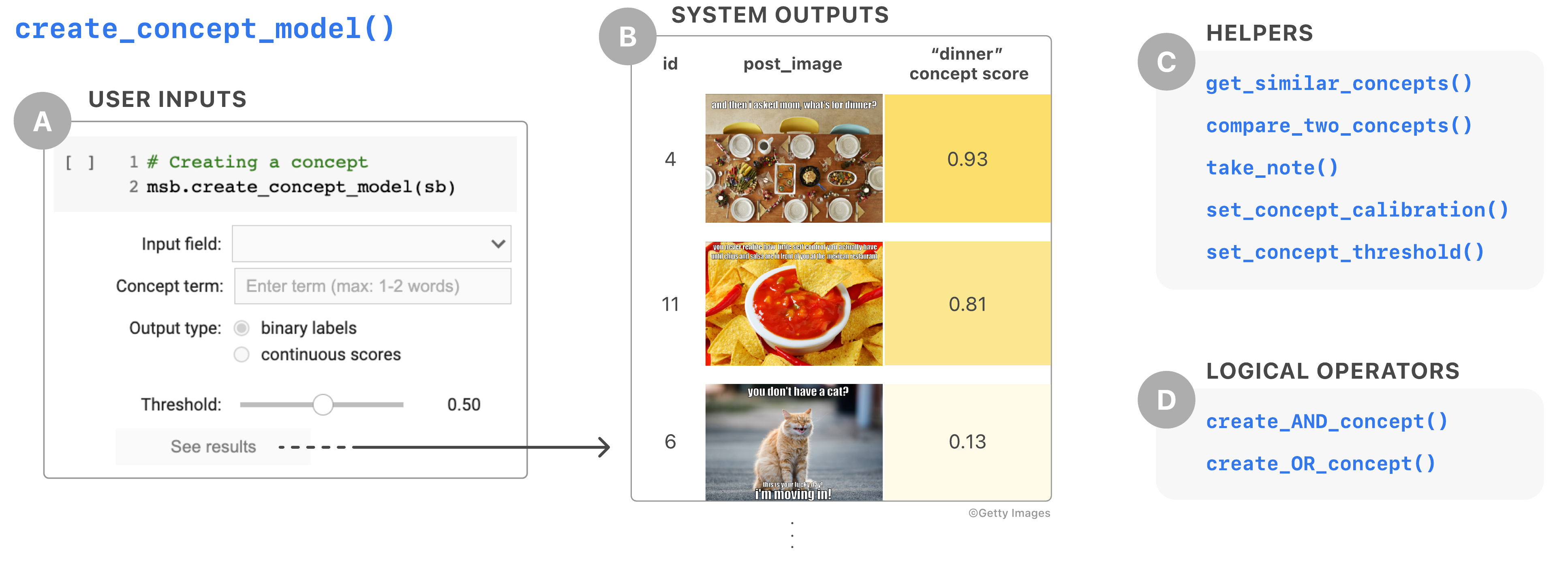}
  \caption{
      Creating concepts with {ModelSketchBook} in a computational notebook. 
      (A)~The \texttt{create\_concept\_model()} function accepts user inputs for the data input field, concept term, and other settings. 
      (B)~Then, the API outputs a dataframe displaying all training examples sorted by their concept score. 
      (C)~We provide helper functions for finetuning, making sense of existing concepts, and brainstorming new concepts. 
      (D)~We also support logical \texttt{AND} and \texttt{OR} operators to create compound concepts. 
  }
  \label{fig:sys_api_concepts}
  \Description{Figure demonstrating the process of creating a concept model in the ModelSketchBook API, showing the user inputs and the dataframe visualization provided by the system output. Additionally, there are helper functions and logical operators available which assist the user to create the best concept model possible.}
\end{figure*}

\subsubsection{Creating concepts with the API}
After setting up the sketchbook, users can freely author concepts. While we also provide raw API functions under the hood, for convenience, we surface UI elements via ipywidgets.\footnote{\url{http://ipywidgets.readthedocs.io}} 
When the user makes a call to \texttt{create \_concept\_model()}, a set of widgets will appear for them to specify the input field, a concept term, and other details like the output type of the concept score~(\autoref{fig:sys_api_concepts}A).

Once the user enters in those details, they can execute the cell; after several seconds, the API returns a dataframe visualization that displays a sorted view of the examples in the training dataset along with their concept score~(\autoref{fig:sys_api_concepts}B). The user can browse through the concept score results to determine whether the model's understanding of the concept sufficiently aligns with their own. If not, they can try out other concept terms or experiment with other input fields to work towards better alignment. The API also includes additional features for users to binarize continuous scores using a custom threshold value, auto-normalize continuous scores to a 0-1 score range, or custom-calibrate continuous scores based on a specified minimum and maximum score value. 

We provide functions that assist users as they brainstorm and evaluate tradeoffs among different concepts. A brainstorming helper function surfaces synonyms or antonyms of a provided word to spur ideas when users seek to refine their concept. Meanwhile, a concept comparison helper function allows users to specify concept terms and compare their correlation with the input data and ground-truth ratings. To scaffold users' sensemaking process, we also provide a note-taking function that allows them to save scratch notes on concepts and sketches for future reference.

Finally, to allow for further control and refinement of concepts, we enable users to link existing concepts together with logical \texttt{AND} and \texttt{OR} operators to create compound concepts. With a call to \texttt{create\_AND\_concept()}, users can take the logical \texttt{AND} of any number of binary concepts (and can similarly link arbitrary concepts with an \texttt{OR} relation using the \texttt{create\_OR\_concept()} function). This grants users the ability to express more complex concepts, such as a single concept that can manifest in several ways or a multimodal concept that requires both image and text signals.

\subsubsection{Creating sketches with the API}
\begin{figure*}[!tb]
  \includegraphics[width=\textwidth]{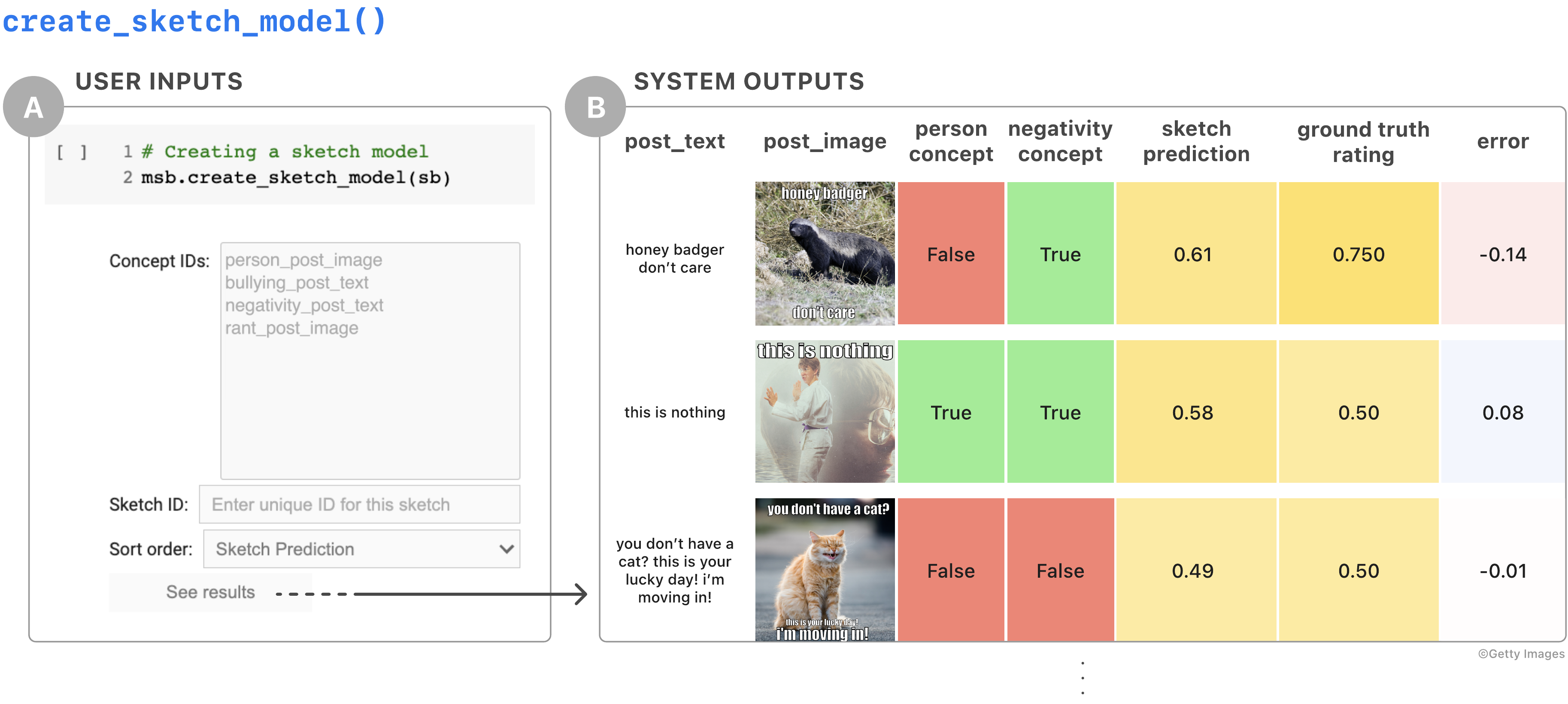}
  \caption{
      Authoring sketches with {ModelSketchBook} in a computational notebook. 
      (A)~The \texttt{create\_sketch\_model()} function accepts user inputs on the concepts to aggregate, with optional parameters to set the aggregation method and sort-column.
      (B)~Then, the API outputs a dataframe displaying all training examples sorted by the specified column. This view shows all input fields, all concept scores, and the sketch model's prediction. It also displays the ground truth rating and the error between the sketch prediction and ground truth rating to aid error analysis.
  }
  \label{fig:sys_api_sketches}
  \Description{Figure demonstrating the process of creating a sketch model in the ModelSketchBook API, showing the user inputs and the dataframe visualization of the sketch model’s prediction provided by the system output.}
\end{figure*}

Once a user has authored several concepts, they can proceed to create a sketch model that aggregates those concepts by calling \texttt{create\_sketch\_model()}. This function displays widgets for the user to select the concepts and aggregation method (default: linear regression) for their sketch model~(\autoref{fig:sys_api_sketches}A).

After executing the cell, the API creates the new sketch model and returns a dataframe visualization that displays all of the input fields of the training examples, all of the relevant concept scores, and the overall prediction produced by the sketch model~(\autoref{fig:sys_api_sketches}B). This view also includes the ground-truth ratings and the difference between the sketch prediction and these ground-truth ratings. Users can choose to sort by various columns to aid their understanding of the model outputs.
The sketch model output also displays traditional performance metrics for both regression and classification formulations of the task based on the ground-truth labels for the training examples; we display Mean Absolute Error (MAE), Classification Accuracy, F1 Score, Precision, and Recall.

Finally, the API provides functions that help the user to assess their sketch models. The \texttt{compare\_sketches()} function allows the user to compare performance metrics for specified sketches by displaying the performance metrics side by side in a plot similar to Figure~\ref{fig:mean_metric_plot}. The \texttt{test\_sketch()} function allows users to test out sketch models on a labeled or unlabeled validation set (if one is provided) to get a sense for how well it generalizes to a broader set of examples. After the user selects a sketch model and a dataset, this function produces a dataframe similar to \autoref{fig:sys_api_sketches}B, only including the ground truth rating and error if present.
\revFinal{Please see Appendix~\ref{appendix:msb-process-overview} for an overview of the three function calls required to instantiate a sketchbook, concept, and sketch model}.

\subsubsection{Implementation}
The {ModelSketchBook} API was authored in Python. We use OpenAI's API for GPT-3 to perform zero-shot prompt-based text classification and use the \texttt{text-davinci-002} model, the state-of-the-art GPT-3 variant at the time of development. We use OpenCLIP~\cite{open_clip}, an open-source version of CLIP, to perform zero-shot image-text alignment; we use a Vision Transformer (ViT) model (\texttt{ViT-B-32-quickgelu}) pre-trained on the \texttt{laion400m\_e32} dataset. We use scikit-learn~\cite{scikit_learn} model implementations for our sketch model aggregators. 
All training and inference takes place on the local machine on which the notebook is running (i.e., a user's own machine if they are running a local Juypter notebook, or a Google cloud server if they are running a Colab notebook). 
When loaded in a computational notebook environment, the input interface elements are rendered using the ipywidgets package. The output visualizations are rendered using Pandas dataframes with custom CSS styling.

\section{Demonstration Case studies}
To illustrate the breadth of machine learning tasks to which we can apply our model sketching approach, we present example case studies that we authored to demonstrate recommendation, unsupervised clustering, search, and bias audit applications~(Figure~\ref{fig:case_studies}; details in Appendix~\autoref{table:case_studies}).

\subsubsection*{Travel planning---personalized recommendation} 
In our travel planning example, given a particular user, we aimed to recommend an Airbnb listing that would best match their preferences. 
Here, our sketches could capture key aspects of a user's taste in Airbnbs: based on images, we expressed preferred aesthetics (``clean,'' ``bright,'' ``designer'') and red flags (``outdated,'' ``ugly''); based on text descriptions, we conveyed desirable attributes of the neighborhood (``good vibes,'' ``central,'' ``trendy''). 
The sketching process helped us to discover gaps in the modeling setup such as the potential value of customer reviews to provide an objective assessment of factors like cleanliness and the importance of proximity and specific physical location-related information for the travel use case.

\subsubsection*{Creative inspiration---unsupervised embedding and clustering} 
Model sketching can also serve unsupervised learning tasks by allowing users to create customized sketch-like embedding spaces to perform item similarity, clustering, and analogical retrieval tasks. Building on a dataset of museum artwork, we authored sketch models for text and image fields related to bold visual style (``colorful,'' ``surreal,'' ``fantasy,'' ``abstract''), social justice (``social justice themed,'' ``race themed,'' ``feminist,'' ``queer themed''), and political movements (``politics themed,'' ``revolution themed,'' ``labor movement themed''). Then, we applied our sketch models to new artworks to retrieve their concept score-based ``embeddings.'' We could then cluster artworks and gather creative inspiration by surfacing analogous works of art in disparate genres: by selecting an artwork in a painting genre and selecting a target genre of sculpture, we could retrieve analogous artworks from the sculpture genre that were most similar to the original artwork.

\subsubsection*{Political candidate research---personalized search}
Our approach can also support search-related tasks. For this example, we assisted a user in researching political candidates before an election to discover which candidates were most aligned with them on the issues they cared about most. Drawing from third-party candidate information sources including platforms, political statements, and speeches, we authored sketches that captured issues related to the environment (``climate change,'' ``environment,'' ``eco-friendly''), gun control (``gun control,'' ``anti-gun), and other civil rights (``equal rights,'' ``equality,'' ``voting rights,'' ``pro-choice''). Model sketching was especially beneficial for iteratively exploring nuanced candidate stances that weren't explicitly highlighted as core issues in standard voting guides.

\subsubsection*{Restaurant reviews---reviewer bias auditing}
We can also apply model sketching for auditing tasks by working backwards from decision-making criteria to uncover latent weights from human or algorithmic decision-makers. For this example, we assisted a food reviewer in assessing their own potential biases. Based on the reviewer's self-reported decision-making factors (ordered by importance: taste, quality, presentation, vibes, service, and price), we created analogous concepts. Then, we authored sketch models to aggregate these concepts, using the reviewer's overall restaurant review ratings as ground truth. 
Our model inferred high positive coefficients for the ``tasty'' and ``quality'' concepts, which aligned with the reviewer's top two decision-making factors. However, the model inferred high negative coefficients for the ``presentation'' and ``vibes'' concepts, which corresponded to the reviewer's next most important factors. This revealed to the reviewer that these factors had less of a sway on their overall ratings than they intended.


\section{Evaluation}

Our study aims to evaluate the impact of applying a model sketching approach to a realistic model authoring task with experienced machine learning practitioners. Since our goal is to shift the cognitive frame with which ML practitioners approach model authoring, our evaluation centers on these main research questions:


\aptLtoX[graphic=no,type=html]{\begin{enumerate}
    \item[\textbf{RQ1:}] \textit{How does engaging in model sketching impact the cognitive focus of ML practitioners during model development?} Using this approach, do they successfully shift away from technical and implementation-oriented thinking and towards higher-level, concept-oriented thinking?

    \item[\textbf{RQ2:}] \textit{What are the outcomes of a model sketching workflow for ML models and ML practitioners?}
    When ML practitioners shift to higher-level thinking, what kinds of models do they produce? How do they anticipate modeling gaps and consider value tradeoffs in model design? 
\end{enumerate}}{\begin{enumerate}[label=\textbf{RQ\arabic*:}]
    \itemsep=10pt
    \item \textit{How does engaging in model sketching impact the cognitive focus of ML practitioners during model development?} Using this approach, do they successfully shift away from technical and implementation-oriented thinking and towards higher-level, concept-oriented thinking?

    \item \textit{What are the outcomes of a model sketching workflow for ML models and ML practitioners?}
    When ML practitioners shift to higher-level thinking, what kinds of models do they produce? How do they anticipate modeling gaps and consider value tradeoffs in model design? 
\end{enumerate}}

To address these research questions, we designed a field evaluation for participants with significant ML experience to use our {ModelSketchBook} API to author models that would detect hateful memes posted on a social platform.

\subsection{Study design}

\subsubsection{Study format}
Our evaluation consisted of an hour-long study session. Participants were sent a brief 10-minute labeling task to complete before the study session so that we could use their personal ground-truth ratings for the model-authoring task. We asked them to provide labels for 40 examples; 20 examples were used as a training set during the study session, and 20 were reserved as a test set. During the hour-long study session, the first 15-20 minutes were allocated to consent, Colab notebook setup, and a tutorial and video demo of the {ModelSketchBook} API. Then, the participant was given about 30 minutes to work on the model authoring task using the {ModelSketchBook} API in a Colab notebook. Finally, the last 10-15 minutes were split between a brief interview on their experience and a post-study survey questionnaire. \rev{Our full written response questions, survey questions, and interview scripts are included in Appendix Section~\ref{section:appendix_task_questions}-\ref{section:appendix_interview_questions}. We describe our qualitative analysis method in Appendix Section~\ref{section:qualitative_analysis}.}

\subsubsection{The Hateful Memes detection task}
\label{section:hateful_memes_data}
Our dataset originated from the Hateful Memes Challenge~\cite{kiela_hatefulMemes} launched in 2020 by Meta AI to bring attention to the complex task of identifying multimodal hate speech. The dataset holds over 10,000 multimodal meme examples (combining text and images) based on real-world memes.
The scoring scheme that we provided participants for their labeling task was: $0.00$ = Benign (keep); $0.25$ = Slightly problematic (may be questionable, but keep); $0.50$ = Threshold (problematic enough to be removed); $0.75$ = Problematic (remove); $1.00$ = Very problematic (remove). 
Our formulation differs from the original Hateful Memes Challenge; we frame the labels around content removal rather than hate speech to orient the modeling task around the users' preferred decision-making logic rather than a centralized definition. Additionally, we filtered our dataset to examples that were labeled as ``non-hateful'' in the original dataset to reduce participants' exposure to unpleasant content and to skew towards more ambiguous, grey-area cases where users may make different value tradeoffs.

We selected this task for our user evaluation because we sought to evaluate model sketching in a difficult, realistic setting, and this task represented an open challenge that ML practitioners were actively working on.
In addition, we wanted our task to be relatively subjective and not have clear right-or-wrong answers to provide participants a sufficiently broad model design space. 
Decisions around multimodal hate speech and content moderation commonly involve difficult value tradeoffs.


\subsubsection{Pre- and Post-task written responses}
Before receiving an introduction to model sketching, we asked participants to write out their modeling approach for the Hateful Memes task based on their current knowledge. After participants viewed the tutorial and demo, but before they started using the {ModelSketchBook} API, we asked them to brainstorm an initial set of at least three concepts that they might want to use in the task. At the end of the task period, we asked participants to once again describe their planned modeling approach for the Hateful Memes task, where they could assume that they had access to a tool like the {ModelSketchBook} API, but they did not have to use such a tool. We also asked participants to write down any learnings they had after engaging in this task, and we asked them to specify which sketch model they viewed as the ``best sketch'' among those they authored.

\cut{In a post-task interview, we asked participants to walk through their notebook and explain their thought process. Then, we asked them to compare this model sketching experience with their prior experience authoring machine learning models. Lastly, we asked them what they learned from the model sketching process.
Our final post-task survey asked participants about their satisfaction with their concepts and sketch models, their perception of the comfort and expressivity of prior modeling tools and of {ModelSketchBook}, and other aspects that they liked and disliked about the model sketching experience. 
The full text of our interview script and survey is included in Appendix Section~\ref{section:appendix_task_questions}-\ref{section:appendix_interview_questions}. We describe our qualitative analysis method in Appendix Section~\ref{section:qualitative_analysis}.}


\subsection{Participant recruitment}
We sought participants who had significant experience with machine learning and IPython notebooks. We recruited participants by sending emails to university mailing lists for Computer Science and AI-related groups, and we sent recruitment materials to personal contacts working in the AI industry to share with their networks. 
We selected from among participants who answered either ``rather much'' or ``very much'' (the two highest options on a 5-point Likert scale) to indicate their amount of experience with IPython notebooks and their experience with machine learning. In total, 17 users participated in our study (see Appendix Section~\ref{section:participant_dem} for participant demographics). 
We compensated participants with a \$60 Amazon gift card for completing the hour-long study session. All studies were conducted remotely over video conferencing.

\section{Results}

\begin{figure}[!tb]
  \includegraphics[width=0.9\linewidth]{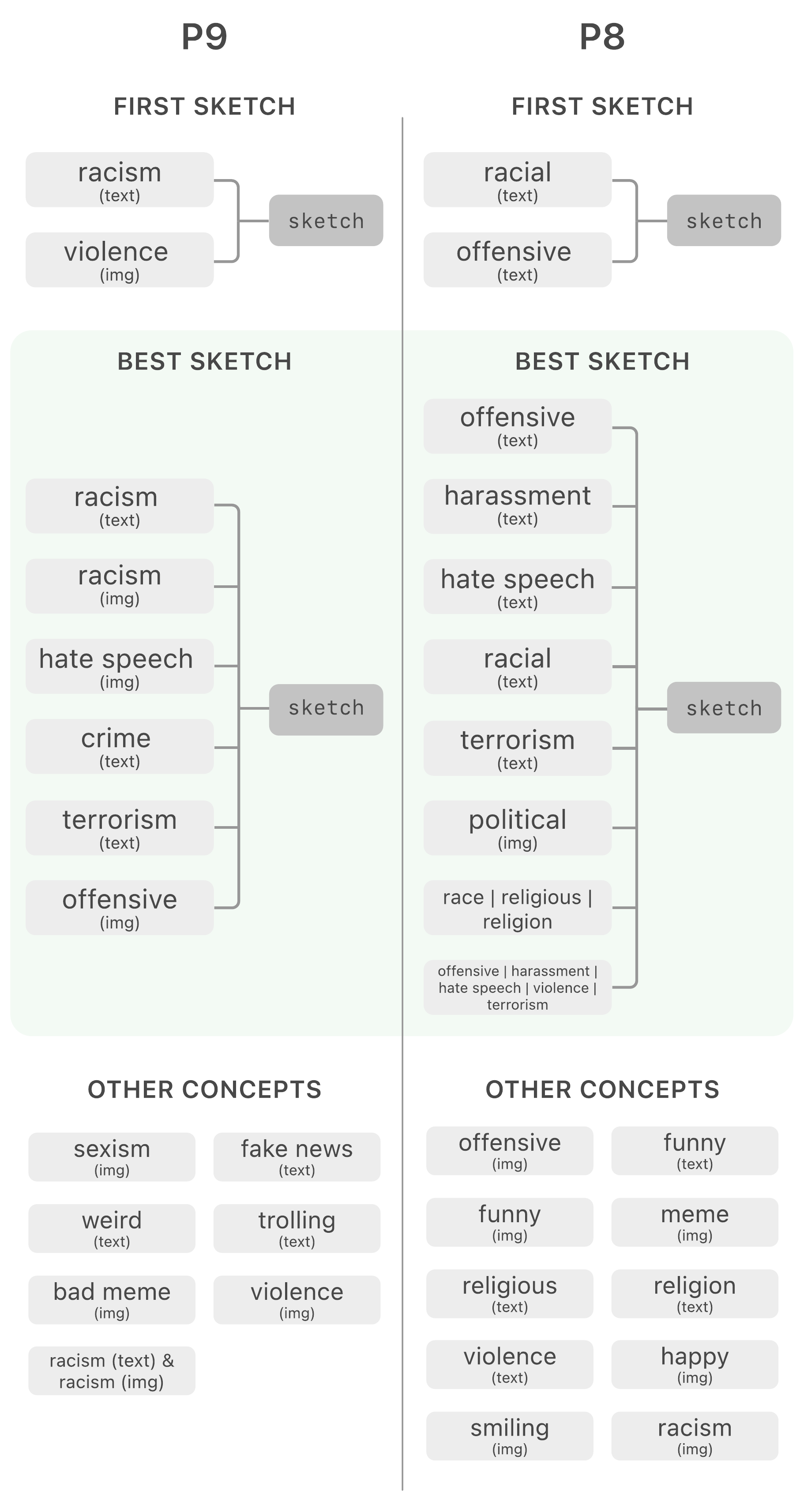}
  \caption{
    Examples of participant sketches. For illustrative purposes, we display the first sketch, best sketch, and other concepts for two study participants. Participants explored a broad range of concepts and expanded from their initial sketch models.
  }
  \label{fig:eval_sketches}
  \Description{Figure showing the first and best sketches of two participants (P9 and P8) and the concepts they used to create the particular sketch. Additionally, the figure shows other concepts that each participant created, but did not use in the two shown sketches.}
\end{figure}

Using our {ModelSketchBook} API, participants successfully authored an average of 12.2 (SD=4.7) concepts and 4.1 (SD=2.0) different sketch models that built on those concepts for the Hateful Memes task. Figure~\ref{fig:eval_sketches} displays sample concepts and sketch models authored by participants, \rev{and Table~\ref{table:concept_summary} summarizes all participants' concepts.}

Of the concepts authored in the study, $53.6\%$ were text concepts, $32.4\%$ were image concepts, and $14.0\%$ were compound concepts involving a logical operator. Participants each authored on average $6.53$ text concepts (SD=$3.00$), $4.19$ image concepts (SD=$2.14$), and $2.07$ compound concepts (SD=$1.53$). Of the compound concepts, $62.1\%$ used an \texttt{AND} operator while $37.9\%$ used an \texttt{OR} operator, and $79.3\%$ combined only text concepts while the remaining $20.7\%$ combined both text and image concepts.
There were 136 distinct concepts authored by the study participants. On an individual level, participants each authored an average of $6.29$ unique concepts (SD=$3.77$) that no other participants had authored.

We aimed for our modeling task to involve a high amount of subjectivity to provide a sufficiently broad model design space among participants. This was indeed the case: there were substantial differences in how our participants chose to label the provided dataset. Across participants and data examples, the mean standard deviation in labels (on a 0-1 scale) was $0.23$. For the binarized versions of those same scores, this meant that $75\%$ of the examples had at least one participant who disagreed with the majority label, and $37.5\%$ of the examples had a quarter or more of the participants who disagreed with the majority label.

\subsection{Changes in the cognitive process of model authoring}
Addressing RQ1, model sketching aided ML practitioners in shifting their thinking from technical, implementation-oriented factors toward higher-level, conceptual, design-oriented factors of model development. 

\begin{table*}[!tb]
  \centering
  \footnotesize
    \begin{tabular}{p{0.08\textwidth} p{0.18\textwidth} p{0.5\textwidth}}
    \toprule
    \textbf{Study Phase} &
    \textbf{Modeling Plan Theme} &
    \textbf{Quotes}\\
    \midrule
    \multirow{4}{0.1\textwidth}{\textbf{Before model sketching}}
    & {Specific model architectures or techniques} &
    {
        ``I would use a convolutional neural network model [...] probably use transformers in conjunction with CNNs to track long-term dependencies and help contextualize the text data.'' (P1),
        ``Two types of models: sentiment based and another that is a simple CNN, and then combine the two models with a simple fully connected layer.'' (P13)
    } \\[0.2cm]
    
    & 
    {Evaluating on training and validation sets} &
    {
        ``Ensure that the dataset is divided into train, test and validation slices. Experiment with several modeling techniques while evaluating on the validation set.'' (P3)
    } \\[0.2cm]

    & 
    {Standard performance metrics} &
    {
        ``Measure various metrics (accuracy, bias, variance etc).'' (P1),
        ``Figure out the right set of metrics for evaluating any model that is trained. Choices could be accuracy, F1 scores and other metrics such as AUC or AuPR depending on relative label weights.'' (P3)
    } \\[0.2cm]

    & 
    {Building low-level features stemming from data} &
    {
        ``Can use features from the posts and user interactions. Can use general user information along with the post information.'' (P4),
        ``Build features representing intuitions for things relevant to removal; like topic metadata, language choice, likes/dislikes/controversiality.'' (P5)
    } \\
    
    \midrule

    \multirow{4}{0.1\textwidth}{\textbf{After model sketching}}
    & {Spending time brainstorming and curating concepts} &
    {
        ``I would likely lean into feature engineering carefully to have an interpretable model. Spending a few days tinkering with concepts could produce a fairly effective model that would be easier to debug.'' (P15)
    } \\[0.2cm]
    
    & 
    {Iterating on concepts by inspecting sketch model outputs} &
    {
        ``[I would] see mislabelled (and correctly labeled) examples from the aggregator and brainstorm more specific features that might account for that.'' (P5)
    } \\[0.2cm]

    & 
    {Authoring concepts informed by different categories of harm} &
    {
        ``I think filtering on hate (race/gender/religion) speech and political speech would be good first steps. There are some examples of graphic/violent posts that might not be captured by those categories, so I would add more filters for those specifically.'' (P2)
    } \\[0.2cm]

    & 
    {Combining a concept-based approach with traditional approaches} &
    {
        ``Also can use these labels as annotations to the datapoints for a more complex model (nonlinear). Seems like having the original data is still good so wouldn't rely on the CLIP/GPT annotations alone.'' (P8),
        ``I will probably still collect data manually and then instead of using a pretrained model and build the pipeline from scratch, I will use the ModelSketchBook API.'' (P11)
    } \\

    \bottomrule
    \end{tabular}
    \caption{
    Participants' modeling plans before and after the study. Participants shifted from more technical, implementation-related plans to more procedural, value-oriented plans and chose to adopt model sketching in their future modeling plans.
    }
    \label{table:modeling_plan}
\end{table*}

\subsubsection{Changes in planned modeling approach}
In their pre-task modeling plans, participants tended to focus on implementation details such as specific modeling methods or architectures (e.g., transformer models, convolutional neural networks, supervised models, linear classifiers), ML conventions like train-test splits and performance metrics, and data-oriented plans for feature extraction (summarized in Table~\ref{table:modeling_plan}). 
For example, P13 described their approach as ``two types of models: [one] sentiment based and another that is a simple CNN, and then combine the two models with a simple fully connected layer,'' and P8 planned to ``create test/val/train sets with equal distributions of data'' and then ``train some model and inspect how it performs on a held out test set.''

However, after exposure to model sketching, participants described that they would spend time on brainstorming and curating concepts relevant to the decision-making task and that they would iterate on their concepts by inspecting sketch model outputs. P15 described that ``Spending a few days tinkering with concepts could produce a fairly effective model that would be easier to debug.'' Participants also branched out to mention that they would like to consider different subclasses of harm and described how they might combine a concept-based approach with more traditional modeling approaches. For example, while P8 still preferred to use a more complex, non-linear model, they wished to use concepts from ModelSketchBook to augment their datapoints. \rev{Participants also described changes to their mode of thinking during model sketching, which we qualitatively summarize in Appendix Section~\ref{section:participant_thought_processes}.}

\subsubsection{Concept evolution}
One window into a participant's mindset is the way that their concepts evolve over time while using ModelSketchBook.
We observed that on average, participants engaged in 8.5 (SD=2.7) instances of \textit{concept innovation}, which we define as the number of distinct concept ideas; on average 5.0 (SD=2.8) of those concept innovations were \textit{not} in the participant's original brainstormed list. Concept evolution often stemmed from participants discovering errors in their sketch models: P7 noted that a sketch model combining racism and sex concepts ran into false positives with humorous posts, which inspired an exploration of humor-related concepts. These progressions often built on each other: P4 initially started with a profanity concept that worked well, so they then brainstormed other dimensions of hate and decided to explore racism-related concepts; they then discovered that race and ethnicity-related hate in this dataset was often targeted at Muslims, so they got the idea to explore concepts related to Islamophobia.
Participants on average displayed 3.2 (SD=2.3) instances of \textit{concept execution}, which we define as the number of concepts authored that were continuations of the same concept idea (for example, switching from an image-based racism concept to a text-based racism concept, or changing from a concept term of ``racism'' to one of ``race''). Participants more frequently changed modalities, but rarely modified concept terms.
Then, participants underwent \textit{concept abandonment} for an average of 4.2 (SD=2.6) concept ideas; we define this phenomenon as the number of concept ideas that were not ultimately used in the user's self-designated ``best'' sketch model. This abandonment is also a valuable part of sketching---to identify and discard paths that appear to be dead-ends. For example, P14 tried a ``harmful'' image-based concept, but found that this surfaced noisy results; they concluded that this was because the concept was too vague and didn't closely relate to the images in the dataset, and they decided to explore more grounded, specific concepts for images.
Thus, participants appear to engage in a variety of different generative, iterative, and reflective modes as they author concepts.

\subsection{Modeling outcomes}
Regarding RQ2, we observed that the high-level cognitive shift in focus during model development brought about concrete benefits for the models themselves---from the breadth of concepts that sketch models explored to the gaps that participants uncovered in the modeling setup.

\subsubsection{Identifying modeling gaps}
\label{section:modeling_gaps}
Model sketching helped participants to gain insights beyond just the model itself.
Participants uncovered major categories of harm that were not sufficiently represented in the provided sample of the dataset, \rev{such as homophobia and transphobia, and they noted that memes in the dataset appeared unrealistic compared to current online memes, so models built on this data may fail in practice due to distribution shift}. Participants also identified limitations of the labeling approach such as class imbalance and inconsistent ratings. They gained intuitions about the task itself: participants discovered that for these memes, text was often more meaningful since the images were usually quite benign, and they started to form a sense of cases where multimodal approaches may reap more benefits. 
\rev{See a more detailed summary in Appendix Section~\ref{section:appendix_modeling_gaps}.}

\rev{
After the study concluded, we sought quantitative evidence to support the modeling gaps that participants raised. 

\textit{Dataset representativity}: First, we explored whether the concepts that participants claimed were insufficiently represented in their 40-example dataset were indeed underrepresented in the full 8.5k-example training dataset. Since the full dataset was not annotated for these concepts, a member of our team performed manual labeling for the three participant-identified concepts (``homophobia,'' ``transphobia,'' and ``nudity'') for both the 40-example study dataset and a random 2\% sample of the full training set (170 examples). As a point of comparison, we additionally performed manual annotation for three participant-authored, higher prevalence concepts---``religion,'' ``political,'' and ``racism''---which had 17.5\%, 20.0\%, and 7.5\% prevalence in the study dataset, respectively. We found that indeed homophobia, transphobia, and nudity, which had 0\%, 0\%, and 5\% prevalence in the study dataset, also appear underrepresented in the full training dataset with 1.8\%, 1.2\%, and 10\% prevalence~(Figure~\ref{fig:data_rep_modeling_gap}). 
Overall, we see that concept representativity in the small dataset used for model sketching can provide helpful warning signals on representativity in the full dataset.

\begin{figure}[!tb]
  \includegraphics[width=\linewidth]{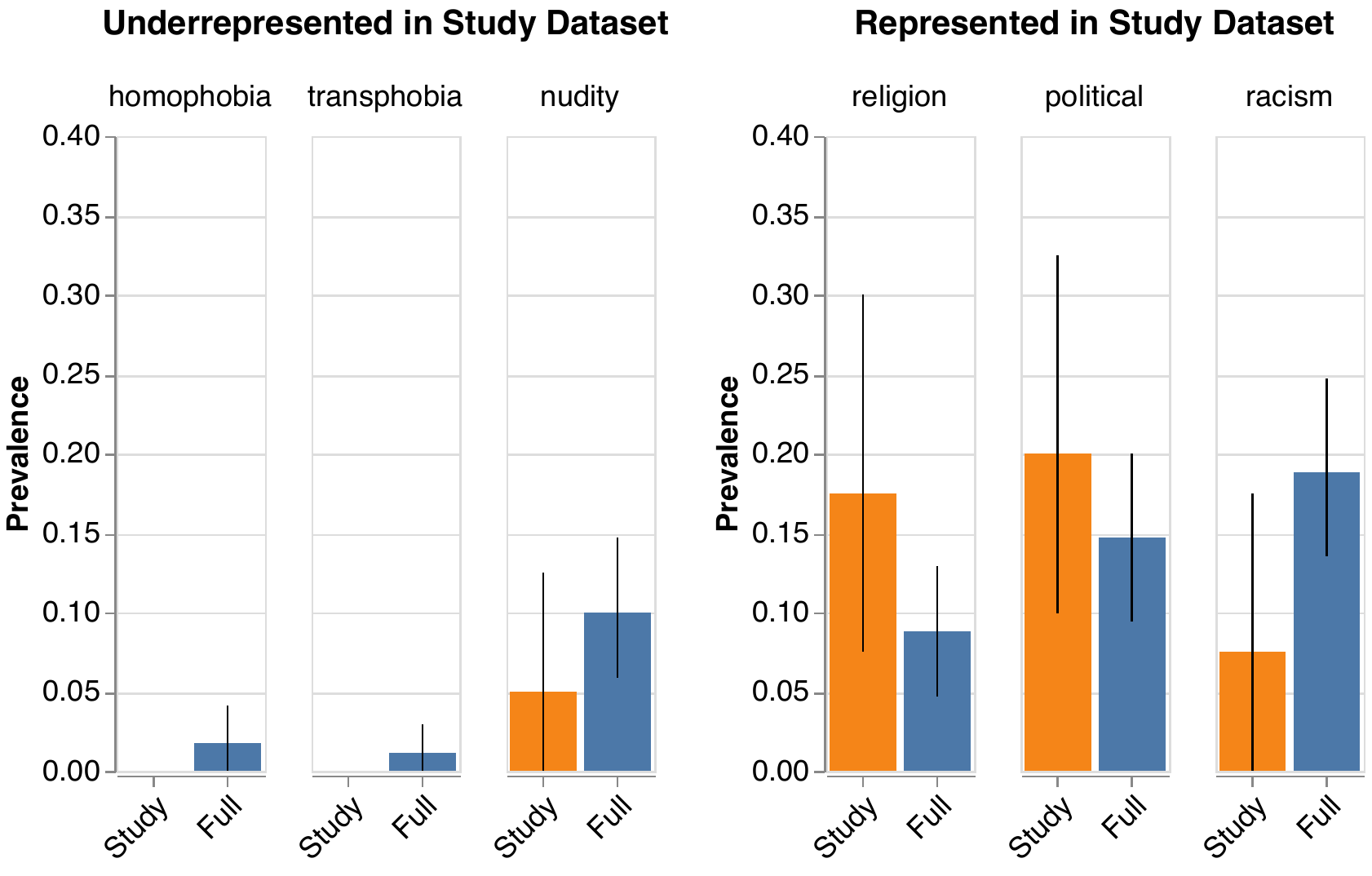}
  \caption{
  Concepts that participants noted as underrepresented based on their sketching process with the study dataset (left) were generally underrepresented in the full dataset. Already-prevalent concepts (right) maintained substantial representation in the full dataset. 
  }
  \label{fig:data_rep_modeling_gap}
  \Description{Bar chart comparing the prevalence of a set of concepts that was underrepresented in the study dataset versus concepts that were represented in the study dataset. For each concept, there is a pair of bars, one indicating prevalence for the study dataset and the other for the full dataset.}
\end{figure}

\textit{Class imbalance}: Then, we characterized the extent of class imbalance among participants' labels. During the study, four participants raised the issue of class imbalance; all of these participants indeed had a high level of imbalance, and all were skewed toward negative labels (mean=$79.3\%$ negative). Two of these participants displayed the highest class imbalance among all participants (with $92.5\%$ and $87.5\%$ negative labels). In addition, outside of the participants who noted the issue, three other participants also had more than $75\%$ negative labels. Thus, it appears that indeed class imbalance was a common challenge for this task.

We were encouraged to find that participants were able to rapidly identify potential modeling gaps before gathering a large-scale dataset or implementing a full model. 
}

\subsubsection{Concept diversity}
\begin{figure*}[!tb]
  \includegraphics[width=0.8\linewidth]{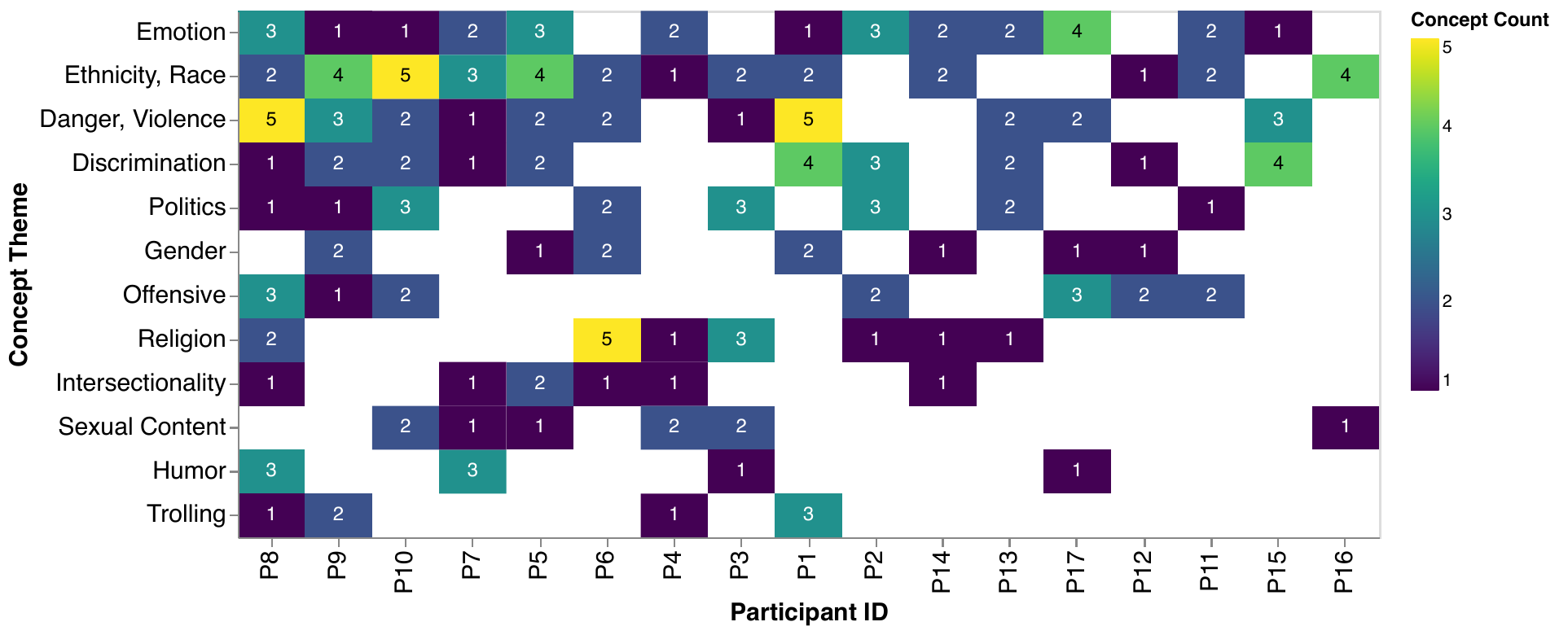}
  \caption{
  \rev{A heatmap matrix mapping from participants (x-axis) to concept themes (y-axis) depicts substantial spread between participants and themes, indicating that participants each explored diverse sets of concepts. Concept themes are sorted from top to bottom by the unique participant count, and participants are sorted from left to right by the unique concept theme count.
  }
  }
  \label{fig:concept_diversity_matrix}
  \Description{Heatmap matrix that lists the twelve concept themes along the y-axis and all participant IDs along the x-axis (P1-P17). Each cell in the matrix indicates the number of concepts authored by one participant for one concept theme with a count label in text and a background color according to a continuous gradient.}
\end{figure*}

Participants experimented with a broad variety of concept terms---ranging from concepts related to race and ethnicity to concepts related to politics, religion, and violence~(summarized by theme in Table~\ref{table:concept_summary}). 
Notably, all participants authored concepts that they had not considered in their initial brainstorm; the process of authoring concepts and sketches helped them to discover important and relevant concepts to incorporate. 
\rev{
Mapping participants' individual concepts to the 12 common themes that we identified, we found that participants created concepts that spanned on average $5.65$ of these themes (SD=$1.90$). In the reverse direction, each theme was covered by an average of $8.0$ participants (SD=$3.10$). We thus see that participants were able to explore a diverse set of concepts in their sketching sessions, and many concept themes were investigated from a variety of angles by different participants~(Figure~\ref{fig:concept_diversity_matrix}).
}

\begin{table}[!tb]
  \centering
  \footnotesize 
    \begin{tabular}{p{0.25\linewidth} p{0.5\linewidth} p{0.1\linewidth}}
    \toprule
    \textbf{Theme} & \textbf{Sample Participant Concepts} & \textbf{Count}\\
    \midrule
    \textbf{Ethnicity, Race} & 
    {racism, racist, race} &
    {34}\\[0.1cm]
    
    \textbf{Danger, Violence} & 
    {violence, terrorism, harm} &
    {28}\\[0.1cm]

    \textbf{Emotion} & 
    {rant, scary, anger} &
    {27}\\[0.1cm]
    
    \textbf{Discrimination} & 
    {hate, hate speech, discrimination} &
    {22}\\[0.1cm]
    
    \textbf{Politics} & 
    {political, politics, fake news} &
    {16}\\[0.1cm]
    
    \textbf{Offensive} & 
    {offensive, vulgar, obscene} &
    {15}\\[0.1cm]
    
    \textbf{Religion} & 
    {religion, religious, muslim} &
    {14}\\[0.1cm]
    
    \textbf{Gender} & 
    {sexist, sexism, gender} &
    {10}\\[0.1cm]
    
    \textbf{Sexual Content} & 
    {nudity, sexual, sex} &
    {9}\\[0.1cm]
    
    \textbf{Humor} & 
    {funny, humor, joke} &
    {8}\\[0.1cm]
    
    \textbf{Trolling} & 
    {trolling, meme, bad meme} &
    {7}\\[0.1cm]
    
    \textbf{Intersectionality} & 
    {color \& gender, islamophobia | sexism} &
    {7}\\
    \bottomrule
    \end{tabular}
    \caption{Summary of participant concepts for themes covered by 3 or more participants. We display the most common concept terms for each theme and the total number of participant concepts that fell within the theme.
    }
    \label{table:concept_summary}
\end{table}

\subsubsection{Time required}
Finally, we note that participants were able to achieve these positive modeling outcomes in a fraction of the time that would be required to author a full model for this task. Participants' time estimates for executing on their pre-task modeling plans for the Hateful Memes task varied drastically from half a day to several weeks or months. However, even considering the most optimistic estimate of several hours (provided by 3 of 17 participants), our model sketching approach was able to deliver functional model prototypes in less than 30 minutes. 
\rev{Beyond the speed of model sketching, we document participant experience outcomes in detail in Appendix Section~\ref{section:participant_experience}.}


\rev{
\subsection{Benchmarking concept performance}
While we found in our case study demonstrations and field evaluation that concepts worked sufficiently well to support users' sketch modeling process, we sought to characterize the accuracy of our zero-shot concept scores compared to human annotations. We selected the most frequently-instantiated concepts among our study participants, which had been independently used by four or more participants. These included four text-based concepts (``religion,'' ``political,'' ``racism,'' and ``violence'') and three image-based concepts (``racism,'' ``nudity,'' and ``violence''). Four members of our team independently provided ratings for these concepts on all 40 items shown to participants in our study. 
Based on these manual labels, the prevalence for the image-based concepts was too low to provide a reliable estimate, with a mean of $4.2\%$ prevalence (including 0 examples for the ``nudity'' concept) in participants' datasets.\footnote{\rev{This relates to the same issues that participants raised in Section~\ref{section:modeling_gaps} about the need for greater concept representation in the Hateful Memes dataset.}}
We thus repeated the same procedure with the next most frequently-instantiated image-based concepts among our study participants (``muslim,'' ``historical,'' ``political,'' and ``animal'').}

\begin{figure}[!tb]
    \includegraphics[width=\linewidth]{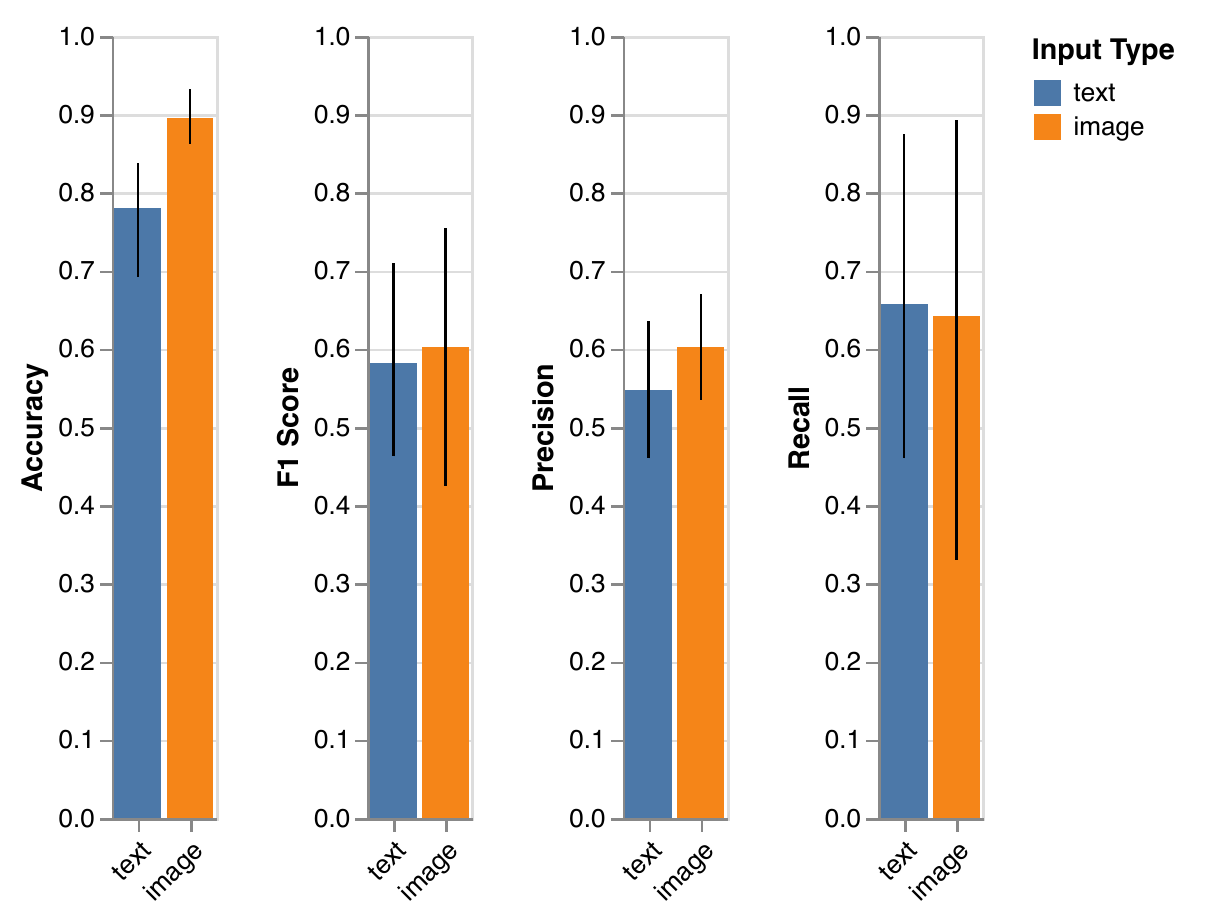}
    \caption{
    Performance for participants' text and image concepts is moderately high based on manual annotations.
    }
    \label{fig:concept_perf}
    \Description{Four bar charts indicating the Accuracy, F1 score, precision, and recall on the y-axis of each chart, with bars for text concepts and image concepts in each chart.}
\end{figure}

\begin{table}[!tb]
    \centering  
    \small 
    \rev{  
    \begin{tabular}{r r | c c}
    \toprule
    \textbf{} & \textbf{Concept Type} & \textbf{Text} & \textbf{Image}\\
    \midrule
    \multirow{2}{0.2\linewidth}{\raggedleft\textit{Annotation metrics}}&
    \textbf{Mean Prevalence} & 
    {19.4\%} &
    {13.3\%}\\[0.1cm]
    &\textbf{Fleiss' Kappa} & 
    {0.68} &
    {0.60}\\[0.1cm]
    \midrule
    \multirow{4}{0.2\linewidth}{\raggedleft\textit{Concept performance}}&
    \textbf{Accuracy} & 
    {0.78} &
    {0.90}\\[0.1cm]
    &\textbf{F1 Score} & 
    {0.58} &
    {0.60}\\[0.1cm]
    &\textbf{Recall} & 
    {0.66} &
    {0.64}\\[0.1cm]
    &\textbf{Precision} & 
    {0.55} &
    {0.60}\\
    \bottomrule
    \end{tabular}
    }
    \caption{
    Performance estimates were based on concepts that were sufficiently prevalent in the dataset and that displayed substantial inter-rater agreement.
    }
    \label{table:concept_perf}
\end{table}

We observe comparable levels of performance for both text and image concepts, as summarized in Figure~\ref{fig:concept_perf} and Table~\ref{table:concept_perf}. Both sets of concepts displayed substantial levels of inter-rater agreement as measured by Fleiss' Kappa and had sufficient prevalence levels based on these manual labels.
For the text-based concepts, we found a mean accuracy of $0.78$ (SD=$0.09$) and a mean F1 score of $0.58$ (SD=$0.14$), with a higher recall than precision (Recall=$0.66$, Precision=$0.55$). 
We found that for image-based concepts, there was a mean accuracy of $0.90$ (SD=$0.04$) and a mean F1 score of $0.60$ (SD=$0.20$), with higher recall than precision again (Recall=$0.64$, Precision=$0.60$).
Thus, for an approach that requires no data collection, concept labeling, or training time, the zero-shot concept scores provide users with a reasonable starting point for further iteration and refinement.

\subsection{Benchmarking sketch model performance}
While the goal of model sketching is not to produce full-scale models, we sought to understand whether sketch models achieve reasonable functionality relative to full-scale models.

\subsubsection{Study participants' sketch models}
\begin{figure*}[!tb]
  \includegraphics[width=\textwidth]{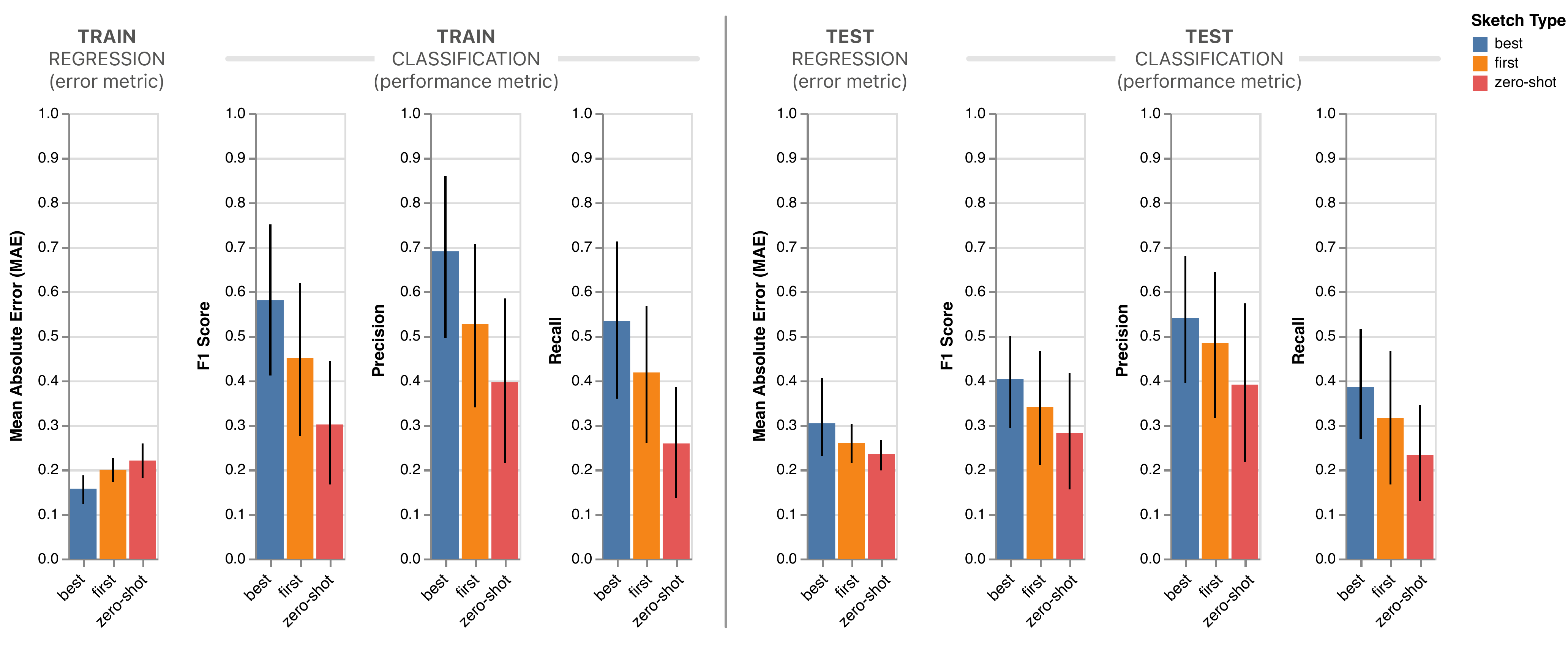}
  \caption{
  \rev{
  For the training set (left), we observe consistently better performance (MAE, F1 score, precision, and recall) for participants' self-designated ``best'' sketches compared to baselines of their first sketch and a zero-shot model.
  For the test set (right), we observe better classification performance (F1 score, precision, and recall) for the ``best'' sketch, but higher MAE.
  }
  }
  \label{fig:mean_metric_plot}
  \Description{Two sets of bar charts, one the training data and one for the test data. In each set are four bar charts of the mean absolute error (MAE), F1 score, precision and recall on the y-axis of each respective chart. Each chart has three sketch types on the x-axis: best, first, and zero-shot.}
\end{figure*}

First, we investigated the sketch models authored by our study participants. 
We compared all participants' self-determined ``best sketch model'' against two kinds of baselines: 
1)~\textit{first sketch}, the very first sketch that a participant authored with the {ModelSketchBook} API and 
2)~\rev{\textit{zero-shot}, a baseline that used a single GPT-3 prompt to predict whether each example is ``hateful'' or not. For each model variant, performance was measured against the participant's own ground truth labels on their 20-example test set.
We found that participants' best sketches outperformed both baselines on classification metrics (F1, precision, recall), but the best sketches achieved slightly lower performance than the baselines on the regression version of the task, as measured by Mean Absolute Error (MAE)~(Figure~\ref{fig:mean_metric_plot}). Looking back at performance on the 20-example training set upon which users were actively iterating, participants' best sketches more strongly outperformed the first sketch and zero-shot baselines on all metrics.
} 

These results suggest that participants are able to iteratively improve their sketch models to achieve higher alignment with their modeling goals and that sketch models may generalize better for classification tasks than for fine-grained regression tasks.
\rev{Given that sketch model performance metrics declined between the training set and test set, participants may have overfit to the training dataset during their sketch model development. Meanwhile, the zero-shot baseline displayed a generally worse---but consistent---level of performance between the train and test set as expected, precisely because this model was given no opportunity to overfit to the training set. These two extremes fall in line with classic bias-variance tradeoffs in machine learning. Our sketch model results mirror prior literature on the common challenge of overfitting tendencies in IML~\cite{wu2019localDecisionPitfallsIML, daee2019userModelingAvoidingOverfitting}, especially given that we did not take steps to explicitly steer users away from overfitting behavior.}
Lastly, we note that for this task, the classification metrics were relatively low overall due to class imbalance \rev{as discussed in Section~\ref{section:modeling_gaps}}. Many participants had substantial skew in labels toward the negative class (``non-hateful'') and thus ended up producing models that always predicted the negative class (behavior which results in scores of 0.0 for recall, precision, and F1).

\subsubsection{Investigating the impact of task difficulty on sketch model performance}
Then, we sought to further investigate the performance of sketch models in cases without such strong class imbalance and with many more than 20 training examples. We prepared three additional tasks: an IMDB movie review sentiment analysis task (easiest: per-example ratings and balanced classes)~\cite{imdb_sota}; a comment toxicity task (harder: per-annotator ratings and class skew)~\cite{kumar2021designing}; and the original Hateful Memes task (hardest: balanced classes, but nuanced and multimodal task)~\cite{kiela_hatefulMemes}. For each of these, we compared sketch models against published state-of-the-art (SOTA) models, using the same headline performance metric and train/test datasets as were used by the SOTA model. 
Please see Appendix Section~\ref{section:sota_details} for additional details on \revFinal{the benchmarks and our sketch models.}
For the IMDB task, the SOTA model reported 0.93 accuracy and our sketch model achieved 0.91 accuracy, only slightly lower.
On the comment toxicity task (which used a 0-4 score range), the SOTA model reported 0.90 MAE while our sketch model achieved an MAE of 1.39.
On the Hateful Memes task, the SOTA model reported 0.845 AUROC, and 0.714 AUROC was the SOTA result when the challenge was first released; our sketch model achieved 0.584 AUROC.
Thus, model sketching demonstrates strong or reasonable performance on the easy and medium difficulty tasks, but displays weaker performance results for the third, significantly harder task.

\subsection{Failure modes}
While our model sketching approach achieved its aims of shifting model developers' cognitive frame and presented a number of benefits for the models they developed, we observed several failure modes.
First, though participants had some promising concept ideas, they were not always able to actualize them with our zero-shot model-based approach. 
Participants often relied heavily on the results of the first concept term that they attempted. If that failed to return satisfactory results, they often abandoned the concept idea since they assumed that the model was not sophisticated enough to understand that concept. Participants would benefit from scaffolding to understand whether the model lacks an understanding of a given concept or just requires different terminology.

Second, we found that some participants heavily tailored their concept terms based on what they thought zero-shot models would be able to understand, especially using concrete, specific concept terms.
We observed that several of these participants eventually branched out and found that broader terms could work well. However, the downside still remains that since our concepts rely on pretrained models, these models can be limiting factors \rev{(further discussed in Section~\ref{section:limits_pretrained_models})}. Users might constrain their thinking to areas where the pretrained model already excels rather than thinking more creatively about what decisionmaking factors may be valuable.

Finally, we noticed that because our approach emphasized a ``small data'' approach to help users stay focused, some participants lost sight of the overall goal to build a robust, general model. Participants sometimes fell into overfitting approaches by scrutinizing cases where their sketch model made errors and patching errors with very specific concepts that were not relevant to the larger decision-making task. Participants would benefit from tools that would help them to assess the generalizability of their sketch models. We believe that a curated dataset is worth this tradeoff in the sketching context because it encourages ML practitioners to avoid detailed data analysis and instead focus on broad-strokes concepts. 

\section{Discussion}
We have introduced model sketching, instantiated the approach in a tool designed for ML practitioners, and demonstrated that the approach successfully redirects attention to the high-level model design questions that are critical to get right early on. Here, we discuss the opportunities afforded by model sketching as well as limitations and areas for future work.

\subsection{Opportunities for model sketching}
We envision several exciting implications for bringing a sketching practice to machine learning.


\subsubsection{Expanding the purview of sketch-inspired ML model development approaches}
While our work supports compound concepts with logical operators to add multimodality and more nuanced concepts, these are more cumbersome to author and may enforce too much rule-based rigidity.
Future work might explore multiple nested levels of model sketching to express complex concepts. For example, a concept like \textit{racism} is extremely multifaceted; racism has many forms and can be experienced in many different ways. Thus, rather than just accept a generic definition as given, a model designer must take care to specify the nuanced notion of racism that their particular problem formulation and model deployment context requires. Through nested sketch models, ML practitioners could iteratively ``zoom in'' to refine concepts and subconcepts as needed. 

\cut{Next, while concepts are a sufficiently generic abstraction to cover a broad range of existing ML modeling approaches, other abstractions may also enable sketch-like interaction. 
For example, if the ML practitioner's modeling goal is to build a model that captures the opinion of a diverse population, they may prefer to use individual voices in that population as their core abstraction~\cite{gordon_juryLearning}. If their goal is to reconcile multiple competing metrics or objectives~\cite{yu_keepingDesignersInTheLoop}, then they may prefer a primary abstraction based on distinct metrics to help them to reason about how the metrics might trade off. To reap the benefits of a sketching practice in machine learning, we need to build tools to support different \textit{ways of thinking} about modeling that are suited to varying modeling contexts.}

\subsubsection{Bringing sketching to model-adjacent tasks}
Participants in our evaluation uncovered gaps not just in their models, but in the larger modeling ecosystem of problem formulation, data, and labeling methods. Data work in particular tends to be neglected and undervalued relative to modeling work~\cite{sambasivan_everyoneWantsToDoModelWork, hohman2020dataIteration}. 
The model sketching approach could be extended to assist with some of these tasks. For example, in data annotation tasks, a major challenge lies in iteratively developing a refined labeling policy~\cite{kulesza_conceptEvolution}. Model sketching could iteratively simulate the results of different labeling policies, exposing areas where further concept refinement is needed \textit{before} ML practitioners launch costly human labeling efforts.


\subsubsection{Empowering non-technical users to participate in ML prototyping}
Currently, stakeholders who lack ML experience (whether end users, domain experts, designers, or product managers) have little ability to take ownership of ML model design decisions and are largely limited to the information that ML experts convey to them~\cite{piorkowski_AIDevelopersCommChallenges, lam2022endUserAudits}. Model sketching could enable non-experts in ML to build functional sketch models that incorporate their domain expertise, a highly valuable contribution given that ML practitioners typically lack expertise in particular deployment domains. Several participants in our user evaluation explicitly brought up that model sketching would be very useful for non-technical users.

\subsection{Limitations and future work}
\subsubsection{Expanding beyond the constraints of pretrained models and zero-shot modeling}
\label{section:limits_pretrained_models}
One limitation of our technical approach based on zero-shot modeling using pretrained models is that these models may not be flexible enough to support a broad range of modeling tasks. Given that the GPT-3 and CLIP models were trained on Internet text and image data, these models have a limited view of the world and cannot provide high-quality concept scores for domains that are not well represented in this training data. Future work might explore the efficacy of other specialized pretrained models or alternative zero-shot modeling approaches that could overcome these gaps.

While we restricted ourselves to zero-shot concept instantiation to preserve a sketch-like interaction, further work might take hybrid approaches that allow for example-based concept specification. As we found with our concept generalization benchmarking, zero-shot concepts achieve reasonable performance on average for the purpose of sketching, but they are not consistently accurate and may benefit from more manual control. Through few-shot modeling or fine-tuning on pretrained models, users might be able to refine their concepts with annotated examples in cases where that refinement is worth the additional time investment.

\subsubsection{Ensuring proper use and interpretation of sketch models}
We must take care to communicate to users the intended purpose of our tool. Since our approach bridges between human-understandable concepts and model predictions, ill-informed users might attempt to use our method for interpretability or explainability: purposes that our tool was not designed to support. Trading off accuracy for speed and flexibility, our model sketching approach was designed to assist with early-stage model design exploration. Meanwhile, an interpretability use case requires a much higher bar of concept and model accuracy. A model sketching tool must make these distinctions clear. By limiting the dataset size and orienting our tool around novel model authoring rather than post-hoc analysis of production models, we also steer users away from these unsupported use cases.

Additionally, as with any technology that builds on top of large language models and other pretrained models, sketch models will be subject to biases embedded in these models~\cite{kurita-etal-2019-measuring, sheng_biasesInLangGeneration}. It is crucial that ML practitioners are aware of these risks, so we encourage others who adopt model sketching to make clear to users that these underlying models can display harmful biases that may manifest in their sketch models. Though these biases in themselves are negative, one positive side effect is that encountering them early in the model development process forces ML practitioners to consider how the same kinds of biases might impact their full-scale models. Early encounters with such failures may prompt model developers to prioritize fairness-related work, and our approach allows them to experiment with different strategies---whether with the modeling approach, upstream data curation, or downstream score processing and bias checks---that might mitigate the harmful biases they encounter.


\subsubsection{Bridging from sketches to full-scale models}
While sketch models are early explorations rather than final versions of a model, they still need to provide useful signal on a full-scale model's behavior. As detailed in Section~\ref{section:sketch_to_production}, sketch models might inform the design of production models by guiding decisions on how to decompose large modeling tasks and refine data requirements. Our technical benchmark against state-of-the-art models suggests that model sketching can achieve substantial performance on some tasks, but may struggle to generalize on more challenging tasks.
The model sketching process still provides value in facilitating reflection and iteration on high-level modeling approaches even if performance metrics hit a ceiling. However, important areas of future work will be to support the bridging work between a model sketch and its production deployment and to provide stronger guarantees on the mapping between sketch and full-scale performance metrics. 


\section{Conclusion}
In a space dominated by high-fidelity technical implementation and engineering effort, we argue for low-fidelity, quick, and expressive explorations \rev{of a model's decision-making logic} with model sketching. 
Just as work in end-user programming and programming by demonstration for AI systems has long worked towards enabling users to focus on \textit{what} the system should do rather than \textit{how}~\cite{eager_progByExample, watchWhatIDo_PBD}, we seek to enable model authors to focus on \textit{what} their models should reason over rather than dive too early into \textit{how} their models are implemented when facing the task of problem formulation.
Instead of waiting weeks or months until a model is implemented before evaluating in the real world and discovering that \rev{a problem formulation encodes harmful biases}, we hope that ML practitioners can, in the course of a few days, instantiate multiple high-level model ideas and test them with users. Then, if they discover that their modeling approach is prone to bias in several ways, they can experiment with rounds of sketch models that might counter this bias, and only after this early-stage iteration would they move on to implement full production models. 
With model sketching, ML practitioners can move away from the mode of reactive implementation fixes and instead engage in the kind of proactive and creative model design exploration that can produce better models from the start.

\begin{acks}
\revFinal{We thank our anonymous reviewers as well as Ranjay Krishna, Matthew J\"{o}rke, Mitchell Gordon, Lindsay Popowski, and Hancheng Cao for their valuable feedback on our paper.
This work was partially supported by IBM as a founding member of the Stanford Institute for Human-centered Artificial Intelligence (HAI).
Michelle S. Lam was supported by Stanford HAI and the Brown Institute for Media Innovation at the Stanford School of Engineering.}
\end{acks}



\bibliographystyle{ACM-Reference-Format}
\bibliography{references}


\begin{thebibliography}{69}


\ifx \showCODEN    \undefined \def \showCODEN     #1{\unskip}     \fi
\ifx \showDOI      \undefined \def \showDOI       #1{#1}\fi
\ifx \showISBNx    \undefined \def \showISBNx     #1{\unskip}     \fi
\ifx \showISBNxiii \undefined \def \showISBNxiii  #1{\unskip}     \fi
\ifx \showISSN     \undefined \def \showISSN      #1{\unskip}     \fi
\ifx \showLCCN     \undefined \def \showLCCN      #1{\unskip}     \fi
\ifx \shownote     \undefined \def \shownote      #1{#1}          \fi
\ifx \showarticletitle \undefined \def \showarticletitle #1{#1}   \fi
\ifx \showURL      \undefined \def \showURL       {\relax}        \fi
\providecommand\bibfield[2]{#2}
\providecommand\bibinfo[2]{#2}
\providecommand\natexlab[1]{#1}
\providecommand\showeprint[2][]{arXiv:#2}

\bibitem[Amershi et~al\mbox{.}(2014)]%
        {amershi_powerToThePeople}
\bibfield{author}{\bibinfo{person}{Saleema Amershi}, \bibinfo{person}{Maya
  Cakmak}, \bibinfo{person}{William~Bradley Knox}, {and} \bibinfo{person}{Todd
  Kulesza}.} \bibinfo{year}{2014}\natexlab{}.
\newblock \showarticletitle{Power to the People: The Role of Humans in
  Interactive Machine Learning}.
\newblock \bibinfo{journal}{\emph{AI Magazine}} \bibinfo{volume}{35},
  \bibinfo{number}{4} (\bibinfo{date}{Dec.} \bibinfo{year}{2014}),
  \bibinfo{pages}{105--120}.
\newblock
\urldef\tempurl%
\url{https://doi.org/10.1609/aimag.v35i4.2513}
\showDOI{\tempurl}


\bibitem[Apple(2022)]%
        {create_ml}
\bibfield{author}{\bibinfo{person}{Apple}.} \bibinfo{year}{2022}\natexlab{}.
\newblock \bibinfo{title}{Create ML}.
\newblock
\newblock
\urldef\tempurl%
\url{https://developer.apple.com/machine-learning/create-ml/}
\showURL{%
\tempurl}


\bibitem[Aquilina(2022)]%
        {reddit_moderator_tools}
\bibfield{author}{\bibinfo{person}{Phil Aquilina}.}
  \bibinfo{year}{2022}\natexlab{}.
\newblock \bibinfo{title}{Building Better Moderator Tools}.
\newblock
\newblock
\urldef\tempurl%
\url{https://www.reddit.com/r/RedditEng/comments/uly8s4/building_better_moderator_tools/}
\showURL{%
\tempurl}


\bibitem[Brooks et~al\mbox{.}(2015)]%
        {brooks2015featureInsight}
\bibfield{author}{\bibinfo{person}{Michael Brooks}, \bibinfo{person}{Saleema
  Amershi}, \bibinfo{person}{Bongshin Lee}, \bibinfo{person}{Steven~M.
  Drucker}, \bibinfo{person}{Ashish Kapoor}, {and} \bibinfo{person}{Patrice
  Simard}.} \bibinfo{year}{2015}\natexlab{}.
\newblock \showarticletitle{FeatureInsight: Visual support for error-driven
  feature ideation in text classification}. In \bibinfo{booktitle}{\emph{2015
  IEEE Conference on Visual Analytics Science and Technology (VAST)}}.
  \bibinfo{pages}{105--112}.
\newblock
\urldef\tempurl%
\url{https://doi.org/10.1109/VAST.2015.7347637}
\showDOI{\tempurl}


\bibitem[Brown et~al\mbox{.}(2020)]%
        {GPT-3}
\bibfield{author}{\bibinfo{person}{Tom Brown}, \bibinfo{person}{Benjamin Mann},
  \bibinfo{person}{Nick Ryder}, \bibinfo{person}{Melanie Subbiah},
  \bibinfo{person}{Jared~D Kaplan}, \bibinfo{person}{Prafulla Dhariwal},
  \bibinfo{person}{Arvind Neelakantan}, \bibinfo{person}{Pranav Shyam},
  \bibinfo{person}{Girish Sastry}, \bibinfo{person}{Amanda Askell},
  \bibinfo{person}{Sandhini Agarwal}, \bibinfo{person}{Ariel Herbert-Voss},
  \bibinfo{person}{Gretchen Krueger}, \bibinfo{person}{Tom Henighan},
  \bibinfo{person}{Rewon Child}, \bibinfo{person}{Aditya Ramesh},
  \bibinfo{person}{Daniel Ziegler}, \bibinfo{person}{Jeffrey Wu},
  \bibinfo{person}{Clemens Winter}, \bibinfo{person}{Chris Hesse},
  \bibinfo{person}{Mark Chen}, \bibinfo{person}{Eric Sigler},
  \bibinfo{person}{Mateusz Litwin}, \bibinfo{person}{Scott Gray},
  \bibinfo{person}{Benjamin Chess}, \bibinfo{person}{Jack Clark},
  \bibinfo{person}{Christopher Berner}, \bibinfo{person}{Sam McCandlish},
  \bibinfo{person}{Alec Radford}, \bibinfo{person}{Ilya Sutskever}, {and}
  \bibinfo{person}{Dario Amodei}.} \bibinfo{year}{2020}\natexlab{}.
\newblock \showarticletitle{Language Models are Few-Shot Learners}. In
  \bibinfo{booktitle}{\emph{Advances in Neural Information Processing
  Systems}}, \bibfield{editor}{\bibinfo{person}{H.~Larochelle},
  \bibinfo{person}{M.~Ranzato}, \bibinfo{person}{R.~Hadsell},
  \bibinfo{person}{M.F. Balcan}, {and} \bibinfo{person}{H.~Lin}} (Eds.),
  Vol.~\bibinfo{volume}{33}. \bibinfo{publisher}{Curran Associates, Inc.},
  \bibinfo{pages}{1877--1901}.
\newblock
\urldef\tempurl%
\url{https://proceedings.neurips.cc/paper/2020/file/1457c0d6bfcb4967418bfb8ac142f64a-Paper.pdf}
\showURL{%
\tempurl}


\bibitem[Buxton(2010)]%
        {buxton2010sketching}
\bibfield{author}{\bibinfo{person}{Bill Buxton}.}
  \bibinfo{year}{2010}\natexlab{}.
\newblock \bibinfo{booktitle}{\emph{Sketching User Experiences: Getting the
  Design Right and the Right Design}}.
\newblock \bibinfo{publisher}{Morgan Kaufmann}.
\newblock


\bibitem[Chang et~al\mbox{.}(2017)]%
        {chang_revolt}
\bibfield{author}{\bibinfo{person}{Joseph~Chee Chang}, \bibinfo{person}{Saleema
  Amershi}, {and} \bibinfo{person}{Ece Kamar}.}
  \bibinfo{year}{2017}\natexlab{}.
\newblock \showarticletitle{Revolt: Collaborative Crowdsourcing for Labeling
  Machine Learning Datasets}. In \bibinfo{booktitle}{\emph{Proceedings of the
  2017 CHI Conference on Human Factors in Computing Systems}} (Denver,
  Colorado, USA) \emph{(\bibinfo{series}{CHI '17})}.
  \bibinfo{publisher}{Association for Computing Machinery},
  \bibinfo{address}{New York, NY, USA}, \bibinfo{pages}{2334–2346}.
\newblock
\showISBNx{9781450346559}
\urldef\tempurl%
\url{https://doi.org/10.1145/3025453.3026044}
\showDOI{\tempurl}


\bibitem[Charmaz(2006)]%
        {charmaz2006constructing}
\bibfield{author}{\bibinfo{person}{Kathy Charmaz}.}
  \bibinfo{year}{2006}\natexlab{}.
\newblock \bibinfo{booktitle}{\emph{Constructing grounded theory: A practical
  guide through qualitative analysis}}.
\newblock \bibinfo{publisher}{Sage}.
\newblock


\bibitem[Cypher(1991)]%
        {eager_progByExample}
\bibfield{author}{\bibinfo{person}{Allen Cypher}.}
  \bibinfo{year}{1991}\natexlab{}.
\newblock \showarticletitle{EAGER: Programming Repetitive Tasks by Example}. In
  \bibinfo{booktitle}{\emph{Proceedings of the SIGCHI Conference on Human
  Factors in Computing Systems}} (New Orleans, Louisiana, USA)
  \emph{(\bibinfo{series}{CHI '91})}. \bibinfo{publisher}{Association for
  Computing Machinery}, \bibinfo{address}{New York, NY, USA},
  \bibinfo{pages}{33–39}.
\newblock
\showISBNx{0897913833}
\urldef\tempurl%
\url{https://doi.org/10.1145/108844.108850}
\showDOI{\tempurl}


\bibitem[Cypher et~al\mbox{.}(1993)]%
        {watchWhatIDo_PBD}
\bibfield{editor}{\bibinfo{person}{Allen Cypher}, \bibinfo{person}{Daniel~C.
  Halbert}, \bibinfo{person}{David Kurlander}, \bibinfo{person}{Henry
  Lieberman}, \bibinfo{person}{David Maulsby}, \bibinfo{person}{Brad~A. Myers},
  {and} \bibinfo{person}{Alan Turransky}} (Eds.).
  \bibinfo{year}{1993}\natexlab{}.
\newblock \bibinfo{booktitle}{\emph{Watch What I Do: Programming by
  Demonstration}}.
\newblock \bibinfo{publisher}{MIT Press}, \bibinfo{address}{Cambridge, MA,
  USA}.
\newblock
\showISBNx{0262032139}


\bibitem[Daee et~al\mbox{.}(2018)]%
        {daee2019userModelingAvoidingOverfitting}
\bibfield{author}{\bibinfo{person}{Pedram Daee}, \bibinfo{person}{Tomi
  Peltola}, \bibinfo{person}{Aki Vehtari}, {and} \bibinfo{person}{Samuel
  Kaski}.} \bibinfo{year}{2018}\natexlab{}.
\newblock \showarticletitle{User Modelling for Avoiding Overfitting in
  Interactive Knowledge Elicitation for Prediction}. In
  \bibinfo{booktitle}{\emph{23rd International Conference on Intelligent User
  Interfaces}} (Tokyo, Japan) \emph{(\bibinfo{series}{IUI '18})}.
  \bibinfo{publisher}{Association for Computing Machinery},
  \bibinfo{address}{New York, NY, USA}, \bibinfo{pages}{305–310}.
\newblock
\showISBNx{9781450349451}
\urldef\tempurl%
\url{https://doi.org/10.1145/3172944.3172989}
\showDOI{\tempurl}


\bibitem[Dangi(2020)]%
        {linkedin_dwell_time}
\bibfield{author}{\bibinfo{person}{Siddharth Dangi}.}
  \bibinfo{year}{2020}\natexlab{}.
\newblock \bibinfo{title}{Understanding dwell time to improve LinkedIn feed
  ranking}.
\newblock
\newblock
\urldef\tempurl%
\url{https://engineering.linkedin.com/blog/2020/understanding-feed-dwell-time}
\showURL{%
\tempurl}


\bibitem[Dow et~al\mbox{.}(2011)]%
        {dow_parallelPrototyping}
\bibfield{author}{\bibinfo{person}{Steven~P. Dow}, \bibinfo{person}{Alana
  Glassco}, \bibinfo{person}{Jonathan Kass}, \bibinfo{person}{Melissa Schwarz},
  \bibinfo{person}{Daniel~L. Schwartz}, {and} \bibinfo{person}{Scott~R.
  Klemmer}.} \bibinfo{year}{2011}\natexlab{}.
\newblock \showarticletitle{Parallel Prototyping Leads to Better Design
  Results, More Divergence, and Increased Self-Efficacy}.
\newblock \bibinfo{journal}{\emph{ACM Trans. Comput.-Hum. Interact.}}
  \bibinfo{volume}{17}, \bibinfo{number}{4}, Article \bibinfo{articleno}{18}
  (\bibinfo{date}{dec} \bibinfo{year}{2011}), \bibinfo{numpages}{24}~pages.
\newblock
\showISSN{1073-0516}
\urldef\tempurl%
\url{https://doi.org/10.1145/1879831.1879836}
\showDOI{\tempurl}


\bibitem[Eckles(2022)]%
        {eckles_2022}
\bibfield{author}{\bibinfo{person}{Dean Eckles}.}
  \bibinfo{year}{2022}\natexlab{}.
\newblock \bibinfo{title}{Algorithmic transparency and assessing effects of
  algorithmic ranking}.
\newblock
\newblock
\urldef\tempurl%
\url{https://doi.org/10.31235/osf.io/c8za6}
\showDOI{\tempurl}


\bibitem[Fails and Olsen(2003)]%
        {fails_olsen_IML}
\bibfield{author}{\bibinfo{person}{Jerry~Alan Fails} {and}
  \bibinfo{person}{Dan~R. Olsen}.} \bibinfo{year}{2003}\natexlab{}.
\newblock \showarticletitle{Interactive Machine Learning}. In
  \bibinfo{booktitle}{\emph{Proceedings of the 8th International Conference on
  Intelligent User Interfaces}} (Miami, Florida, USA)
  \emph{(\bibinfo{series}{IUI '03})}. \bibinfo{publisher}{Association for
  Computing Machinery}, \bibinfo{address}{New York, NY, USA},
  \bibinfo{pages}{39–45}.
\newblock
\showISBNx{1581135866}
\urldef\tempurl%
\url{https://doi.org/10.1145/604045.604056}
\showDOI{\tempurl}


\bibitem[Fallman(2003)]%
        {fallman_designOrientedHCI}
\bibfield{author}{\bibinfo{person}{Daniel Fallman}.}
  \bibinfo{year}{2003}\natexlab{}.
\newblock \showarticletitle{Design-Oriented Human-Computer Interaction}. In
  \bibinfo{booktitle}{\emph{Proceedings of the SIGCHI Conference on Human
  Factors in Computing Systems}} (Ft. Lauderdale, Florida, USA)
  \emph{(\bibinfo{series}{CHI '03})}. \bibinfo{publisher}{Association for
  Computing Machinery}, \bibinfo{address}{New York, NY, USA},
  \bibinfo{pages}{225–232}.
\newblock
\showISBNx{1581136307}
\urldef\tempurl%
\url{https://doi.org/10.1145/642611.642652}
\showDOI{\tempurl}


\bibitem[Fiebrink et~al\mbox{.}(2009)]%
        {wekinator}
\bibfield{author}{\bibinfo{person}{Rebecca Fiebrink}, \bibinfo{person}{Dan
  Trueman}, {and} \bibinfo{person}{Perry~R. Cook}.}
  \bibinfo{year}{2009}\natexlab{}.
\newblock \showarticletitle{A Meta-Instrument for Interactive, On-the-fly
  Machine Learning}. In \bibinfo{booktitle}{\emph{International Conference on
  New Interfaces for Musical Expression}} (Pittsburgh, PA)
  \emph{(\bibinfo{series}{NIME '09})}.
\newblock


\bibitem[Fogarty et~al\mbox{.}(2008)]%
        {cueflik}
\bibfield{author}{\bibinfo{person}{James Fogarty}, \bibinfo{person}{Desney
  Tan}, \bibinfo{person}{Ashish Kapoor}, {and} \bibinfo{person}{Simon Winder}.}
  \bibinfo{year}{2008}\natexlab{}.
\newblock \showarticletitle{CueFlik: Interactive Concept Learning in Image
  Search}. In \bibinfo{booktitle}{\emph{Proceedings of the SIGCHI Conference on
  Human Factors in Computing Systems}} (Florence, Italy)
  \emph{(\bibinfo{series}{CHI '08})}. \bibinfo{publisher}{Association for
  Computing Machinery}, \bibinfo{address}{New York, NY, USA},
  \bibinfo{pages}{29–38}.
\newblock
\showISBNx{9781605580111}
\urldef\tempurl%
\url{https://doi.org/10.1145/1357054.1357061}
\showDOI{\tempurl}


\bibitem[Fran\c{c}oise et~al\mbox{.}(2021)]%
        {francoise2021marcelle}
\bibfield{author}{\bibinfo{person}{Jules Fran\c{c}oise},
  \bibinfo{person}{Baptiste Caramiaux}, {and} \bibinfo{person}{T\'{e}o
  Sanchez}.} \bibinfo{year}{2021}\natexlab{}.
\newblock \showarticletitle{Marcelle: Composing Interactive Machine Learning
  Workflows and Interfaces}. In \bibinfo{booktitle}{\emph{The 34th Annual ACM
  Symposium on User Interface Software and Technology}} (Virtual Event, USA)
  \emph{(\bibinfo{series}{UIST '21})}. \bibinfo{publisher}{Association for
  Computing Machinery}, \bibinfo{address}{New York, NY, USA},
  \bibinfo{pages}{39–53}.
\newblock
\showISBNx{9781450386357}
\urldef\tempurl%
\url{https://doi.org/10.1145/3472749.3474734}
\showDOI{\tempurl}


\bibitem[Goel(1995)]%
        {goel1995sketches}
\bibfield{author}{\bibinfo{person}{Vinod Goel}.}
  \bibinfo{year}{1995}\natexlab{}.
\newblock \bibinfo{booktitle}{\emph{Sketches of Thought}}.
\newblock \bibinfo{publisher}{MIT Press}.
\newblock


\bibitem[Google(2022)]%
        {teachable_machine}
\bibfield{author}{\bibinfo{person}{Google}.} \bibinfo{year}{2022}\natexlab{}.
\newblock \bibinfo{title}{Teachable Machine}.
\newblock
\newblock
\urldef\tempurl%
\url{https://teachablemachine.withgoogle.com/}
\showURL{%
\tempurl}


\bibitem[Gordon et~al\mbox{.}(2022)]%
        {gordon_juryLearning}
\bibfield{author}{\bibinfo{person}{Mitchell~L. Gordon},
  \bibinfo{person}{Michelle~S. Lam}, \bibinfo{person}{Joon~Sung Park},
  \bibinfo{person}{Kayur Patel}, \bibinfo{person}{Jeff Hancock},
  \bibinfo{person}{Tatsunori Hashimoto}, {and} \bibinfo{person}{Michael~S.
  Bernstein}.} \bibinfo{year}{2022}\natexlab{}.
\newblock \showarticletitle{Jury Learning: Integrating Dissenting Voices into
  Machine Learning Models}. In \bibinfo{booktitle}{\emph{Proceedings of the
  2022 CHI Conference on Human Factors in Computing Systems}} (New Orleans, LA,
  USA) \emph{(\bibinfo{series}{CHI '22})}. \bibinfo{publisher}{Association for
  Computing Machinery}, \bibinfo{address}{New York, NY, USA}, Article
  \bibinfo{articleno}{115}, \bibinfo{numpages}{19}~pages.
\newblock
\showISBNx{9781450391573}
\urldef\tempurl%
\url{https://doi.org/10.1145/3491102.3502004}
\showDOI{\tempurl}


\bibitem[Hartmann et~al\mbox{.}(2006)]%
        {hartmann_dTools}
\bibfield{author}{\bibinfo{person}{Bj\"{o}rn Hartmann},
  \bibinfo{person}{Scott~R. Klemmer}, \bibinfo{person}{Michael Bernstein},
  \bibinfo{person}{Leith Abdulla}, \bibinfo{person}{Brandon Burr},
  \bibinfo{person}{Avi Robinson-Mosher}, {and} \bibinfo{person}{Jennifer Gee}.}
  \bibinfo{year}{2006}\natexlab{}.
\newblock \showarticletitle{Reflective Physical Prototyping through Integrated
  Design, Test, and Analysis}. In \bibinfo{booktitle}{\emph{Proceedings of the
  19th Annual ACM Symposium on User Interface Software and Technology}}
  (Montreux, Switzerland) \emph{(\bibinfo{series}{UIST '06})}.
  \bibinfo{publisher}{Association for Computing Machinery},
  \bibinfo{address}{New York, NY, USA}, \bibinfo{pages}{299–308}.
\newblock
\showISBNx{1595933131}
\urldef\tempurl%
\url{https://doi.org/10.1145/1166253.1166300}
\showDOI{\tempurl}


\bibitem[Hohman et~al\mbox{.}(2020)]%
        {hohman2020dataIteration}
\bibfield{author}{\bibinfo{person}{Fred Hohman}, \bibinfo{person}{Kanit
  Wongsuphasawat}, \bibinfo{person}{Mary~Beth Kery}, {and}
  \bibinfo{person}{Kayur Patel}.} \bibinfo{year}{2020}\natexlab{}.
\newblock \showarticletitle{Understanding and Visualizing Data Iteration in
  Machine Learning}. In \bibinfo{booktitle}{\emph{Proceedings of the 2020 CHI
  Conference on Human Factors in Computing Systems}} (Honolulu, HI, USA)
  \emph{(\bibinfo{series}{CHI '20})}. \bibinfo{publisher}{Association for
  Computing Machinery}, \bibinfo{address}{New York, NY, USA},
  \bibinfo{pages}{1–13}.
\newblock
\showISBNx{9781450367080}
\urldef\tempurl%
\url{https://doi.org/10.1145/3313831.3376177}
\showDOI{\tempurl}


\bibitem[Hou and Wang(2017)]%
        {hou2017hacking}
\bibfield{author}{\bibinfo{person}{Youyang Hou} {and} \bibinfo{person}{Dakuo
  Wang}.} \bibinfo{year}{2017}\natexlab{}.
\newblock \showarticletitle{Hacking with NPOs: Collaborative Analytics and
  Broker Roles in Civic Data Hackathons}.
\newblock \bibinfo{journal}{\emph{Proc. ACM Hum.-Comput. Interact.}}
  \bibinfo{volume}{1}, \bibinfo{number}{CSCW}, Article \bibinfo{articleno}{53}
  (\bibinfo{date}{dec} \bibinfo{year}{2017}), \bibinfo{numpages}{16}~pages.
\newblock
\urldef\tempurl%
\url{https://doi.org/10.1145/3134688}
\showDOI{\tempurl}


\bibitem[Ilharco et~al\mbox{.}(2021)]%
        {open_clip}
\bibfield{author}{\bibinfo{person}{Gabriel Ilharco}, \bibinfo{person}{Mitchell
  Wortsman}, \bibinfo{person}{Ross Wightman}, \bibinfo{person}{Cade Gordon},
  \bibinfo{person}{Nicholas Carlini}, \bibinfo{person}{Rohan Taori},
  \bibinfo{person}{Achal Dave}, \bibinfo{person}{Vaishaal Shankar},
  \bibinfo{person}{Hongseok Namkoong}, \bibinfo{person}{John Miller},
  \bibinfo{person}{Hannaneh Hajishirzi}, \bibinfo{person}{Ali Farhadi}, {and}
  \bibinfo{person}{Ludwig Schmidt}.} \bibinfo{year}{2021}\natexlab{}.
\newblock \bibinfo{booktitle}{\emph{OpenCLIP}}.
\newblock
\urldef\tempurl%
\url{https://doi.org/10.5281/zenodo.5143773}
\showDOI{\tempurl}
\newblock
\shownote{If you use this software, please cite it as below.}.


\bibitem[Jacobs and Wallach(2021)]%
        {jacobs_wallach_measurementAndFairness}
\bibfield{author}{\bibinfo{person}{Abigail~Z. Jacobs} {and}
  \bibinfo{person}{Hanna Wallach}.} \bibinfo{year}{2021}\natexlab{}.
\newblock \showarticletitle{Measurement and Fairness}. In
  \bibinfo{booktitle}{\emph{Proceedings of the 2021 ACM Conference on Fairness,
  Accountability, and Transparency}} (Virtual Event, Canada)
  \emph{(\bibinfo{series}{FAccT '21})}. \bibinfo{publisher}{Association for
  Computing Machinery}, \bibinfo{address}{New York, NY, USA},
  \bibinfo{pages}{375–385}.
\newblock
\showISBNx{9781450383097}
\urldef\tempurl%
\url{https://doi.org/10.1145/3442188.3445901}
\showDOI{\tempurl}


\bibitem[Jiang et~al\mbox{.}(2022)]%
        {jiang_PromptMaker}
\bibfield{author}{\bibinfo{person}{Ellen Jiang}, \bibinfo{person}{Kristen
  Olson}, \bibinfo{person}{Edwin Toh}, \bibinfo{person}{Alejandra Molina},
  \bibinfo{person}{Aaron Donsbach}, \bibinfo{person}{Michael Terry}, {and}
  \bibinfo{person}{Carrie~J Cai}.} \bibinfo{year}{2022}\natexlab{}.
\newblock \showarticletitle{PromptMaker: Prompt-Based Prototyping with Large
  Language Models}. In \bibinfo{booktitle}{\emph{Extended Abstracts of the 2022
  CHI Conference on Human Factors in Computing Systems}} (New Orleans, LA, USA)
  \emph{(\bibinfo{series}{CHI EA '22})}. \bibinfo{publisher}{Association for
  Computing Machinery}, \bibinfo{address}{New York, NY, USA}, Article
  \bibinfo{articleno}{35}, \bibinfo{numpages}{8}~pages.
\newblock
\showISBNx{9781450391566}
\urldef\tempurl%
\url{https://doi.org/10.1145/3491101.3503564}
\showDOI{\tempurl}


\bibitem[Jun et~al\mbox{.}(2022)]%
        {jun_hypothesisFormalization}
\bibfield{author}{\bibinfo{person}{Eunice Jun}, \bibinfo{person}{Melissa
  Birchfield}, \bibinfo{person}{Nicole De~Moura}, \bibinfo{person}{Jeffrey
  Heer}, {and} \bibinfo{person}{Ren\'{e} Just}.}
  \bibinfo{year}{2022}\natexlab{}.
\newblock \showarticletitle{Hypothesis Formalization: Empirical Findings,
  Software Limitations, and Design Implications}.
\newblock \bibinfo{journal}{\emph{ACM Trans. Comput.-Hum. Interact.}}
  \bibinfo{volume}{29}, \bibinfo{number}{1}, Article \bibinfo{articleno}{6}
  (\bibinfo{date}{Jan} \bibinfo{year}{2022}), \bibinfo{numpages}{28}~pages.
\newblock
\showISSN{1073-0516}
\urldef\tempurl%
\url{https://doi.org/10.1145/3476980}
\showDOI{\tempurl}


\bibitem[Jung et~al\mbox{.}(2020)]%
        {jung_goel_simpleRulesExpertClassification}
\bibfield{author}{\bibinfo{person}{Jongbin Jung}, \bibinfo{person}{Connor
  Concannon}, \bibinfo{person}{Ravi Shroff}, \bibinfo{person}{Sharad Goel},
  {and} \bibinfo{person}{Daniel~G. Goldstein}.}
  \bibinfo{year}{2020}\natexlab{}.
\newblock \showarticletitle{Simple rules to guide expert classifications}.
\newblock \bibinfo{journal}{\emph{Journal of the Royal Statistical Society:
  Series A (Statistics in Society)}} \bibinfo{volume}{183}, \bibinfo{number}{3}
  (\bibinfo{year}{2020}), \bibinfo{pages}{771--800}.
\newblock
\urldef\tempurl%
\url{https://doi.org/10.1111/rssa.12576}
\showDOI{\tempurl}
\showeprint{https://rss.onlinelibrary.wiley.com/doi/pdf/10.1111/rssa.12576}


\bibitem[Kiela et~al\mbox{.}(2020)]%
        {kiela_hatefulMemes}
\bibfield{author}{\bibinfo{person}{Douwe Kiela}, \bibinfo{person}{Hamed
  Firooz}, \bibinfo{person}{Aravind Mohan}, \bibinfo{person}{Vedanuj Goswami},
  \bibinfo{person}{Amanpreet Singh}, \bibinfo{person}{Pratik Ringshia}, {and}
  \bibinfo{person}{Davide Testuggine}.} \bibinfo{year}{2020}\natexlab{}.
\newblock \showarticletitle{The Hateful Memes Challenge: Detecting Hate Speech
  in Multimodal Memes}. In \bibinfo{booktitle}{\emph{Advances in Neural
  Information Processing Systems}},
  \bibfield{editor}{\bibinfo{person}{H.~Larochelle},
  \bibinfo{person}{M.~Ranzato}, \bibinfo{person}{R.~Hadsell},
  \bibinfo{person}{M.F. Balcan}, {and} \bibinfo{person}{H.~Lin}} (Eds.),
  Vol.~\bibinfo{volume}{33}. \bibinfo{publisher}{Curran Associates, Inc.},
  \bibinfo{pages}{2611--2624}.
\newblock
\urldef\tempurl%
\url{https://proceedings.neurips.cc/paper/2020/file/1b84c4cee2b8b3d823b30e2d604b1878-Paper.pdf}
\showURL{%
\tempurl}


\bibitem[Kim et~al\mbox{.}(2018)]%
        {kim2018TCAV}
\bibfield{author}{\bibinfo{person}{Been Kim}, \bibinfo{person}{Martin
  Wattenberg}, \bibinfo{person}{Justin Gilmer}, \bibinfo{person}{Carrie Cai},
  \bibinfo{person}{James Wexler}, \bibinfo{person}{Fernanda Viegas}, {and}
  \bibinfo{person}{Rory sayres}.} \bibinfo{year}{2018}\natexlab{}.
\newblock \showarticletitle{Interpretability Beyond Feature Attribution:
  Quantitative Testing with Concept Activation Vectors ({TCAV})}. In
  \bibinfo{booktitle}{\emph{Proceedings of the 35th International Conference on
  Machine Learning}} \emph{(\bibinfo{series}{Proceedings of Machine Learning
  Research}, Vol.~\bibinfo{volume}{80})},
  \bibfield{editor}{\bibinfo{person}{Jennifer Dy} {and}
  \bibinfo{person}{Andreas Krause}} (Eds.). \bibinfo{publisher}{PMLR},
  \bibinfo{pages}{2668--2677}.
\newblock
\urldef\tempurl%
\url{https://proceedings.mlr.press/v80/kim18d.html}
\showURL{%
\tempurl}


\bibitem[Koh et~al\mbox{.}(2020)]%
        {koh2020ConceptBottleneck}
\bibfield{author}{\bibinfo{person}{Pang~Wei Koh}, \bibinfo{person}{Thao
  Nguyen}, \bibinfo{person}{Yew~Siang Tang}, \bibinfo{person}{Stephen
  Mussmann}, \bibinfo{person}{Emma Pierson}, \bibinfo{person}{Been Kim}, {and}
  \bibinfo{person}{Percy Liang}.} \bibinfo{year}{2020}\natexlab{}.
\newblock \showarticletitle{Concept Bottleneck Models}. In
  \bibinfo{booktitle}{\emph{Proceedings of the 37th International Conference on
  Machine Learning}} \emph{(\bibinfo{series}{Proceedings of Machine Learning
  Research}, Vol.~\bibinfo{volume}{119})},
  \bibfield{editor}{\bibinfo{person}{Hal~Daumé III} {and}
  \bibinfo{person}{Aarti Singh}} (Eds.). \bibinfo{publisher}{PMLR},
  \bibinfo{pages}{5338--5348}.
\newblock
\urldef\tempurl%
\url{https://proceedings.mlr.press/v119/koh20a.html}
\showURL{%
\tempurl}


\bibitem[Kulesza et~al\mbox{.}(2014)]%
        {kulesza_conceptEvolution}
\bibfield{author}{\bibinfo{person}{Todd Kulesza}, \bibinfo{person}{Saleema
  Amershi}, \bibinfo{person}{Rich Caruana}, \bibinfo{person}{Danyel Fisher},
  {and} \bibinfo{person}{Denis Charles}.} \bibinfo{year}{2014}\natexlab{}.
\newblock \showarticletitle{Structured Labeling for Facilitating Concept
  Evolution in Machine Learning}. In \bibinfo{booktitle}{\emph{Proceedings of
  the SIGCHI Conference on Human Factors in Computing Systems}} (Toronto,
  Ontario, Canada) \emph{(\bibinfo{series}{CHI '14})}.
  \bibinfo{publisher}{Association for Computing Machinery},
  \bibinfo{address}{New York, NY, USA}, \bibinfo{pages}{3075–3084}.
\newblock
\showISBNx{9781450324731}
\urldef\tempurl%
\url{https://doi.org/10.1145/2556288.2557238}
\showDOI{\tempurl}


\bibitem[Kumar et~al\mbox{.}(2021)]%
        {kumar2021designing}
\bibfield{author}{\bibinfo{person}{Deepak Kumar}, \bibinfo{person}{Patrick~Gage
  Kelley}, \bibinfo{person}{Sunny Consolvo}, \bibinfo{person}{Joshua Mason},
  \bibinfo{person}{Elie Bursztein}, \bibinfo{person}{Zakir Durumeric},
  \bibinfo{person}{Kurt Thomas}, {and} \bibinfo{person}{Michael Bailey}.}
  \bibinfo{year}{2021}\natexlab{}.
\newblock \showarticletitle{Designing Toxic Content Classification for a
  Diversity of Perspectives}. In \bibinfo{booktitle}{\emph{Seventeenth
  Symposium on Usable Privacy and Security (SOUPS 2021)}}.
  \bibinfo{publisher}{USENIX Association}, \bibinfo{pages}{299--318}.
\newblock
\showISBNx{978-1-939133-25-0}
\urldef\tempurl%
\url{https://www.usenix.org/conference/soups2021/presentation/kumar}
\showURL{%
\tempurl}


\bibitem[Kurita et~al\mbox{.}(2019)]%
        {kurita-etal-2019-measuring}
\bibfield{author}{\bibinfo{person}{Keita Kurita}, \bibinfo{person}{Nidhi Vyas},
  \bibinfo{person}{Ayush Pareek}, \bibinfo{person}{Alan~W Black}, {and}
  \bibinfo{person}{Yulia Tsvetkov}.} \bibinfo{year}{2019}\natexlab{}.
\newblock \showarticletitle{Measuring Bias in Contextualized Word
  Representations}. In \bibinfo{booktitle}{\emph{Proceedings of the First
  Workshop on Gender Bias in Natural Language Processing}}.
  \bibinfo{publisher}{Association for Computational Linguistics},
  \bibinfo{address}{Florence, Italy}, \bibinfo{pages}{166--172}.
\newblock
\urldef\tempurl%
\url{https://doi.org/10.18653/v1/W19-3823}
\showDOI{\tempurl}


\bibitem[Lam et~al\mbox{.}(2022)]%
        {lam2022endUserAudits}
\bibfield{author}{\bibinfo{person}{Michelle~S. Lam},
  \bibinfo{person}{Mitchell~L. Gordon}, \bibinfo{person}{Dana\"{e} Metaxa},
  \bibinfo{person}{Jeffrey~T. Hancock}, \bibinfo{person}{James~A. Landay},
  {and} \bibinfo{person}{Michael~S. Bernstein}.}
  \bibinfo{year}{2022}\natexlab{}.
\newblock \showarticletitle{End-User Audits: A System Empowering Communities to
  Lead Large-Scale Investigations of Harmful Algorithmic Behavior}.
\newblock \bibinfo{journal}{\emph{Proc. ACM Hum.-Comput. Interact.}}
  \bibinfo{volume}{6}, \bibinfo{number}{CSCW2}, Article
  \bibinfo{articleno}{512} (\bibinfo{date}{Nov} \bibinfo{year}{2022}),
  \bibinfo{numpages}{34}~pages.
\newblock
\urldef\tempurl%
\url{https://doi.org/10.1145/3555625}
\showDOI{\tempurl}


\bibitem[Landay and Myers(1995)]%
        {landay_interactiveSketching}
\bibfield{author}{\bibinfo{person}{James~A. Landay} {and}
  \bibinfo{person}{Brad~A. Myers}.} \bibinfo{year}{1995}\natexlab{}.
\newblock \showarticletitle{Interactive Sketching for the Early Stages of User
  Interface Design}. In \bibinfo{booktitle}{\emph{Proceedings of the SIGCHI
  Conference on Human Factors in Computing Systems}} (Denver, Colorado, USA)
  \emph{(\bibinfo{series}{CHI '95})}. \bibinfo{publisher}{ACM
  Press/Addison-Wesley Publishing Co.}, \bibinfo{address}{USA},
  \bibinfo{pages}{43–50}.
\newblock
\showISBNx{0201847051}
\urldef\tempurl%
\url{https://doi.org/10.1145/223904.223910}
\showDOI{\tempurl}


\bibitem[Lim et~al\mbox{.}(2008)]%
        {lim_anatomyOfPrototypes}
\bibfield{author}{\bibinfo{person}{Youn-Kyung Lim}, \bibinfo{person}{Erik
  Stolterman}, {and} \bibinfo{person}{Josh Tenenberg}.}
  \bibinfo{year}{2008}\natexlab{}.
\newblock \showarticletitle{The Anatomy of Prototypes: Prototypes as Filters,
  Prototypes as Manifestations of Design Ideas}.
\newblock \bibinfo{journal}{\emph{ACM Trans. Comput.-Hum. Interact.}}
  \bibinfo{volume}{15}, \bibinfo{number}{2}, Article \bibinfo{articleno}{7}
  (\bibinfo{date}{jul} \bibinfo{year}{2008}), \bibinfo{numpages}{27}~pages.
\newblock
\showISSN{1073-0516}
\urldef\tempurl%
\url{https://doi.org/10.1145/1375761.1375762}
\showDOI{\tempurl}


\bibitem[Maas et~al\mbox{.}(2011)]%
        {imdb_dataset}
\bibfield{author}{\bibinfo{person}{Andrew~L. Maas}, \bibinfo{person}{Raymond~E.
  Daly}, \bibinfo{person}{Peter~T. Pham}, \bibinfo{person}{Dan Huang},
  \bibinfo{person}{Andrew~Y. Ng}, {and} \bibinfo{person}{Christopher Potts}.}
  \bibinfo{year}{2011}\natexlab{}.
\newblock \showarticletitle{Learning Word Vectors for Sentiment Analysis}. In
  \bibinfo{booktitle}{\emph{Proceedings of the 49th Annual Meeting of the
  Association for Computational Linguistics: Human Language Technologies}}.
  \bibinfo{publisher}{Association for Computational Linguistics},
  \bibinfo{address}{Portland, Oregon, USA}, \bibinfo{pages}{142--150}.
\newblock
\urldef\tempurl%
\url{https://aclanthology.org/P11-1015}
\showURL{%
\tempurl}


\bibitem[Mao et~al\mbox{.}(2019)]%
        {mao2019data}
\bibfield{author}{\bibinfo{person}{Yaoli Mao}, \bibinfo{person}{Dakuo Wang},
  \bibinfo{person}{Michael Muller}, \bibinfo{person}{Kush~R. Varshney},
  \bibinfo{person}{Ioana Baldini}, \bibinfo{person}{Casey Dugan}, {and}
  \bibinfo{person}{Aleksandra Mojsilovi\'{c}}.}
  \bibinfo{year}{2019}\natexlab{}.
\newblock \showarticletitle{How Data Scientists Work Together With Domain
  Experts in Scientific Collaborations: To Find The Right Answer Or To Ask The
  Right Question?}
\newblock \bibinfo{journal}{\emph{Proc. ACM Hum.-Comput. Interact.}}
  \bibinfo{volume}{3}, \bibinfo{number}{GROUP}, Article
  \bibinfo{articleno}{237} (\bibinfo{date}{dec} \bibinfo{year}{2019}),
  \bibinfo{numpages}{23}~pages.
\newblock
\urldef\tempurl%
\url{https://doi.org/10.1145/3361118}
\showDOI{\tempurl}


\bibitem[McDonald et~al\mbox{.}(2019)]%
        {mcDonald2019IRRQualitativeResearch}
\bibfield{author}{\bibinfo{person}{Nora McDonald}, \bibinfo{person}{Sarita
  Schoenebeck}, {and} \bibinfo{person}{Andrea Forte}.}
  \bibinfo{year}{2019}\natexlab{}.
\newblock \showarticletitle{Reliability and Inter-Rater Reliability in
  Qualitative Research: Norms and Guidelines for CSCW and HCI Practice}.
\newblock \bibinfo{journal}{\emph{Proc. ACM Hum.-Comput. Interact.}}
  \bibinfo{volume}{3}, \bibinfo{number}{CSCW}, Article \bibinfo{articleno}{72}
  (\bibinfo{date}{nov} \bibinfo{year}{2019}), \bibinfo{numpages}{23}~pages.
\newblock
\urldef\tempurl%
\url{https://doi.org/10.1145/3359174}
\showDOI{\tempurl}


\bibitem[Muller et~al\mbox{.}(2019)]%
        {muller_HowDataScienceWorkersWorkWithData}
\bibfield{author}{\bibinfo{person}{Michael Muller}, \bibinfo{person}{Ingrid
  Lange}, \bibinfo{person}{Dakuo Wang}, \bibinfo{person}{David Piorkowski},
  \bibinfo{person}{Jason Tsay}, \bibinfo{person}{Q.~Vera Liao},
  \bibinfo{person}{Casey Dugan}, {and} \bibinfo{person}{Thomas Erickson}.}
  \bibinfo{year}{2019}\natexlab{}.
\newblock \showarticletitle{How Data Science Workers Work with Data: Discovery,
  Capture, Curation, Design, Creation}. In
  \bibinfo{booktitle}{\emph{Proceedings of the 2019 CHI Conference on Human
  Factors in Computing Systems}} (Glasgow, Scotland Uk)
  \emph{(\bibinfo{series}{CHI '19})}. \bibinfo{publisher}{Association for
  Computing Machinery}, \bibinfo{address}{New York, NY, USA},
  \bibinfo{pages}{1–15}.
\newblock
\showISBNx{9781450359702}
\urldef\tempurl%
\url{https://doi.org/10.1145/3290605.3300356}
\showDOI{\tempurl}


\bibitem[Ng et~al\mbox{.}(2020)]%
        {ng_IMTknowledgeDecomp}
\bibfield{author}{\bibinfo{person}{Felicia Ng}, \bibinfo{person}{Jina Suh},
  {and} \bibinfo{person}{Gonzalo Ramos}.} \bibinfo{year}{2020}\natexlab{}.
\newblock \showarticletitle{Understanding and Supporting Knowledge
  Decomposition for Machine Teaching}. In \bibinfo{booktitle}{\emph{Proceedings
  of the 2020 ACM Designing Interactive Systems Conference}} (Eindhoven,
  Netherlands) \emph{(\bibinfo{series}{DIS '20})}.
  \bibinfo{publisher}{Association for Computing Machinery},
  \bibinfo{address}{New York, NY, USA}, \bibinfo{pages}{1183–1194}.
\newblock
\showISBNx{9781450369749}
\urldef\tempurl%
\url{https://doi.org/10.1145/3357236.3395454}
\showDOI{\tempurl}


\bibitem[Passi and Barocas(2019)]%
        {passi_barocas_problemFormulationFairness}
\bibfield{author}{\bibinfo{person}{Samir Passi} {and} \bibinfo{person}{Solon
  Barocas}.} \bibinfo{year}{2019}\natexlab{}.
\newblock \showarticletitle{Problem Formulation and Fairness}. In
  \bibinfo{booktitle}{\emph{Proceedings of the Conference on Fairness,
  Accountability, and Transparency}} (Atlanta, GA, USA)
  \emph{(\bibinfo{series}{FAT* '19})}. \bibinfo{publisher}{Association for
  Computing Machinery}, \bibinfo{address}{New York, NY, USA},
  \bibinfo{pages}{39–48}.
\newblock
\showISBNx{9781450361255}
\urldef\tempurl%
\url{https://doi.org/10.1145/3287560.3287567}
\showDOI{\tempurl}


\bibitem[Passi and Jackson(2017)]%
        {passi_jackson_dataVision}
\bibfield{author}{\bibinfo{person}{Samir Passi} {and} \bibinfo{person}{Steven
  Jackson}.} \bibinfo{year}{2017}\natexlab{}.
\newblock \showarticletitle{Data Vision: Learning to See Through Algorithmic
  Abstraction}. In \bibinfo{booktitle}{\emph{Proceedings of the 2017 ACM
  Conference on Computer Supported Cooperative Work and Social Computing}}
  (Portland, Oregon, USA) \emph{(\bibinfo{series}{CSCW '17})}.
  \bibinfo{publisher}{Association for Computing Machinery},
  \bibinfo{address}{New York, NY, USA}, \bibinfo{pages}{2436–2447}.
\newblock
\showISBNx{9781450343350}
\urldef\tempurl%
\url{https://doi.org/10.1145/2998181.2998331}
\showDOI{\tempurl}


\bibitem[Passi and Jackson(2018)]%
        {passi_trustInDataScience}
\bibfield{author}{\bibinfo{person}{Samir Passi} {and}
  \bibinfo{person}{Steven~J. Jackson}.} \bibinfo{year}{2018}\natexlab{}.
\newblock \showarticletitle{Trust in Data Science: Collaboration, Translation,
  and Accountability in Corporate Data Science Projects}.
\newblock \bibinfo{journal}{\emph{Proc. ACM Hum.-Comput. Interact.}}
  \bibinfo{volume}{2}, \bibinfo{number}{CSCW}, Article \bibinfo{articleno}{136}
  (\bibinfo{date}{nov} \bibinfo{year}{2018}), \bibinfo{numpages}{28}~pages.
\newblock
\urldef\tempurl%
\url{https://doi.org/10.1145/3274405}
\showDOI{\tempurl}


\bibitem[Patel et~al\mbox{.}(2010)]%
        {patel2010gestalt}
\bibfield{author}{\bibinfo{person}{Kayur Patel}, \bibinfo{person}{Naomi
  Bancroft}, \bibinfo{person}{Steven~M. Drucker}, \bibinfo{person}{James
  Fogarty}, \bibinfo{person}{Amy~J. Ko}, {and} \bibinfo{person}{James Landay}.}
  \bibinfo{year}{2010}\natexlab{}.
\newblock \showarticletitle{Gestalt: Integrated Support for Implementation and
  Analysis in Machine Learning}. In \bibinfo{booktitle}{\emph{Proceedings of
  the 23nd Annual ACM Symposium on User Interface Software and Technology}}
  (New York, New York, USA) \emph{(\bibinfo{series}{UIST '10})}.
  \bibinfo{publisher}{Association for Computing Machinery},
  \bibinfo{address}{New York, NY, USA}, \bibinfo{pages}{37–46}.
\newblock
\showISBNx{9781450302715}
\urldef\tempurl%
\url{https://doi.org/10.1145/1866029.1866038}
\showDOI{\tempurl}


\bibitem[Pedregosa et~al\mbox{.}(2011)]%
        {scikit_learn}
\bibfield{author}{\bibinfo{person}{F. Pedregosa}, \bibinfo{person}{G.
  Varoquaux}, \bibinfo{person}{A. Gramfort}, \bibinfo{person}{V. Michel},
  \bibinfo{person}{B. Thirion}, \bibinfo{person}{O. Grisel},
  \bibinfo{person}{M. Blondel}, \bibinfo{person}{P. Prettenhofer},
  \bibinfo{person}{R. Weiss}, \bibinfo{person}{V. Dubourg}, \bibinfo{person}{J.
  Vanderplas}, \bibinfo{person}{A. Passos}, \bibinfo{person}{D. Cournapeau},
  \bibinfo{person}{M. Brucher}, \bibinfo{person}{M. Perrot}, {and}
  \bibinfo{person}{E. Duchesnay}.} \bibinfo{year}{2011}\natexlab{}.
\newblock \showarticletitle{Scikit-learn: Machine Learning in {P}ython}.
\newblock \bibinfo{journal}{\emph{Journal of Machine Learning Research}}
  \bibinfo{volume}{12} (\bibinfo{year}{2011}), \bibinfo{pages}{2825--2830}.
\newblock


\bibitem[Piorkowski et~al\mbox{.}(2021)]%
        {piorkowski_AIDevelopersCommChallenges}
\bibfield{author}{\bibinfo{person}{David Piorkowski}, \bibinfo{person}{Soya
  Park}, \bibinfo{person}{April~Yi Wang}, \bibinfo{person}{Dakuo Wang},
  \bibinfo{person}{Michael Muller}, {and} \bibinfo{person}{Felix Portnoy}.}
  \bibinfo{year}{2021}\natexlab{}.
\newblock \showarticletitle{How AI Developers Overcome Communication Challenges
  in a Multidisciplinary Team: A Case Study}.
\newblock \bibinfo{journal}{\emph{Proc. ACM Hum.-Comput. Interact.}}
  \bibinfo{volume}{5}, \bibinfo{number}{CSCW1}, Article
  \bibinfo{articleno}{131} (\bibinfo{date}{apr} \bibinfo{year}{2021}),
  \bibinfo{numpages}{25}~pages.
\newblock
\urldef\tempurl%
\url{https://doi.org/10.1145/3449205}
\showDOI{\tempurl}


\bibitem[Radford et~al\mbox{.}(2021)]%
        {CLIP}
\bibfield{author}{\bibinfo{person}{Alec Radford}, \bibinfo{person}{Jong~Wook
  Kim}, \bibinfo{person}{Chris Hallacy}, \bibinfo{person}{Aditya Ramesh},
  \bibinfo{person}{Gabriel Goh}, \bibinfo{person}{Sandhini Agarwal},
  \bibinfo{person}{Girish Sastry}, \bibinfo{person}{Amanda Askell},
  \bibinfo{person}{Pamela Mishkin}, \bibinfo{person}{Jack Clark},
  \bibinfo{person}{Gretchen Krueger}, {and} \bibinfo{person}{Ilya Sutskever}.}
  \bibinfo{year}{2021}\natexlab{}.
\newblock \bibinfo{title}{Learning Transferable Visual Models From Natural
  Language Supervision}.
\newblock
\newblock
\urldef\tempurl%
\url{https://doi.org/10.48550/ARXIV.2103.00020}
\showDOI{\tempurl}


\bibitem[Ramos et~al\mbox{.}(2020)]%
        {ramos2020IMT}
\bibfield{author}{\bibinfo{person}{Gonzalo Ramos}, \bibinfo{person}{Christopher
  Meek}, \bibinfo{person}{Patrice Simard}, \bibinfo{person}{Jina Suh}, {and}
  \bibinfo{person}{Soroush Ghorashi}.} \bibinfo{year}{2020}\natexlab{}.
\newblock \showarticletitle{Interactive machine teaching: a human-centered
  approach to building machine-learned models}.
\newblock \bibinfo{journal}{\emph{Human--Computer Interaction}}
  \bibinfo{volume}{35}, \bibinfo{number}{5-6} (\bibinfo{year}{2020}),
  \bibinfo{pages}{413--451}.
\newblock


\bibitem[Rombach et~al\mbox{.}(2022)]%
        {Rombach_2022_CVPR}
\bibfield{author}{\bibinfo{person}{Robin Rombach}, \bibinfo{person}{Andreas
  Blattmann}, \bibinfo{person}{Dominik Lorenz}, \bibinfo{person}{Patrick
  Esser}, {and} \bibinfo{person}{Bj\"orn Ommer}.}
  \bibinfo{year}{2022}\natexlab{}.
\newblock \showarticletitle{High-Resolution Image Synthesis With Latent
  Diffusion Models}. In \bibinfo{booktitle}{\emph{Proceedings of the IEEE/CVF
  Conference on Computer Vision and Pattern Recognition (CVPR)}}.
  \bibinfo{pages}{10684--10695}.
\newblock


\bibitem[Rule et~al\mbox{.}(2018)]%
        {rule_explorationExplanationCompNotebooks}
\bibfield{author}{\bibinfo{person}{Adam Rule}, \bibinfo{person}{Aur\'{e}lien
  Tabard}, {and} \bibinfo{person}{James~D. Hollan}.}
  \bibinfo{year}{2018}\natexlab{}.
\newblock \showarticletitle{Exploration and Explanation in Computational
  Notebooks}. In \bibinfo{booktitle}{\emph{Proceedings of the 2018 CHI
  Conference on Human Factors in Computing Systems}} (Montreal QC, Canada)
  \emph{(\bibinfo{series}{CHI '18})}. \bibinfo{publisher}{Association for
  Computing Machinery}, \bibinfo{address}{New York, NY, USA},
  \bibinfo{pages}{1–12}.
\newblock
\showISBNx{9781450356206}
\urldef\tempurl%
\url{https://doi.org/10.1145/3173574.3173606}
\showDOI{\tempurl}


\bibitem[Sambasivan et~al\mbox{.}(2021)]%
        {sambasivan_everyoneWantsToDoModelWork}
\bibfield{author}{\bibinfo{person}{Nithya Sambasivan}, \bibinfo{person}{Shivani
  Kapania}, \bibinfo{person}{Hannah Highfill}, \bibinfo{person}{Diana Akrong},
  \bibinfo{person}{Praveen Paritosh}, {and} \bibinfo{person}{Lora~M Aroyo}.}
  \bibinfo{year}{2021}\natexlab{}.
\newblock \showarticletitle{“Everyone Wants to Do the Model Work, Not the
  Data Work”: Data Cascades in High-Stakes AI}. In
  \bibinfo{booktitle}{\emph{Proceedings of the 2021 CHI Conference on Human
  Factors in Computing Systems}} (Yokohama, Japan) \emph{(\bibinfo{series}{CHI
  '21})}. \bibinfo{publisher}{Association for Computing Machinery},
  \bibinfo{address}{New York, NY, USA}, Article \bibinfo{articleno}{39},
  \bibinfo{numpages}{15}~pages.
\newblock
\showISBNx{9781450380966}
\urldef\tempurl%
\url{https://doi.org/10.1145/3411764.3445518}
\showDOI{\tempurl}


\bibitem[Sheng et~al\mbox{.}(2019)]%
        {sheng_biasesInLangGeneration}
\bibfield{author}{\bibinfo{person}{Emily Sheng}, \bibinfo{person}{Kai-Wei
  Chang}, \bibinfo{person}{Premkumar Natarajan}, {and} \bibinfo{person}{Nanyun
  Peng}.} \bibinfo{year}{2019}\natexlab{}.
\newblock \showarticletitle{The Woman Worked as a Babysitter: On Biases in
  Language Generation}. In \bibinfo{booktitle}{\emph{EMNLP/IJCNLP (1)}}.
  \bibinfo{pages}{3405--3410}.
\newblock
\urldef\tempurl%
\url{https://doi.org/10.18653/v1/D19-1339}
\showURL{%
\tempurl}


\bibitem[Simard et~al\mbox{.}(2017)]%
        {simard2017IMT}
\bibfield{author}{\bibinfo{person}{Patrice Simard}, \bibinfo{person}{Saleema
  Amershi}, \bibinfo{person}{Max Chickering}, \bibinfo{person}{Alicia
  Edelman~Pelton}, \bibinfo{person}{Soroush Ghorashi}, \bibinfo{person}{Chris
  Meek}, \bibinfo{person}{Gonzalo Ramos}, \bibinfo{person}{Jina Suh},
  \bibinfo{person}{Johan Verwey}, \bibinfo{person}{Mo Wang}, {and}
  \bibinfo{person}{John Wernsing}.} \bibinfo{year}{2017}\natexlab{}.
\newblock \bibinfo{booktitle}{\emph{Machine Teaching: A New Paradigm for
  Building Machine Learning Systems}}.
\newblock \bibinfo{type}{{T}echnical {R}eport} MSR-TR-2017-26.
\newblock
\urldef\tempurl%
\url{https://www.microsoft.com/en-us/research/publication/machine-teaching-new-paradigm-building-machine-learning-systems/}
\showURL{%
\tempurl}


\bibitem[Suwa and Tversky(1996)]%
        {suwa_tversky_whatArchitectsSeeInTheirSketches}
\bibfield{author}{\bibinfo{person}{Masaki Suwa} {and} \bibinfo{person}{Barbara
  Tversky}.} \bibinfo{year}{1996}\natexlab{}.
\newblock \showarticletitle{What Architects See in Their Sketches: Implications
  for Design Tools}. In \bibinfo{booktitle}{\emph{Conference Companion on Human
  Factors in Computing Systems}} (Vancouver, British Columbia, Canada)
  \emph{(\bibinfo{series}{CHI '96})}. \bibinfo{publisher}{Association for
  Computing Machinery}, \bibinfo{address}{New York, NY, USA},
  \bibinfo{pages}{191–192}.
\newblock
\showISBNx{0897918320}
\urldef\tempurl%
\url{https://doi.org/10.1145/257089.257255}
\showDOI{\tempurl}


\bibitem[Tohidi et~al\mbox{.}(2006)]%
        {tohidi_buxton_gettingTheRightDesign}
\bibfield{author}{\bibinfo{person}{Maryam Tohidi}, \bibinfo{person}{William
  Buxton}, \bibinfo{person}{Ronald Baecker}, {and} \bibinfo{person}{Abigail
  Sellen}.} \bibinfo{year}{2006}\natexlab{}.
\newblock \showarticletitle{Getting the Right Design and the Design Right}. In
  \bibinfo{booktitle}{\emph{Proceedings of the SIGCHI Conference on Human
  Factors in Computing Systems}} (Montr\'{e}al, Qu\'{e}bec, Canada)
  \emph{(\bibinfo{series}{CHI '06})}. \bibinfo{publisher}{Association for
  Computing Machinery}, \bibinfo{address}{New York, NY, USA},
  \bibinfo{pages}{1243–1252}.
\newblock
\showISBNx{1595933727}
\urldef\tempurl%
\url{https://doi.org/10.1145/1124772.1124960}
\showDOI{\tempurl}


\bibitem[von Werra(2021)]%
        {imdb_sota}
\bibfield{author}{\bibinfo{person}{Leandro von Werra}.}
  \bibinfo{year}{2021}\natexlab{}.
\newblock \bibinfo{title}{distilbert-imdb}.
\newblock
\newblock
\urldef\tempurl%
\url{https://huggingface.co/lvwerra/distilbert-imdb}
\showURL{%
\tempurl}


\bibitem[Walker et~al\mbox{.}(2002)]%
        {walker_hifiOrLofi}
\bibfield{author}{\bibinfo{person}{Miriam Walker}, \bibinfo{person}{Leila
  Takayama}, {and} \bibinfo{person}{James~A. Landay}.}
  \bibinfo{year}{2002}\natexlab{}.
\newblock \showarticletitle{High-Fidelity or Low-Fidelity, Paper or Computer?
  Choosing Attributes when Testing Web Prototypes}.
\newblock \bibinfo{journal}{\emph{Proceedings of the Human Factors and
  Ergonomics Society Annual Meeting}} \bibinfo{volume}{46}, \bibinfo{number}{5}
  (\bibinfo{year}{2002}), \bibinfo{pages}{661--665}.
\newblock
\urldef\tempurl%
\url{https://doi.org/10.1177/154193120204600513}
\showDOI{\tempurl}
\showeprint{https://doi.org/10.1177/154193120204600513}


\bibitem[Wang et~al\mbox{.}(2019)]%
        {wang2019human}
\bibfield{author}{\bibinfo{person}{Dakuo Wang}, \bibinfo{person}{Justin~D.
  Weisz}, \bibinfo{person}{Michael Muller}, \bibinfo{person}{Parikshit Ram},
  \bibinfo{person}{Werner Geyer}, \bibinfo{person}{Casey Dugan},
  \bibinfo{person}{Yla Tausczik}, \bibinfo{person}{Horst Samulowitz}, {and}
  \bibinfo{person}{Alexander Gray}.} \bibinfo{year}{2019}\natexlab{}.
\newblock \showarticletitle{Human-AI Collaboration in Data Science: Exploring
  Data Scientists' Perceptions of Automated AI}.
\newblock \bibinfo{journal}{\emph{Proc. ACM Hum.-Comput. Interact.}}
  \bibinfo{volume}{3}, \bibinfo{number}{CSCW}, Article \bibinfo{articleno}{211}
  (\bibinfo{date}{nov} \bibinfo{year}{2019}), \bibinfo{numpages}{24}~pages.
\newblock
\urldef\tempurl%
\url{https://doi.org/10.1145/3359313}
\showDOI{\tempurl}


\bibitem[Wu et~al\mbox{.}(2022)]%
        {wu_AI_chains}
\bibfield{author}{\bibinfo{person}{Tongshuang Wu}, \bibinfo{person}{Michael
  Terry}, {and} \bibinfo{person}{Carrie~Jun Cai}.}
  \bibinfo{year}{2022}\natexlab{}.
\newblock \showarticletitle{AI Chains: Transparent and Controllable Human-AI
  Interaction by Chaining Large Language Model Prompts}. In
  \bibinfo{booktitle}{\emph{Proceedings of the 2022 CHI Conference on Human
  Factors in Computing Systems}} (New Orleans, LA, USA)
  \emph{(\bibinfo{series}{CHI '22})}. \bibinfo{publisher}{Association for
  Computing Machinery}, \bibinfo{address}{New York, NY, USA}, Article
  \bibinfo{articleno}{385}, \bibinfo{numpages}{22}~pages.
\newblock
\showISBNx{9781450391573}
\urldef\tempurl%
\url{https://doi.org/10.1145/3491102.3517582}
\showDOI{\tempurl}


\bibitem[Wu et~al\mbox{.}(2019)]%
        {wu2019localDecisionPitfallsIML}
\bibfield{author}{\bibinfo{person}{Tongshuang Wu}, \bibinfo{person}{Daniel~S.
  Weld}, {and} \bibinfo{person}{Jeffrey Heer}.}
  \bibinfo{year}{2019}\natexlab{}.
\newblock \showarticletitle{Local Decision Pitfalls in Interactive Machine
  Learning: An Investigation into Feature Selection in Sentiment Analysis}.
\newblock \bibinfo{journal}{\emph{ACM Trans. Comput.-Hum. Interact.}}
  \bibinfo{volume}{26}, \bibinfo{number}{4}, Article \bibinfo{articleno}{24}
  (\bibinfo{date}{jun} \bibinfo{year}{2019}), \bibinfo{numpages}{27}~pages.
\newblock
\showISSN{1073-0516}
\urldef\tempurl%
\url{https://doi.org/10.1145/3319616}
\showDOI{\tempurl}


\bibitem[Yang(2005)]%
        {yang_aStudyOfPrototypes}
\bibfield{author}{\bibinfo{person}{Maria~C. Yang}.}
  \bibinfo{year}{2005}\natexlab{}.
\newblock \showarticletitle{A study of prototypes, design activity, and design
  outcome}.
\newblock \bibinfo{journal}{\emph{Design Studies}} \bibinfo{volume}{26},
  \bibinfo{number}{6} (\bibinfo{year}{2005}), \bibinfo{pages}{649--669}.
\newblock
\showISSN{0142-694X}
\urldef\tempurl%
\url{https://doi.org/10.1016/j.destud.2005.04.005}
\showDOI{\tempurl}


\bibitem[Yang et~al\mbox{.}(2019)]%
        {yang_sketchingNLP}
\bibfield{author}{\bibinfo{person}{Qian Yang}, \bibinfo{person}{Justin
  Cranshaw}, \bibinfo{person}{Saleema Amershi}, \bibinfo{person}{Shamsi~T.
  Iqbal}, {and} \bibinfo{person}{Jaime Teevan}.}
  \bibinfo{year}{2019}\natexlab{}.
\newblock \showarticletitle{Sketching NLP: A Case Study of Exploring the Right
  Things To Design with Language Intelligence}. In
  \bibinfo{booktitle}{\emph{Proceedings of the 2019 CHI Conference on Human
  Factors in Computing Systems}} (Glasgow, Scotland Uk)
  \emph{(\bibinfo{series}{CHI '19})}. \bibinfo{publisher}{Association for
  Computing Machinery}, \bibinfo{address}{New York, NY, USA},
  \bibinfo{pages}{1–12}.
\newblock
\showISBNx{9781450359702}
\urldef\tempurl%
\url{https://doi.org/10.1145/3290605.3300415}
\showDOI{\tempurl}


\bibitem[Yang et~al\mbox{.}(2020)]%
        {yang_reexaminingWhetherWhy}
\bibfield{author}{\bibinfo{person}{Qian Yang}, \bibinfo{person}{Aaron
  Steinfeld}, \bibinfo{person}{Carolyn Ros\'{e}}, {and} \bibinfo{person}{John
  Zimmerman}.} \bibinfo{year}{2020}\natexlab{}.
\newblock \showarticletitle{Re-Examining Whether, Why, and How Human-AI
  Interaction Is Uniquely Difficult to Design}. In
  \bibinfo{booktitle}{\emph{Proceedings of the 2020 CHI Conference on Human
  Factors in Computing Systems}} (Honolulu, HI, USA)
  \emph{(\bibinfo{series}{CHI '20})}. \bibinfo{publisher}{Association for
  Computing Machinery}, \bibinfo{address}{New York, NY, USA},
  \bibinfo{pages}{1–13}.
\newblock
\showISBNx{9781450367080}
\urldef\tempurl%
\url{https://doi.org/10.1145/3313831.3376301}
\showDOI{\tempurl}


\bibitem[Zhang et~al\mbox{.}(2020)]%
        {zhang_howDoDataScienceWorkersCollaborate}
\bibfield{author}{\bibinfo{person}{Amy~X. Zhang}, \bibinfo{person}{Michael
  Muller}, {and} \bibinfo{person}{Dakuo Wang}.}
  \bibinfo{year}{2020}\natexlab{}.
\newblock \showarticletitle{How Do Data Science Workers Collaborate? Roles,
  Workflows, and Tools}.
\newblock \bibinfo{journal}{\emph{Proc. ACM Hum.-Comput. Interact.}}
  \bibinfo{volume}{4}, \bibinfo{number}{CSCW1}, Article \bibinfo{articleno}{22}
  (\bibinfo{date}{may} \bibinfo{year}{2020}), \bibinfo{numpages}{23}~pages.
\newblock
\urldef\tempurl%
\url{https://doi.org/10.1145/3392826}
\showDOI{\tempurl}


\bibitem[Zhu(2020)]%
        {hatefulMemes_winner_alfredLab}
\bibfield{author}{\bibinfo{person}{Ron Zhu}.} \bibinfo{year}{2020}\natexlab{}.
\newblock \bibinfo{title}{Enhance Multimodal Transformer With External Label
  And In-Domain Pretrain: Hateful Meme Challenge Winning Solution}.
\newblock
\newblock
\urldef\tempurl%
\url{https://doi.org/10.48550/ARXIV.2012.08290}
\showDOI{\tempurl}


\end{thebibliography}

\appendix
\section{ModelSketchBook Process Overview}
\label{appendix:msb-process-overview}
\revFinal{Here, we summarize the three commands that a user executes to set up their first sketch model in a computational notebook. 
First, the user sets up their sketchbook, as previously shown in Section~\ref{section:sys-sketchbook-setup}. They only need to perform this operation once to load their data, and they can create as many concepts and sketches as they would like using this single sketchbook.}
\begin{lstlisting}[language=iPython]
    import model_sketch_book as msb

    # Set up the sketchbook
    sb = msb.create_model_sketchbook(
        goal="Detect hateful memes on social media.",
        datasets={  # Pass in datasets to use
            "train": df_train,
            "test": df_test,
        },
        schema = {  # Specify the types of data fields
            "text": InputType.Text,
            "img_url": InputType.Image,
            "overall_rating": InputType.GroundTruth,
        },
    )
\end{lstlisting}

\revFinal{Next, they can create as many concepts as they would like using calls to the \texttt{create\_concept\_model()} function as shown below. The function parameters can be equivalently specified using UI widgets, and the function returns a cell output with a dataframe visualization of the concept results (Figure~\ref{fig:sys_api_concepts}).}
\begin{lstlisting}[language=iPython]
    msb.create_concept_model(
        input_field="post_image",  # Column for input
        concept_term="cartoon",    # Concept description
    )
\end{lstlisting}

\revFinal{Finally, the user can create as many sketches as they would like by making calls to \texttt{create\_sketch\_model()} as shown below; again, the user can alternatively specify the same function parameters using widgets, and the function displays a cell output with a dataframe visualization of the sketch model results (Figure~\ref{fig:sys_api_sketches}).}
\begin{lstlisting}[language=iPython]
    msb.create_sketch_model(
        concepts=["cartoon", "rant"],  # To aggregate
        sketch_id="sketch_version_1",  # Sketch name
    )
\end{lstlisting}

\section{Model Sketching Case Studies}
Table~\ref{table:case_studies} summarizes the task and dataset for each example model sketching application and includes concepts and sample sketches.

\begin{table*}[!tb]
  \centering
  \footnotesize
    \begin{tabular}{p{0.08\textwidth}|p{0.2\textwidth}|p{0.2\textwidth}|p{0.2\textwidth}|p{0.2\textwidth}}
    \toprule
    \textbf{ } &  \textbf{Travel planning} &
    \textbf{Creative inspiration} & \textbf{Political candidate research} & \textbf{Reviewer bias audit}\\
    \midrule
    \textbf{Task} 
    & {Assist travelers with selecting an Airbnb for a trip to New York City.}
    & {Help users to uncover creative connections across artists and artworks.} 
    & {Assist voters with finding a candidate for the 2022 California gubernatorial election who aligns with them on issues that matter most.}
    & {Assist a food reviewer in checking their own biases in reviewing cafes and restaurants on social media.} \\[0.4cm]

    \textbf{Data} 
    & {Airbnb listings in NYC. Includes the name, image, description, and neighborhood overview for each listing.}
    & {Museum artwork pieces. Includes the artist’s name, title, type, style, genre, media, geographic region, image and text description for each artwork.} 
    & {Candidate information aggregated from political statements and speeches by CalMatters, a nonprofit, nonpartisan, third-party group.}
    & {Instagram food reviews. Includes name of the cafe/restaurant, post date, post text, number of likes, number of comments, and overall rating.} \\[0.2cm]

    \textbf{Concepts} 
    & {
        Trendy, Central, Clean,
        Dingy, Tidy, Designed well, Designer, Bright, Outdated, Sunlight natural, Spacious, Ugly,
        Things to do, Activities, Good vibes
    }
    & {
        Colorful, Surreal, Fantasy, Abstract, Realist, Politics themed, Revolution themed, Labor movement themed, Social justice themed, Race themed, Gender themed, Feminist, Queer themed
    } 
    & {
        Environment, Equality, Gun control, Reproductive rights, Voting rights
    }
    & {
        Tasty, Affordable, Creative, Nice vibes, Good service, High quality, Good presentation, Yummy, Cheap, Original, Decent vibes, Sound service
    } \\[0.8cm]

    \textbf{Sample Sketches} 
    & {
        \textit{Aesthetic}: \{Clean, Bright, Designer\}
    }
    & {
        \textit{Bold}: \{Colorful, Surreal, Fantasy, Abstract\}
    } 
    & {
        \textit{High-level issues}: \{Climate change, Gun control\}
    }
    & {
        \textit{Naive concepts}: \{Tasty, High quality, Affordable, Creative, Nice vibes, Good service, Good presentation\}
    } \\[0.2cm]

    \textbf{} 
    & {
        \textit{Fun neighborhood}: \{Good vibes, Central, Trendy\}
    }
    & {
        \textit{Social justice}: \{Social justice themed, Race themed, Feminist, Queer themed\}
    } 
    & {
        \textit{Issues \& opinion}: \{Eco-friendly, Anti gun, Equal rights\}
    }
    & {
        \textit{Refined concepts}: \{Yummy, High quality, Cheap, Original\}
    } \\[0.5cm]

    \textbf{} 
    & {
        \textit{Modern beauty}: \{Outdated, Ugly, Clean\}
    }
    & {
        \textit{Political movements}: \{Politics themed, Revolution themed, Labor movement themed\}
    } 
    & {
        \textit{Issues \& detailed opinion}: \{Eco-friendly, Anti gun, Pro choice, Equality, Voting Rights\}
    }
    & {
        \textit{Refined concepts, extended}: \{Yummy, High quality, Cheap, Original, Decent vibes, Sound service, Good presentation\}
    } \\
    \bottomrule
    \end{tabular}
    \caption{
        A summary of several model sketching applications. We explored a task that assisted users in selecting an Airbnb for travel planning, a task aimed at surfacing creative connections between artists and artworks, a task designed to assist voters in discovering their alignment with political candidates, and a task that allowed a food reviewer to investigate the factors they value most and potential biases they might hold.
    }
    \label{table:case_studies}
\end{table*}

\section{Qualitative analysis}
\label{section:qualitative_analysis}
For the free-text written responses, we sought to summarize the high-level themes that emerged from our participants, so the codes were not the product, but our process~\cite{mcDonald2019IRRQualitativeResearch}. The first author conducted an inductive analysis to summarize participants' model authoring experiences. The first step of this process was to read through all written responses multiple times. Then, the qualitative open coding~\cite{charmaz2006constructing} process was iterative and took place in two phases: the first phase involved line-by-line response coding to closely reflect the original data (e.g., ``noticed minimal representation of violence in the dataset,'' ``plans to use a transformer model,'' or ``there was inconsistency among the labels''). Then, the second phase synthesized codes from the first phase into higher level themes (e.g., ``data representativity'' or ``labeling methodology''). After this theme generation, written responses were coded based on the themes to characterize participants' planned modeling approaches and their learnings and takeaways from the task.

\section{Pre- and Post-task Written Response Questions}
\label{section:appendix_task_questions}
\subsection{Pre-task}
Participants were asked to provide written responses for the following questions prior to the main model sketching task.

\begin{enumerate}
    \item (Before tutorial) \textbf{Describe your modeling approach.} Please write a few sentences describing your ideas on how you might build a model to detect posts that should be removed from a social media platform.
    \begin{itemize}
      \item Approach:
      \item Time estimate (how long do you think it would take to complete this planned approach?):
    \end{itemize}
    \item (After tutorial) \textbf{Initial concept brainstorming.} Please list 3 concepts that you think may be relevant to your task.
\end{enumerate}

\subsection{Post-task}
Participants were asked to provide written responses for the following questions immediately after the main task.

\begin{enumerate}
  \item \textbf{Describe your modeling approach (again).}
After going through today's session, please write a few sentences describing how you would plan to build a model that detects posts that should be removed from a social media platform. You may assume you have access to a tool like ModelSketchBook, but you are not required to use such a tool.
  \item \textbf{Learnings.}
Based on your explorations today, is there anything new that you've learned about the task, the data, or your modeling goals? Are there any gaps you noticed with the data or the task formulation? Are there any changes you'd like to make to address these gaps?
  \item \textbf{Choose your favorite sketch.}
Please run the cell directly below this to see all of the sketches you authored. Then, in the cell below that, please enter the sketch ID of the sketch that you feel performed the best (or most closely matched your modeling goals). Finally, run the last cell of the notebook.
\end{enumerate}

\section{Post-task Survey Questions}
\label{section:appendix_survey_questions}
The following are the questions that participants were asked to respond to in a survey after the main task.
\begin{enumerate}
    \item How \textbf{satisfied} are you with the \textbf{concept models} that you created using the API? (single-select options: Very dissatisfied, Dissatisfied, Somewhat dissatisfied, Neither dissatisfied nor satisfied, Somewhat satisfied, Satisfied, Very satisfied)
    \item How \textbf{satisfied} are you with the \textbf{sketch models} that you created using the API? (single-select options: Very dissatisfied, Dissatisfied, Somewhat dissatisfied, Neither dissatisfied nor satisfied, Somewhat satisfied, Satisfied, Very satisfied)
    \item To what extent do you feel that your approach to the modeling task \textbf{changed} as you experimented with and iterated upon your concepts and sketches? (single-select options: No changes, A few changes, Some changes, Many changes)
    \item To what extent do you feel that your approach to the modeling task \textbf{improved} as you experimented with and iterated upon your concepts and sketches? (No improvement, Slight improvement, Noticeable improvement, Significant improvement)
    \item How would you rate your \textbf{comfort} in using tools like Scikit-learn, PyTorch, or TensorFlow to author ML models? (Very uncomfortable, Uncomfortable, Somewhat uncomfortable, Neither uncomfortable nor comfortable, Somewhat comfortable, Comfortable, Very comfortable)
    \item How would you rate your \textbf{comfort} in using the ModelSketchBook API? (Very uncomfortable, Uncomfortable, Somewhat uncomfortable, Neither uncomfortable nor comfortable, Somewhat comfortable, Comfortable, Very comfortable)
    \item How would you rate your ability to express your voice/intent using tools like Scikit-learn, PyTorch, or TensorFlow to author ML models? (Very inexpressive, Inexpressive, Somewhat inexpressive, Neither inexpressive nor expressive, Somewhat expressive, Expressive, Very expressive)
    \item How would you rate your ability to express your voice/intent when using the ModelSketchBook API? (Very inexpressive, Inexpressive, Somewhat inexpressive, Neither inexpressive nor expressive, Somewhat expressive, Expressive, Very expressive)
    \item What did you \textbf{like} about your model authoring experience with the ModelSketchBook API? Please describe the positives of this API.
    \item What did you \textbf{not like} about your model authoring experience with the ModelSketchBook API? Please describe the negatives of this API or any issues you experienced.
    \item (Optional) Feel free to write any other comments, feedback, or thoughts you may have.
\end{enumerate}

\section{Post-task Interview Questions}
\label{section:appendix_interview_questions}
The following are the questions that we asked participants in an interview after the main task. We also asked follow-up questions as necessary.
\begin{enumerate}
    \item \textbf{Thought process.} Could you walk me through your notebook and explain your thought process as you designed your concepts and sketch models?
    \item \textbf{Contrast.} How did this model authoring process differ from your prior experience authoring other models?
    \begin{itemize}
        \item What did you like most about it?
        \item What did you like least about it?
    \end{itemize}
    \item \textbf{Learnings.} What did you learn from this model sketching process? Do you have any takeaways or perspective changes?
\end{enumerate}

\section{Participant demographics}
\label{section:participant_dem}
We had 3 women, 12 men, 1 non-binary participant, and 1 participant who preferred not to share their gender; we had 14 Asian participants, 2 White participants, and 1 participant who identified as White and Hispanic. We had 9 participants aged 18-24 and 8 aged 25-34. For highest educational attainment, 9 participants had earned a doctorate, master’s or other professional degree, 4 had earned a four-year degree, 3 had completed some college education, and 1 had graduated from high school. There were 9 participants who identified as ML engineers or data scientists working in industry, and there were 8 participants who identified as students who use ML in their work or research. Of the industry practitioners, 3 were employees of startup companies, and 6 were employees of large companies (with 1000+ employees). Participants reported an average of 3.6 years of experience working with machine learning.

\section{Modeling gaps}
\label{section:appendix_modeling_gaps}
\rev{Table~\ref{table:modeling_gaps} summarizes the main categories of modeling gaps that participants identified along with direct participant quotes.}

\begin{table*}[!htb]
  \centering
  \footnotesize
    \begin{tabular}{p{0.075\textwidth}p{0.35\textwidth} p{0.5\textwidth}}
    \toprule
    \textbf{Category} & \textbf{Learning or Takeaway} & \textbf{Quotes}\\
    \midrule
    \multirow{3}{0.075\textwidth}{Task} &
    {Text overall seems more important than images for this task} &
    {
        ``The image could be detached from the text and the text seemed a bit more useful in this exercise.'' (P7)
    }\\[0.2cm]

    &
    {Depending on the case, it may be important to combine the modalities or focus on text or image} &
    {
        ``How text and images interact together is a super important takeaway. You can’t just think about the images alone; you can’t just think about the text alone, it’s together the effect that they create.'' (P16),
        ``One counter to [the greater importance of text] might be images of gore, where text is not necessarily needed for the post to be problematic.'' (P2)
    }\\[0.2cm]

    &
    {Memes are tricky to understand and label 
    } &
    {
        ``Memes are tricky in that the images may be benign but the text may be harmful.'' (P1),
        ``Memes are very vague and tricky to label. The subtext is hard to capture with simple concepts.'' (P6)
    }\\
    \midrule
    \multirow{2}{0.075\textwidth}{Dataset} &
    {There is a need for greater representativity of kinds of harm in the data \rev{(homophobia, transphobia, nudity)}} &
    {
        ``Need more data [to capture] problematic posts of all kinds (not just racism or sexism) [...] Not enough data for homophobia, transphobia etc.'' (P1),
        ``[...] in this case nudity wasn't an issue, but might be for test time.'' (P16)
    }\\[0.2cm]

    &
    {The dataset is may be too unrealistic and contrived to match the online distribution of posts
    } &
    {
        ``The data seems to be far off from what is normally posted online, so the data definitely seems not good for general use.'' (P8)
    }\\[0.2cm]

    \midrule
    \multirow{3}{0.075\textwidth}{Labels} &
    {Class imbalance has a large impact on modeling and makes it difficult to calibrate a model} &
    {
        ``I learned that class imbalance plays a large impact on modeling---it seemed for example that my own labeled dataset was relatively weighted towards not removing posts, so it paid dividends to [try] more specific targeted terms towards things that I found truly offensive like advocating for violence.'' (P5)
    }\\[0.2cm]

    &
    {Their own labels appeared to be internally inconsistent} &
    {
        ``My labels are obviously not very internally consistent either and collecting from larger samples may help to even out the noise.'' (P7)
    }\\[0.2cm]

    &
    {It would be valuable to increase the removal-score granularity beyond binarization} &
    {
        ``One area of improvement could be to increase the task fidelity to actually use the 5 labels instead of binarizing them.'' (P3)
    }\\
    \bottomrule
    \end{tabular}
    \caption{After performing a model sketching task, participants noted takeaways for their modeling approach and identified gaps in the dataset and labeling methodology. 
    }
    \label{table:modeling_gaps}
\end{table*}

\section{Participants' thought process}
\label{section:participant_thought_processes}
\rev{Participants perceived a change in their mode of thinking when they engaged in model sketching. 
Many participants felt that the approach freed them from the tedious tasks that they typically had to work on. P10 described model sketching as ``a different way of thinking about [modeling]'' compared to their day-to-day modeling work, which was ``a lot more manual and technical and a lot more work and a lot more complex,'' so they liked how model sketching ``just does it for you.'' Similarly, P5 expressed that model sketching ``outsourced a lot of the most annoying parts of ML, like creating pipelines to generate features.''

Some participants felt that this approach allowed them to take a more \textit{long-range view} of the modeling process: P1 stated that ``this process helps put into perspective the long term goals and the structure of the ML lifecycle.'' 
Others described the model sketching process as refreshingly \textit{creative} and \textit{artistic}, while their typical model authoring experiences had felt more formulaic. As P12 put it, ``[model sketching] forced me to think creatively in a way I don't normally have to. Normally, my work is pretty boilerplate-y and repetitive. In this case, the model's performance was much more limited by my own creativity and ability, so it felt more like I had artistic direction in how the model did.'' 

Participants also enjoyed the interpretability of model sketching, which afforded them a greater understanding of model behavior compared to past model authoring experiences. P14 shared that ``usually a lot of what I do is just a black box [...] and it's frustrating when it doesn't work,'' but that ``having this other option of trying to build from the bottom up is a pretty good alternative.'' 
This modeling approach also allowed participants to think about decision-making from first principles. P16 described that ``in terms of reasoning at a human level, I really like this idea of breaking it down into smaller concepts. It seems like an intuitive way you’d teach a kid... to me that just seems to be building a model that’s more explainable.''
}

\section{Participant experience outcomes}
\label{section:participant_experience}
Addressing RQ2 in terms of benefits to ML practitioners, we observed that participants found the model sketching approach to be comfortable and expressive. Participants shared a great deal of excitement about this method of model development, expressing that they ``enjoyed the experience of thinking about models in this manner'' (P1) and that the tool ``could be really helpful in the future for similar tasks'' (P11) or for ``showing that something was possible and expanding imaginations'' (P12). 

\subsection{Comparison with existing authoring tools}
\begin{figure*}[!tb]
  \includegraphics[width=1.0\textwidth]{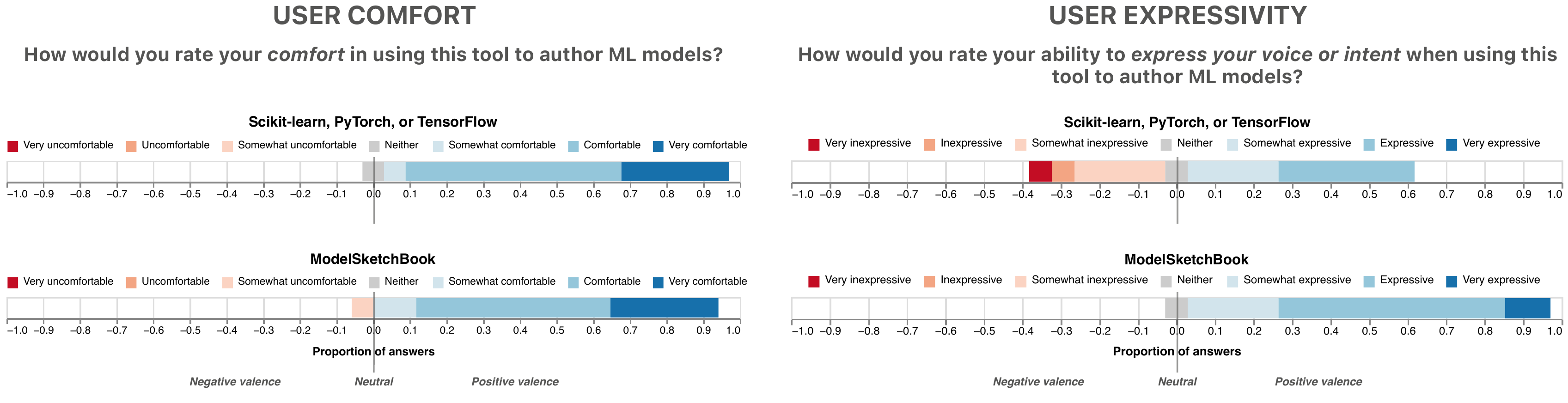}
  \caption{Upon their first-time use of the {ModelSketchBook} API, participants indicated similar levels of comfort as they feel with traditional model authoring tools (left), and they indicated higher levels of expressivity using {ModelSketchBook} (right).
  }
  \label{fig:survey_comfort_expressivity}
  \Description{Four diverging Likert bars with neutrals split; two of which show user comfort and the other two showing user expressivity with traditional ML modeling tools and ModelSketchBook.}
\end{figure*}

Comparing to their experiences with more common model authoring tools like Scikit-learn, PyTorch, and TensorFlow, participants felt similar levels of comfort with the {ModelSketchBook} API~(Figure~\ref{fig:survey_comfort_expressivity}) even though this 30-minute task was their first exposure to our tool, and all participants had significant experience with model authoring using these more traditional tools. At the same time, we saw that participants reported higher levels of expressivity when using the {ModelSketchBook} API than when using those same common model authoring tools. These results present promising support that our model sketching approach can be successfully adopted by ML practitioners as a lightweight, expressive entrypoint to model authoring.

\subsection{Reflections on the model sketching experience}
In our survey, we also probed users' levels of satisfaction with their concepts and sketch models and asked them about the extent to which they felt their modeling approach changed or improved over time. We observed that participants were highly satisfied with both their concepts and sketch models. All but two participants rated their satisfaction with both models as ``somewhat satisfied,'' ``satisfied,'' or ``very satisfied.'' All but one participant responded that their modeling approach changed as they iterated on concepts and sketches, and the vast majority of participants (14 of 17) described a ``slight improvement,'' ``noticeable improvement,'' or ``significant improvement'' in their modeling approach.


In their interview feedback and survey free-responses, participants expressed enthusiasm about the model sketching approach~(Table~\ref{table:summary_participant_experience}). 
The most frequently-cited benefits of model sketching were that it was highly interpretable and explainable, easy to use (especially because it required no code), provided fast feedback loops, and prompted them to take on a different, high-level mode of thought. Meanwhile, the main limitations that participants noted were that the notebook environment often became disorganized, that our example-centric (rather than metric-centric) visualizations required more time to interpret, and that they would appreciate more assistance on brainstorming concepts. Multiple participants were excited about the opportunity to allow non-technical users and domain experts who lack ML knowledge to use ModelSketchBook and leverage their domain-specific expertise. They also suggested that we go beyond our notebook prototype to set up a dedicated user interface, and they suggested helper tools that would address potential issues with bias and overfitting in concepts.

\begin{table*}
  \centering
  \footnotesize
    \begin{tabular}{p{0.1\textwidth} p{0.25\textwidth} p{0.6\textwidth}}
    \toprule
    \textbf{Category} & \textbf{Theme} & \textbf{Quotes}\\
    \midrule
    \multirow{4}{0.1\textwidth}{Likes} &
    {Easy to use and get started} &
    {``...the thing I liked the most was how quickly I was able to get started, like I ran a few lines of code and immediately I was getting some results. So I think ModelSketchBook really makes machine learning more accessible which is really important.'' (P9),
    ``I like that I didn't have to write any code for the task.'' (P5),
    ``It's easy to get started and less prone to error.'' (P7),
    ``This provides a really solid baseline without writing a single line of code.'' (P3)
    }\\[0.1cm]
    
    & 
    {Interpretable, explainable} &
    {
        ``Very interactive and also very interpretable which is something with ML models I usually work with is not the case. So this is super super helpful. It's like a different way of thinking about it.'' (P10),
        ``In terms of interpretability of the decision and explainability, which is also important in this kind of task, it [ModelSketchBook] is very helpful.'' (P4),
        ``I like the semantic idea of a concept because it really puts model interpretability at the forefront and I think from a lot of experience, sometimes you try a bunch of models and you’re not really thinking about them as semantic concepts from the get-go.'' (P16)
    }\\[0.1cm]

    & 
    {Fast feedback loops} &
    {
        ``This modeling process had a lot more quick feedback loops and iteration of feature exploration and then turning that into models pretty fast, which is a good thing for sure.'' (P5),
        ``In my previous models at work, we do ad targeting, so it's very hard to get this kind of real-time feedback.'' (P6)
    }\\[0.1cm]

    & 
    {Different, high-level way of thinking} &
    {
        ``I liked the idea to approach things from how a human will think about things by defining these concepts.'' (P16),
        ``...focused around ideas and thinking about the task rather than hyperparameters or technical ML details.'' (P10),
        ``This process helps put into perspective the long term goals and the structure of the ML lifecycle and the pipeline that you need to create: what kind of data you need, which we prioritize, where textual data can provide more insight than image data and vice versa.'' (P1)
    }\\
    \midrule
    \multirow{3}{0.1\textwidth}{Dislikes} &
    {Notebook interface felt messy at times} &
    {``a bit annoying to copy paste some of the cells.'' (P5)}\\[0.1cm]
    
    & 
    {Takes time to sift through examples to gauge performance} &
    {``hard to go through each example because it's long.'' (P6),
    ``it would be nice to not have all the examples pop up every time just to gauge performance.'' (P16)
    }\\[0.1cm]

    & 
    {Needed more guidance to brainstorm concepts} &
    {
        ``it is hard to make accurate concepts [...] that is more of an issue with choosing relevant features by hand.'' (P9),
        ``Not much experience/guidance on how to come up with concepts creatively.'' (P10),
        ``I wasn't sure how to improve the model.'' (P12)
    }\\
    \midrule
    \multirow{4}{0.1\textwidth}{Suggestions} &
    {Building a dedicated UI} &
    {
        ``I think taking this out of IPython and having a more native UI would be better.'' (P3),
        ``I think having a GUI interface would be nice.'' (P16)
    }\\[0.1cm]
    
    & 
    {A version tailored for non-technical users and domain experts} &
    {
        ``I think the API is great for model developers who do not have a formal background with ML or deep learning. This API is a super useful tool for folks for product engineers and product managers.'' (P3).
        ``I feel like where this could be really relevant and useful is for folks that don't know much about ML, but know a lot about the task at hand, and for them to get something working in their hands.'' (P12)
    }\\[0.1cm]

    & 
    {Providing tools to address overfitting and bias} &
    {
        ``It's easy to fall into the trap of just trying to optimize your performance on the examples you're seeing and then ... forget what the rest of the data looks like.'' (P8),
        ``Human intervention in designing concepts can be biased.'' (P4),
        ``Perhaps I would miss edge cases.'' (P12)
    }\\[0.1cm]

    & 
    {Allowing for more manual control of weights and model types} &
    {
        ``I think a lot of the sketches could be improved by adding more weights to a concept.'' (P13),
        ``Lacking in features a bit (nonlinear models, more types of signal).'' (P12)
    }\\[0.1cm]
    \bottomrule
    \end{tabular}
    \caption{
    In written and interview responses, participants shared that they liked the process and cognitive mode that model sketching unlocked. They felt that the tool could be improved with a dedicated UI and scaffolding to assist concept brainstorming.
    }
    \label{table:summary_participant_experience}
\end{table*}

\section{SOTA model evaluation} \label{section:sota_details}
\subsection{Benchmark Dataset and Task Details}
For the movie review sentiment task, we gathered our data from the IMDB Movie Reviews dataset~\cite{imdb_dataset}, a binary sentiment analysis dataset. Each of the examples in this dataset is a review drawn from the Internet Movie Database (IMDB) that has been labeled as positive or negative sentiment. This dataset is balanced between positive and negative labels, and it was curated to only contain polarizing reviews with negative reviews of 4 or below and positive reviews of 7 or above on a 10-point rating scale. The SOTA model against which we compare our sketch models is \texttt{distilbert-imdb}~\cite{imdb_sota}, which reported 0.928 accuracy on this binary classification task.

For the comment toxicity task, we used the dataset from \citet{kumar2021designing} that gathered diverse perspectives on the toxicity of comments drawn from Reddit, Twitter, and 4chan. The dataset features ratings from 17,280 U.S. survey participants. Each item consists of a comment and a particular participant's toxicity rating for that comment, using a 5-point rating scale from ``Not at all toxic'' to ``Extremely toxic.'' 
This dataset consists of over 100,000 examples and is skewed toward lower-toxicity ratings. A key focus of the \citeauthor{kumar2021designing} work was that there are diverse perspectives on what constitutes toxic content, which drives substantial disagreement among raters in the dataset and results in a challenging toxicity prediction task.
We formulated this task to predict the aggregate rating for each item across annotators since the sketch model was designed to predict at the item level rather than the (item, annotator) level. Thus, the SOTA model for this task is the aggregated model variant from \citet{gordon_juryLearning}, which reported 0.90 MAE.

Finally, for the Hateful Memes task, we drew from the full training set from \citet{kiela_hatefulMemes}, described in Section~\ref{section:hateful_memes_data}.
In contrast to the dataset sample we used in the study, here we did not filter to the examples with ``non-hateful'' labels, and we used the centralized ``hateful meme'' definition provided by Meta AI rather than building on participants' own preferred definitions.
The SOTA model for this task was drawn from first place solution in the Hateful Memes Challenge~\cite{hatefulMemes_winner_alfredLab}, which reported 0.845 AUROC, and we also compare against the 0.714 AUROC SOTA reported by Meta AI at the time of the dataset release.

\subsection{Sketch Model Details}
For each of the task datasets, we randomly sampled up to 500 examples to train our sketch models. In comparison, the SOTA models for these tasks were trained on 25,000 examples for the IMDB task, 100,000 for the comment toxicity, and 8,500 examples for the Hateful Memes task. 
Then, for each task, a member of our team iteratively created sketch models based solely on the sampled training set, attempting to create the most performant sketch models possible based on the training data. 

\revFinal{%
For the movie review sentiment task, this process resulted in a final sketch model with 16 concepts: enjoyable, boring, mediocre, engaging, overrated, underwhelming, award winning, time waste, hate, love, positive sentiment, negative sentiment, sarcasm, praise, compliment, and mocking. Since this dataset consists only of movie review text, all of the concepts were text concepts. This sketch model used linear regression to aggregate the specified concepts.
Then, for the comment toxicity task, there were 14 concepts in the final sketch model: racism, sexism, homophobic, transphobic, ableist, religious discrimination, toxicity, hurtful, threat, misinformation, trolling, xenophobia, kind, and harmless. Again, since all examples in this dataset were text comments, all of the concepts were text concepts. This sketch model also used linear regression to aggregate the concepts.
Lastly, for the Hateful Memes task, there were 7 concepts in the final sketch model: racist, hate speech, vulgar, harmless, religious, sexist, and violence. All of these were text concepts based on the meme text; in line with our study participants, we found that image concepts did not work as well as text concepts for this task. This sketch model used a multi-layer perceptron (MLP) classifier with two hidden layers to aggregate the selected concepts.}

Finally, to evaluate sketch model performance, we used a random sample of 100 examples from the test set (preserving the same label distribution as the full test set). We applied the trained sketch models to this test set to generate sketch model predictions. We then compared these predictions against the ground truth labels to calculate the performance metric matching the headline performance metric reported by each of the tasks' SOTA models.


\end{document}